\documentclass[11pt]{article}

\makeatletter

\usepackage[
hmargin=20mm,vmargin=25mm]{geometry}

\usepackage[mac]{inputenc}
\usepackage[english]{babel}
\usepackage{mdwlist}
\usepackage{graphicx}
\usepackage{hyperref}
\usepackage{appendix}
\usepackage{dsfont}
\usepackage{tikz}
\usepackage{amsmath}
\usepackage{amsbsy}
\usepackage{amsfonts}
\usepackage{amssymb}
\usepackage{amscd}
\usepackage{amsthm}
\usepackage{subfigure}

\newcounter{hypGCounter}

\newcounter{hypSCounter}

\numberwithin{equation}{section}

\newcounter{zut}
\providecommand{\R}{\mathbb{R}}
\providecommand{\RR}{\mathbb{R}}

\providecommand{\1}{\mathds{1}}

\providecommand{\N}{\mathbb{N}}

\newcommand{\bTheta}{\ensuremath{\boldsymbol{\Theta}} }
\newcommand{\btheta}{\ensuremath{\boldsymbol{\theta}} }

\DeclareMathOperator*{\argmin}{\mbox{argmin}}

\makeatother

\begin{document}

\title{Fr残het means of curves for signal averaging and application to ECG data analysis}

\author{J\'er\'emie Bigot  \vspace{0.2cm} \\
Institut de Math\'ematiques de Toulouse \\
Universit\'e de Toulouse et CNRS (UMR 5219) \\
{\small {\tt Jeremie.Bigot@math.univ-toulouse.fr} }}
\maketitle

\thispagestyle{empty}

\begin{abstract}
Signal averaging is  the process that consists in computing a mean shape from a set of noisy signals. In the presence of  geometric variability in time in the data,  the usual Euclidean mean of the raw data yields a mean pattern  that does not reflect the typical shape of the observed signals. In this setting, it is  necessary  to use alignment techniques for a precise synchronization of the signals, and then to average the aligned data to obtain a consistent mean shape. In this paper, we study the numerical performances of  Fr残het means of curves  which are extensions of the usual Euclidean mean to   spaces  endowed with non-Euclidean metrics. This yields a new algorithm for signal averaging without a reference template. We apply this approach to the  estimation of a mean heart cycle from ECG records. 
\end{abstract}

\noindent \emph{Keywords:} Signal averaging; Mean shape; Fr\'echet means;   Curve registration;  Geometric variability; Deformable models; ECG data.

\noindent\emph{AMS classifications:} Primary 62G08; secondary 62P10.

\bigskip

\noindent{\bf Acknowledgements -}   The author acknowledges the support of the French Agence Nationale de la Recherche (ANR) under reference ANR-JCJC-SIMI1 DEMOS. 

\section{Introduction}

In many applications (biology, medicine, road traffic data) one observes a set of $J$ signals that have a similar shape. This may lead to the assumption that such observations are random elements which vary around the same but unknown  mean shape. Signal averaging is then the process that consists in computing a mean curve which reflects the typical shape of the observed signals. This procedure generally amounts to find an appropriate combination of the data to compute an average shape with a better signal-to-noise ratio. In many situations, the observed signals exhibit not only a classical source of random variation in amplitude, but also a less standard geometric source of variability in time. Due to this source of geometric variability, the usual Euclidean mean of the raw data may yield a mean curve  that does not reflect the typical shape of the signals, as illustrated by the following application.

\subsection{Signal averaging in ECG data analysis}

\begin{figure}[htbp]
\centering
\subfigure[]{ \includegraphics[width=15cm,height=5cm]{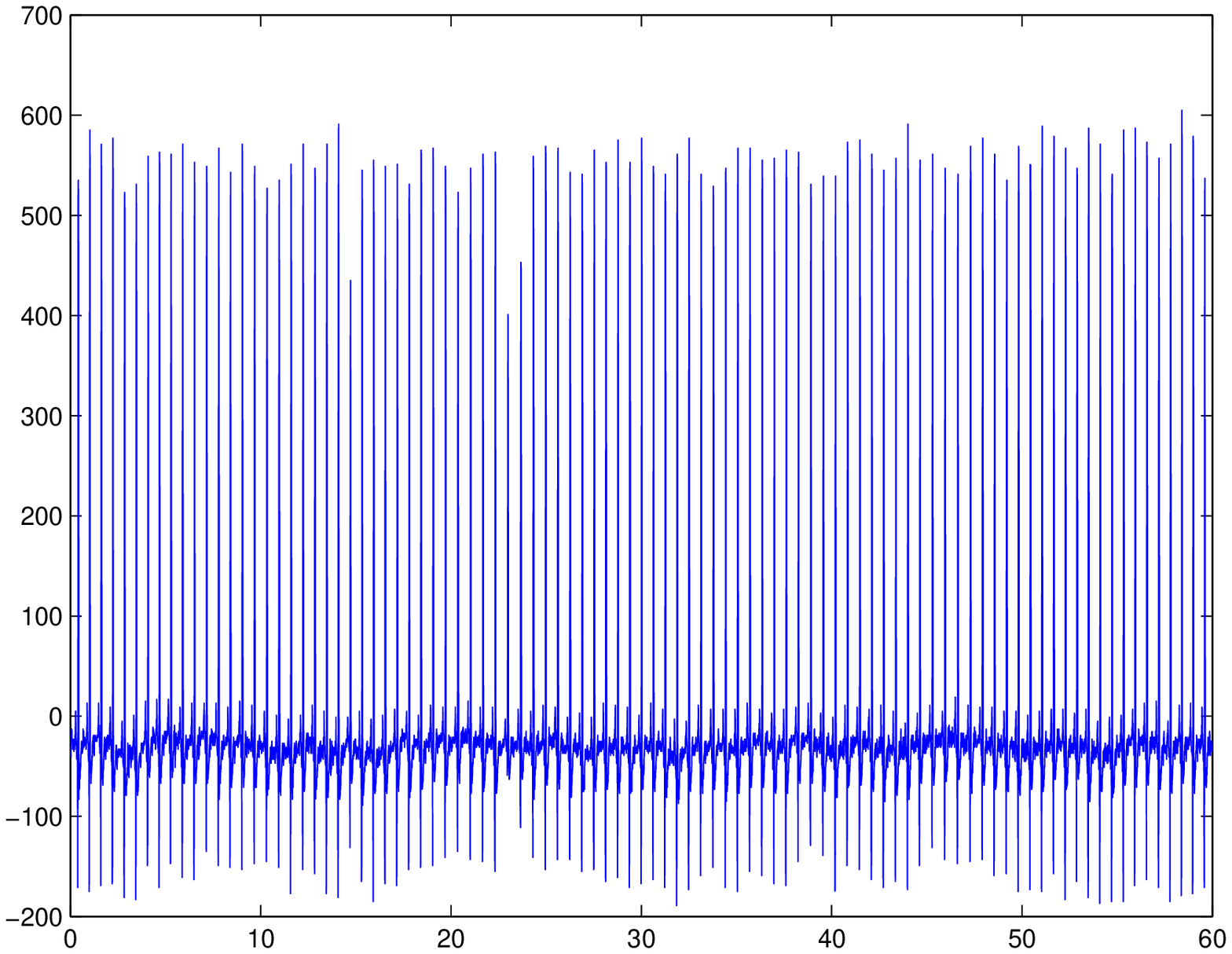} } \\
\subfigure[]{ \includegraphics[width=15cm,height=5cm]{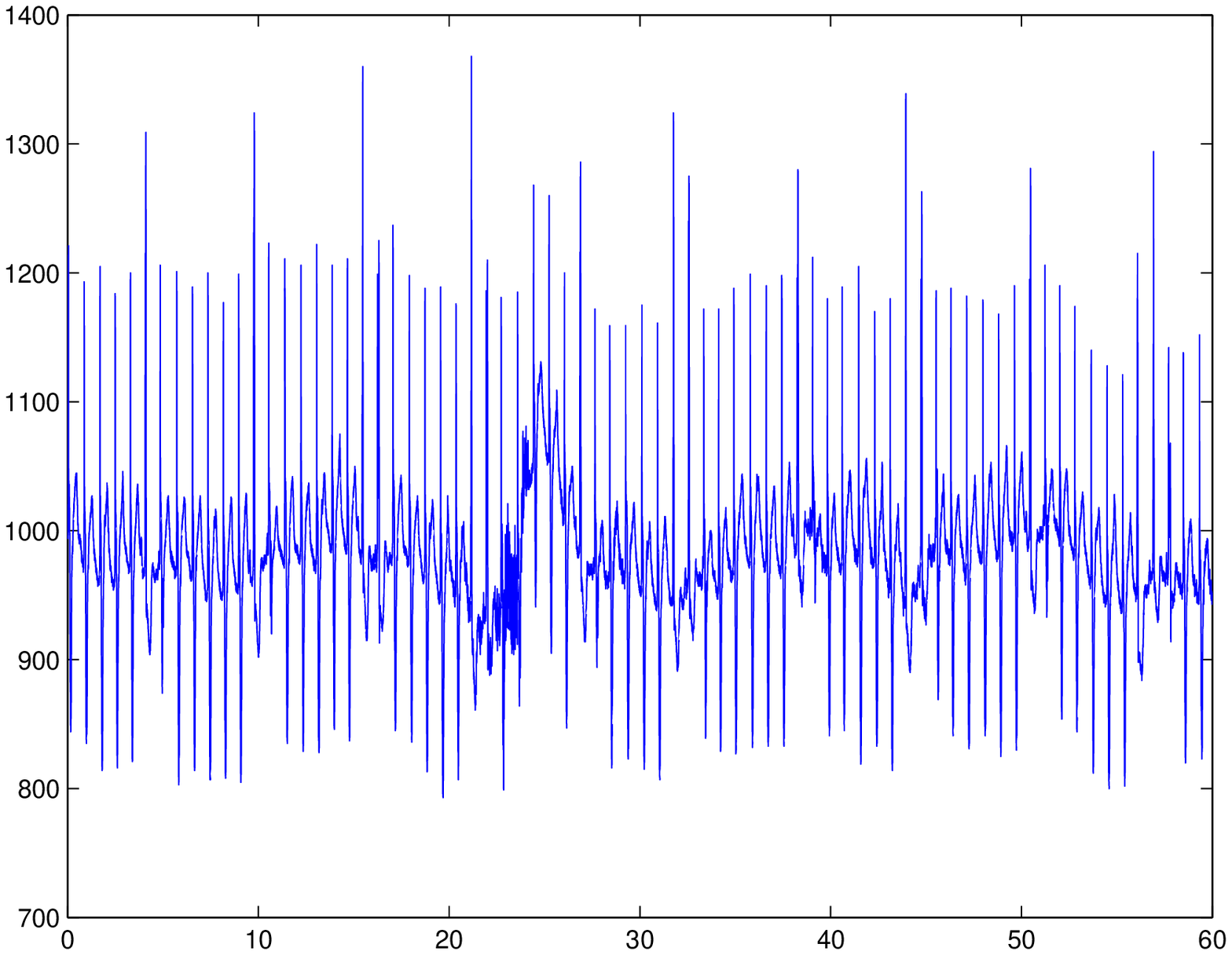} }
\caption{ (a) ECG recording during 60 seconds of a healthy subject  having no  arrhythmia, (b)  ECG recording during 60 seconds of a subject showing evidence of significant arrhythmia.} \label{fig:data}
\end{figure}

An important application of signal averaging is the estimation of a mean heart cycle from electrocardiogram (ECG)  records.  An ECG signal corresponds to the recording of the heart electrical activity. It is a signal, recorded over time, that is composed of the succession of cycles of contraction and release of the heart muscle. Each recorded cycle is a curve composed of a characteristic P-wave, reflecting the atrial depolarization, that is followed by the so-called QRS complex which corresponds to the  depolarization of the  ventricles, and which ends with a T-wave reflecting the repolarization of the heart (see e.g.\ \cite{guyton2006textbook} for a precise description of an ECG recording). In this paper, we present results on two different data sets from the MIT-BIH database \cite{PhysioNet}. In Figure \ref{fig:data}(a), we display data from an ECG record of a healthy  subject  having no  arrhythmia, and in Figure \ref{fig:data}(b) we display data from an ECG record of a subject showing evidence of significant arrhythmia (note that, in all the figures showing ECG data,  units on the vertical axis are in millivolts).

\begin{figure}[htbp]
\centering
\subfigure[]{ \includegraphics[width=3.5cm]{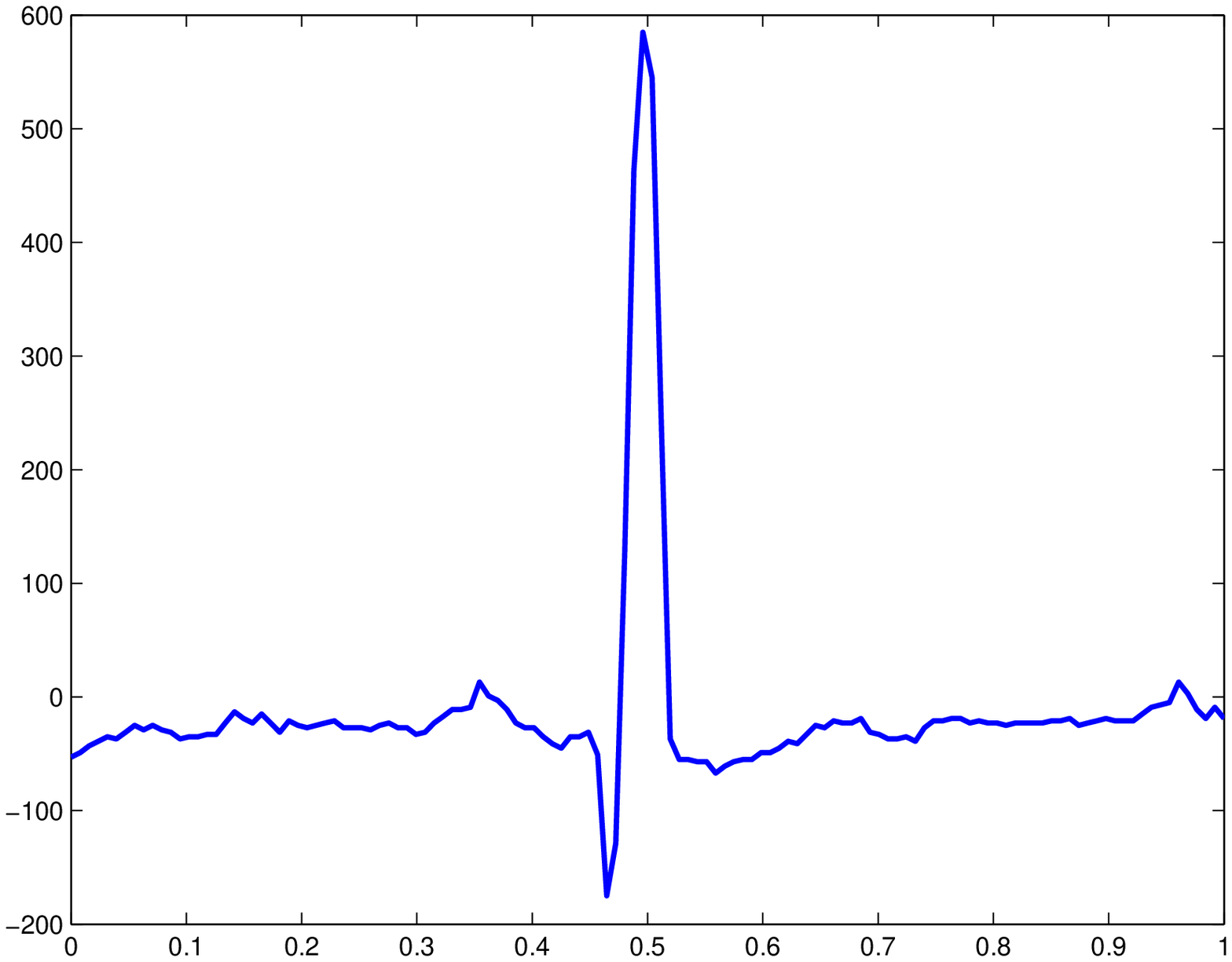} }
\subfigure[]{ \includegraphics[width=3.5cm]{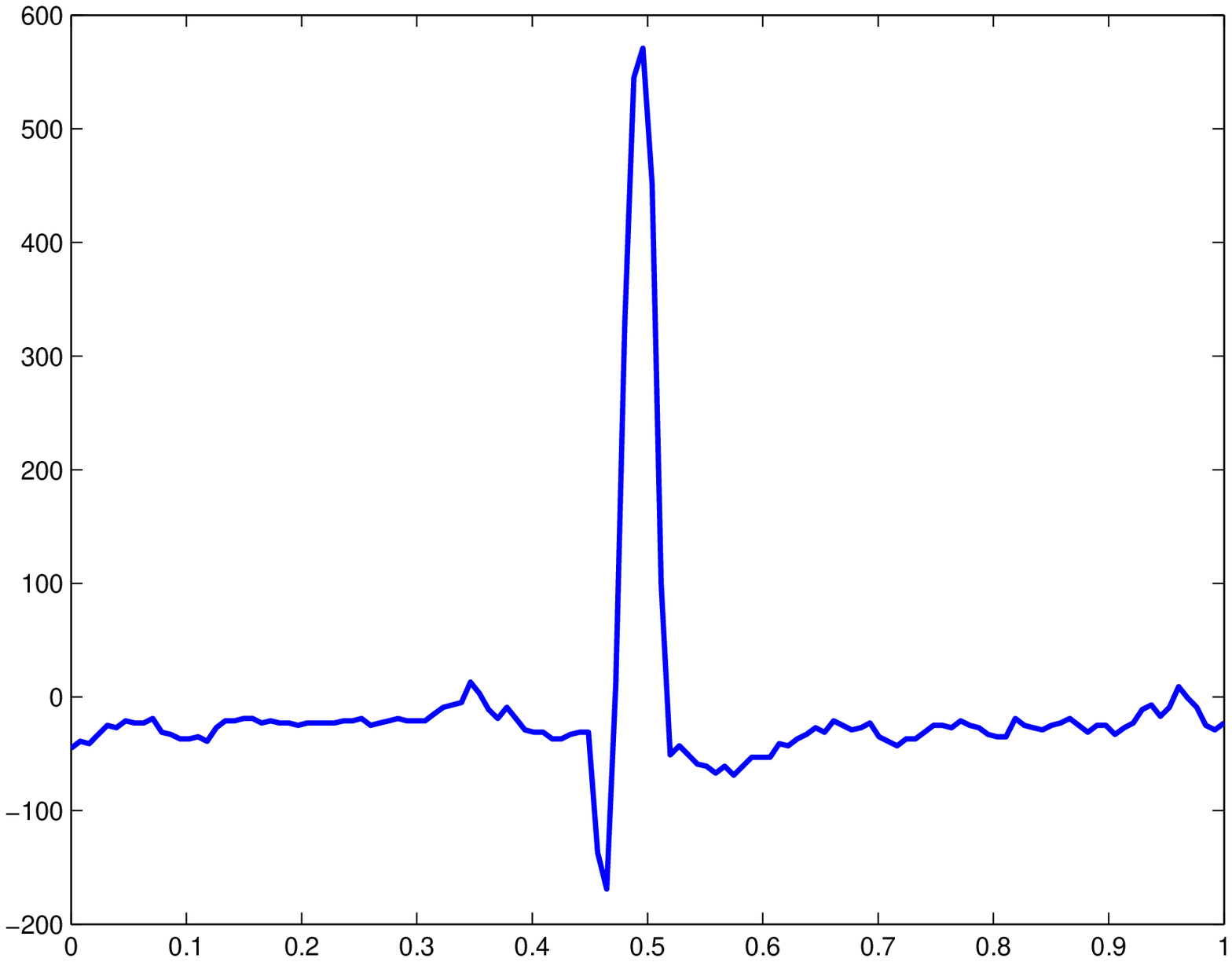} }
\subfigure[]{ \includegraphics[width=3.5cm]{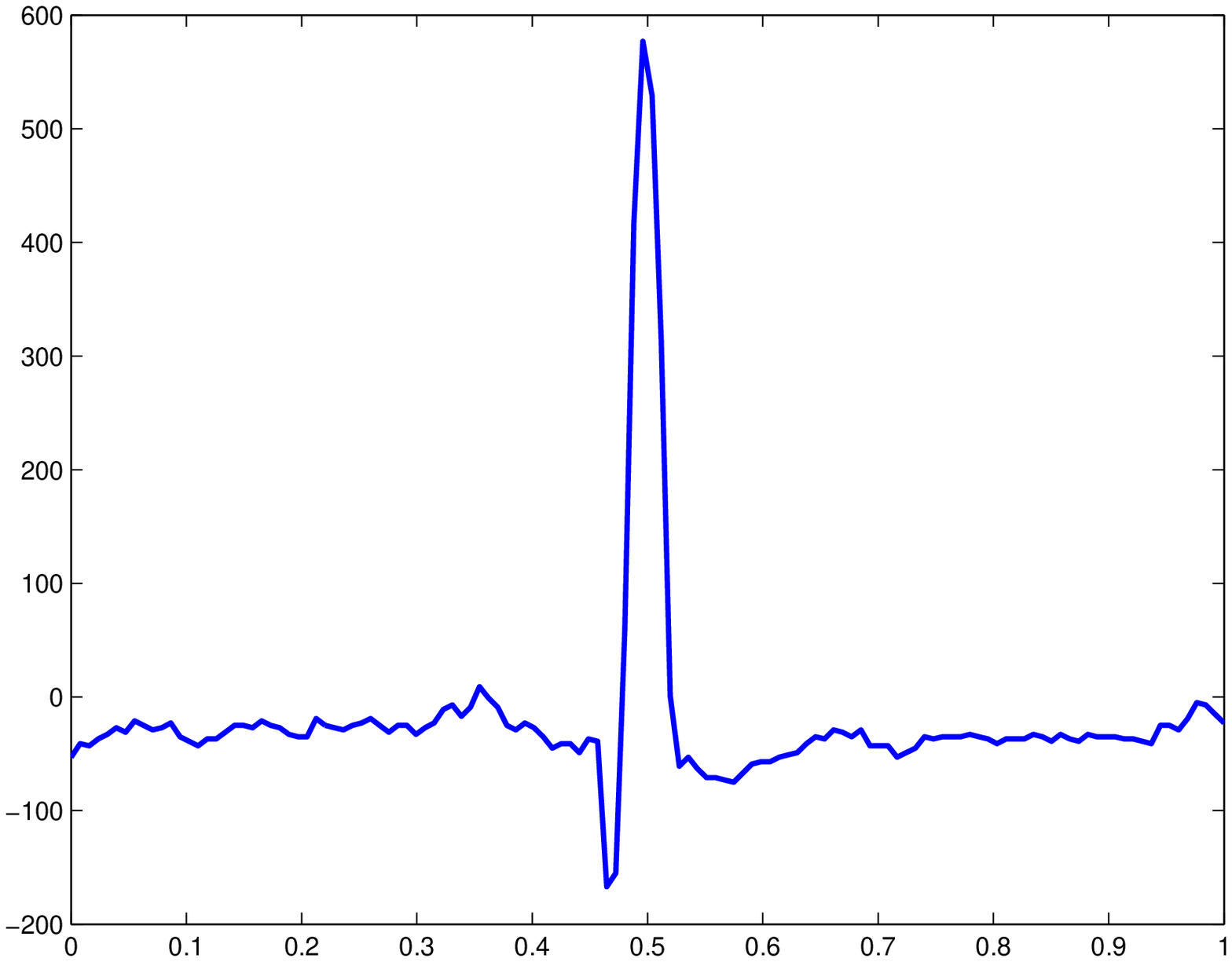} }

\subfigure[]{ \includegraphics[width=3.5cm]{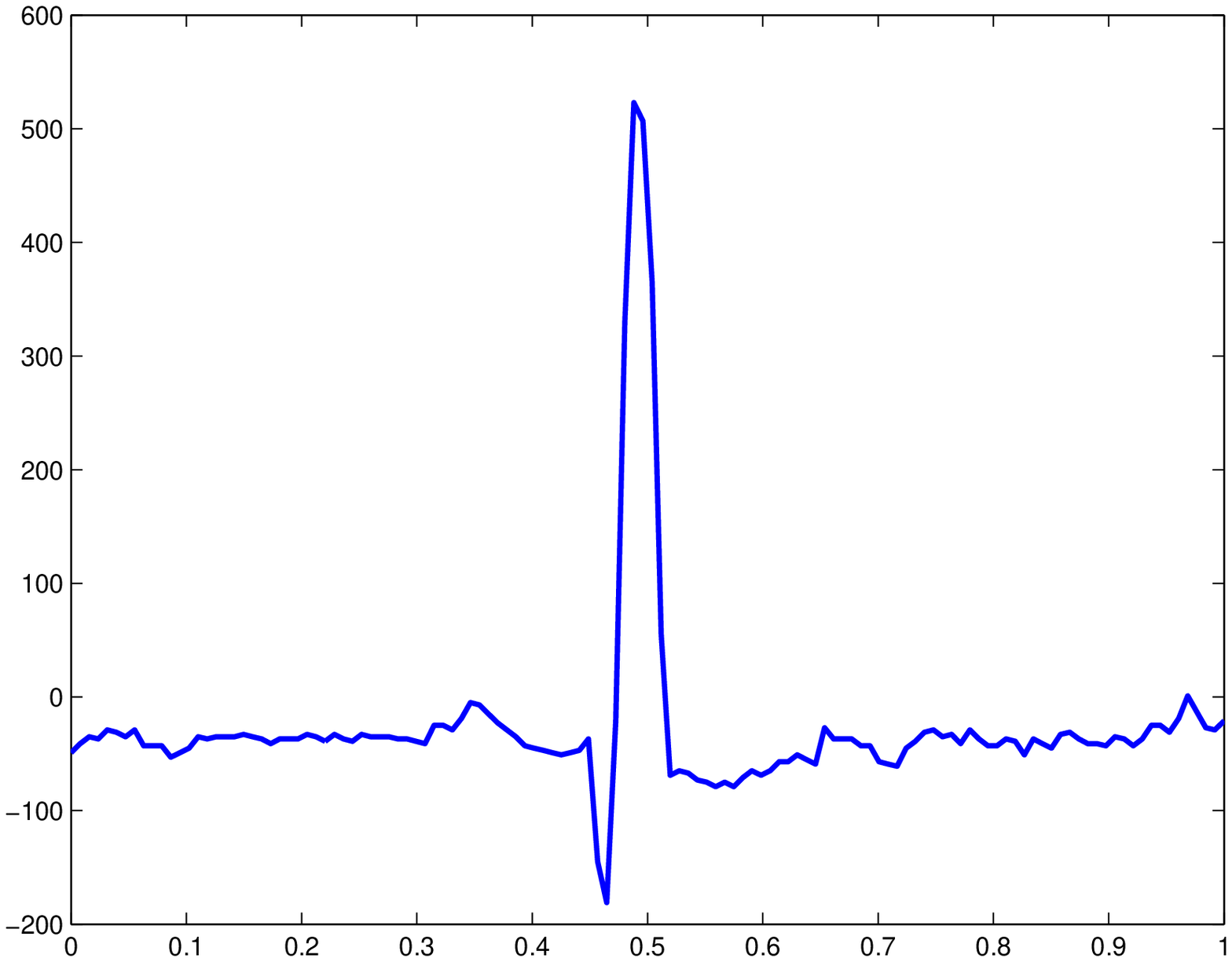} }
\subfigure[]{ \includegraphics[width=3.5cm]{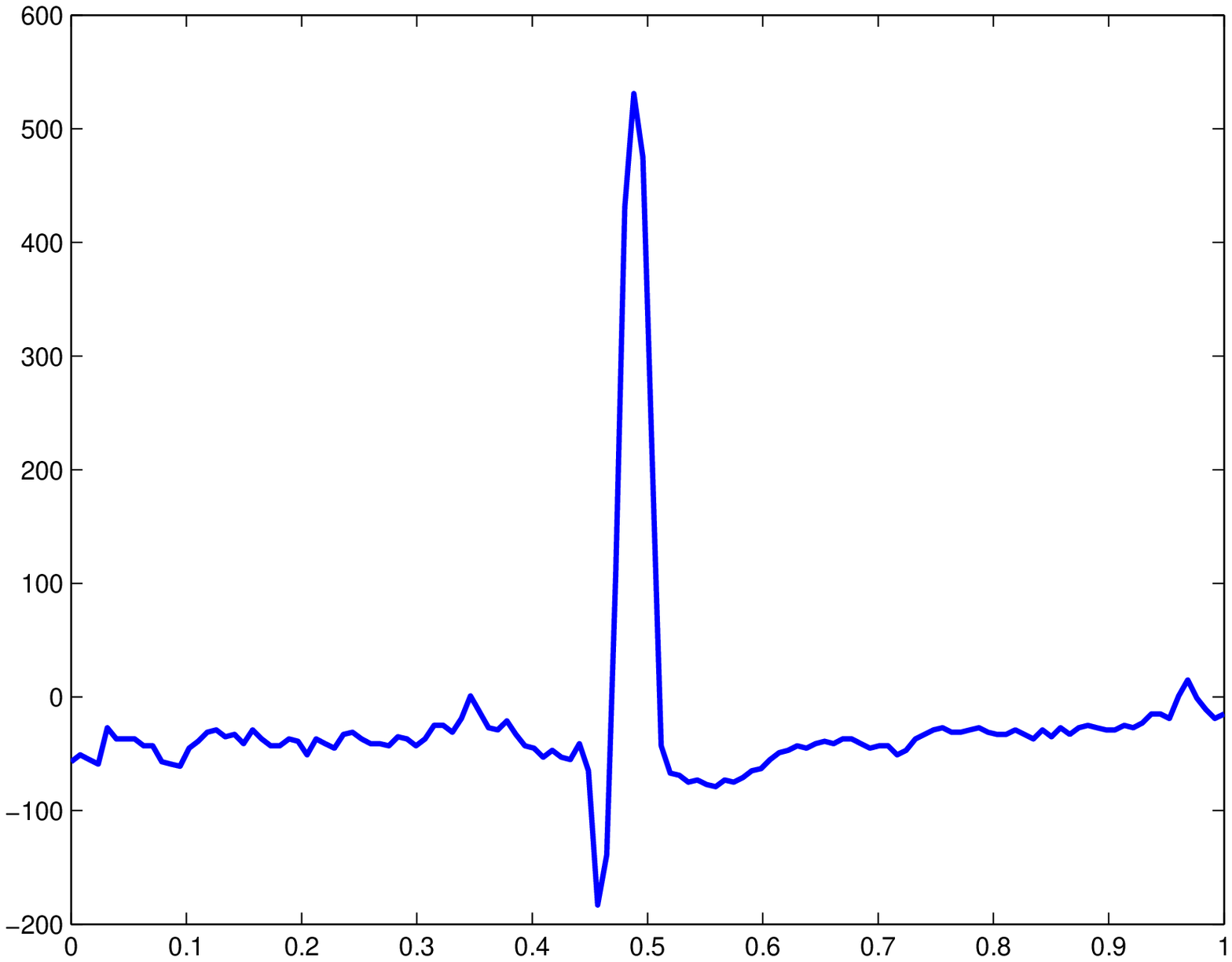} }
\subfigure[]{ \includegraphics[width=3.5cm]{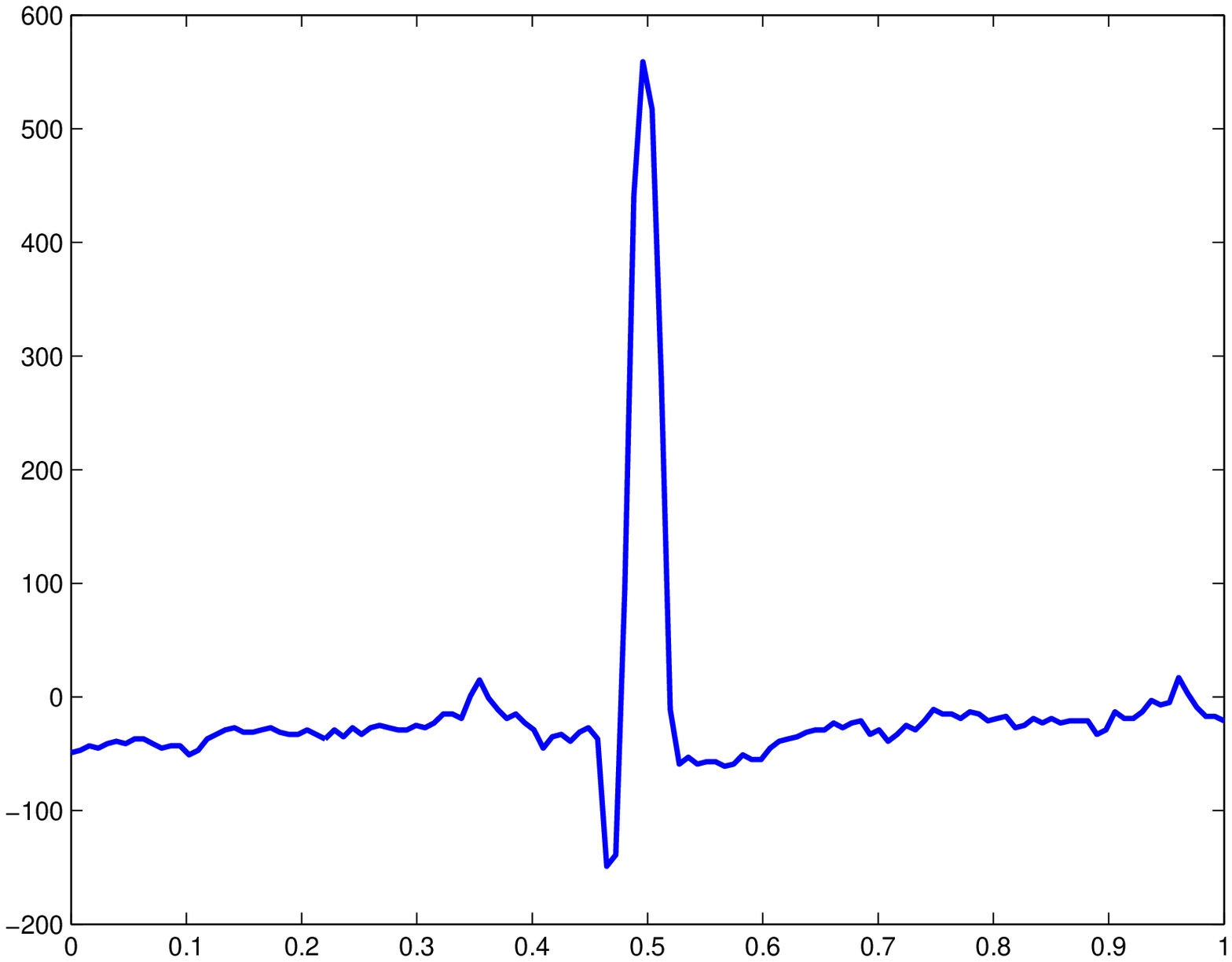} }
\caption{Normal case: six signals containing a single QRS complex out of $J=93$ extracted from the ECG recording displayed in Figure \ref{fig:data}(a). Units on the horizontal axis are arbitrary.  } \label{fig:Normal}
\end{figure}

\begin{figure}[htbp]
\centering
\subfigure[]{ \includegraphics[width=3.5cm]{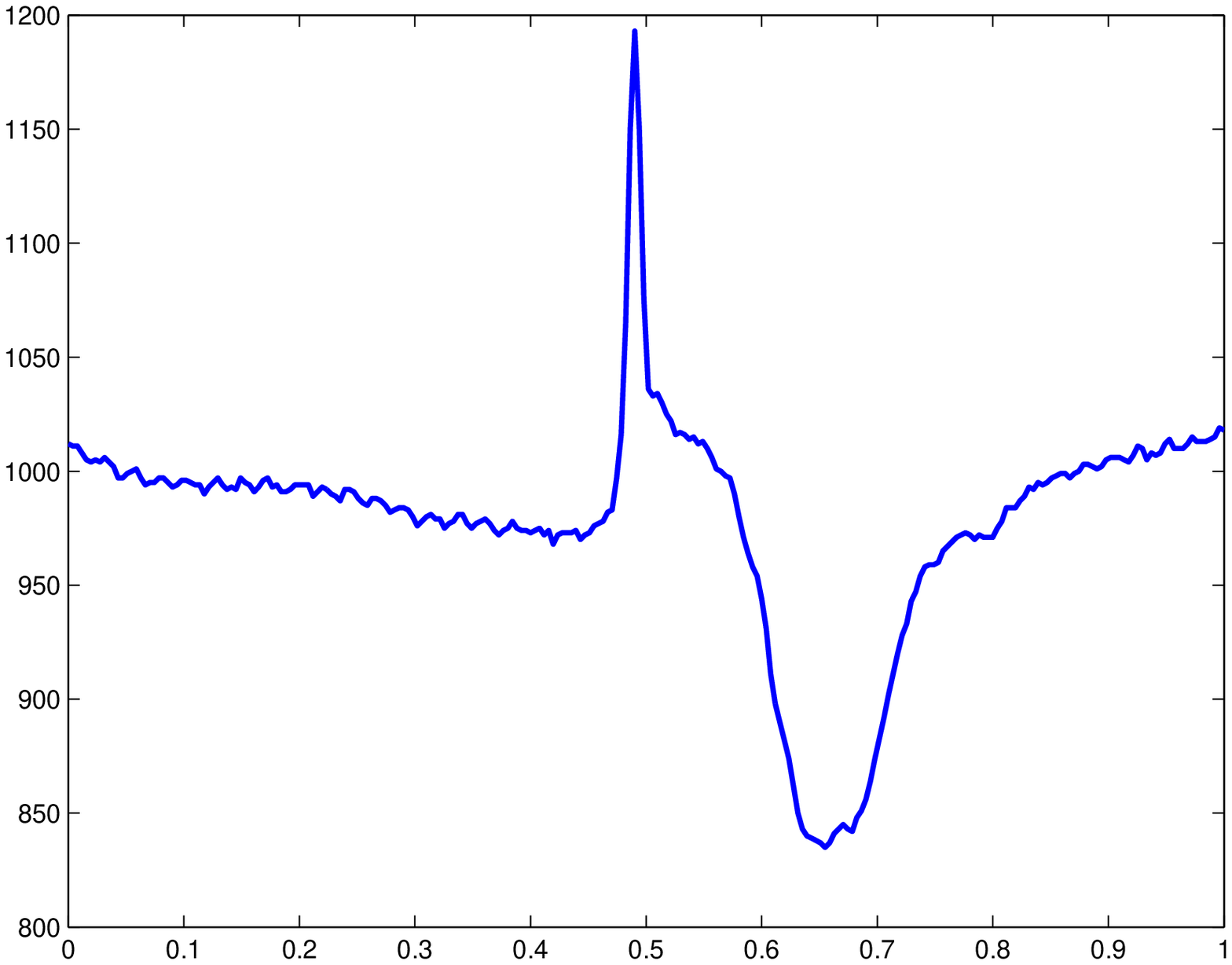} }
\subfigure[]{ \includegraphics[width=3.5cm]{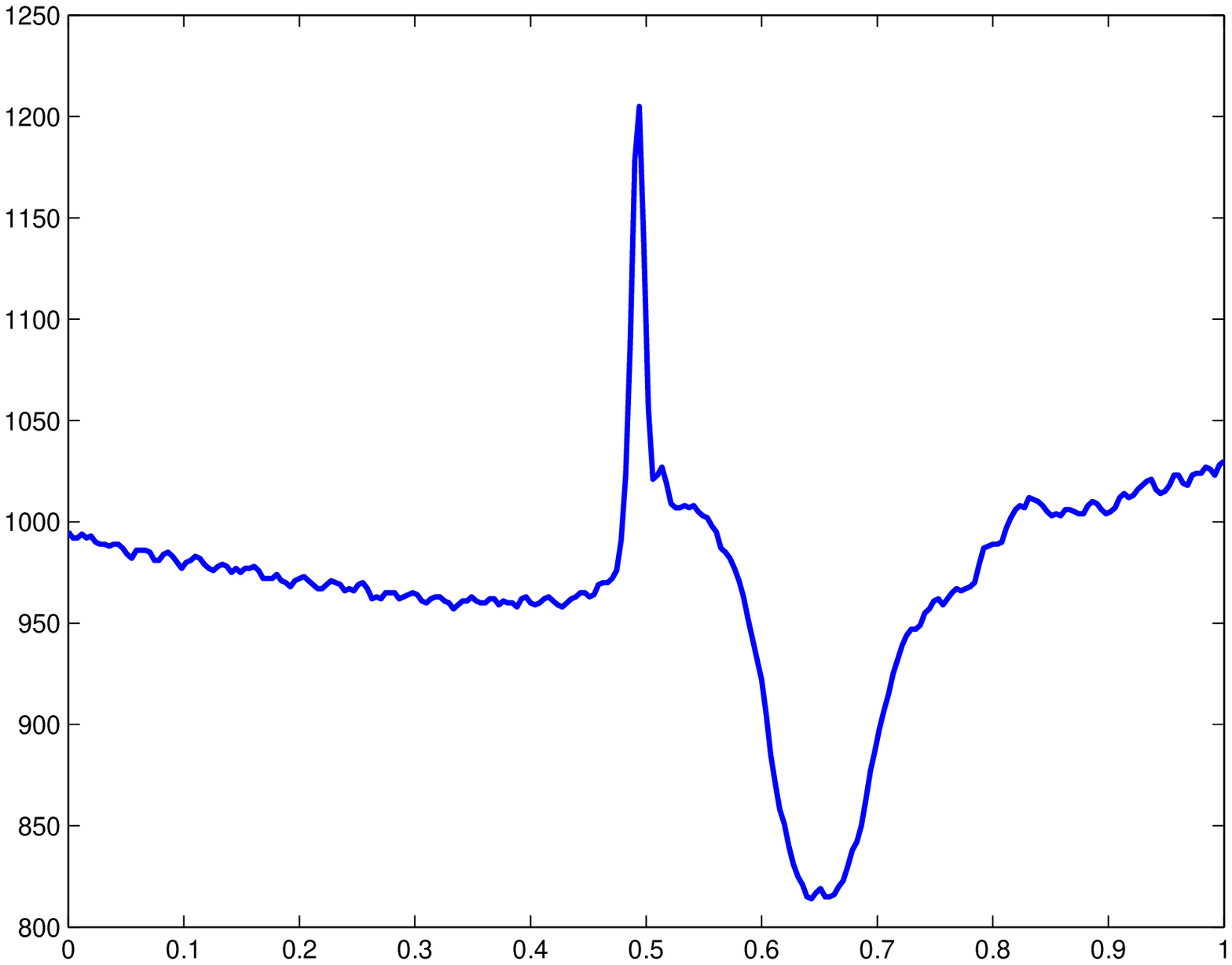} }
\subfigure[]{ \includegraphics[width=3.5cm]{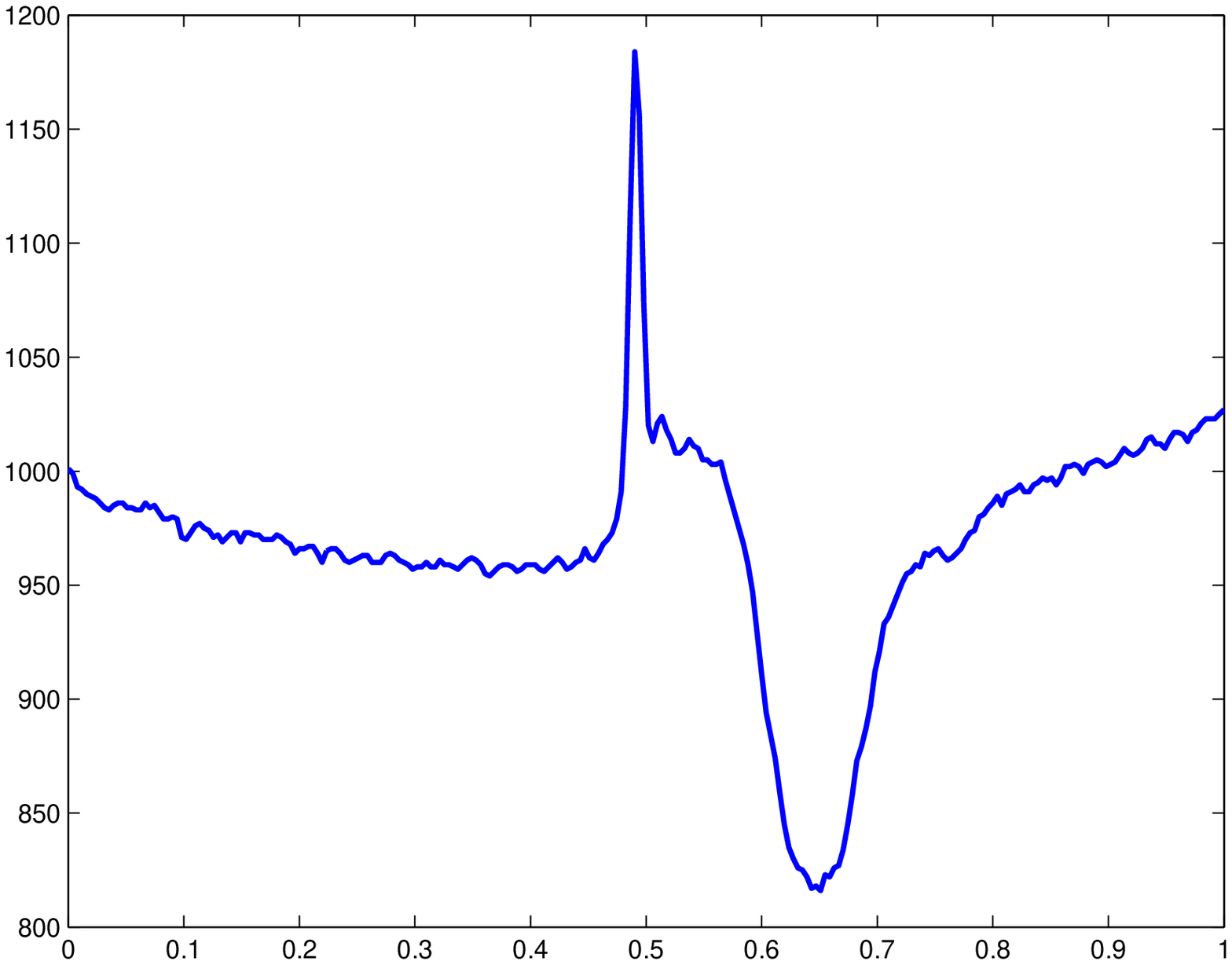} }

\subfigure[]{ \includegraphics[width=3.5cm]{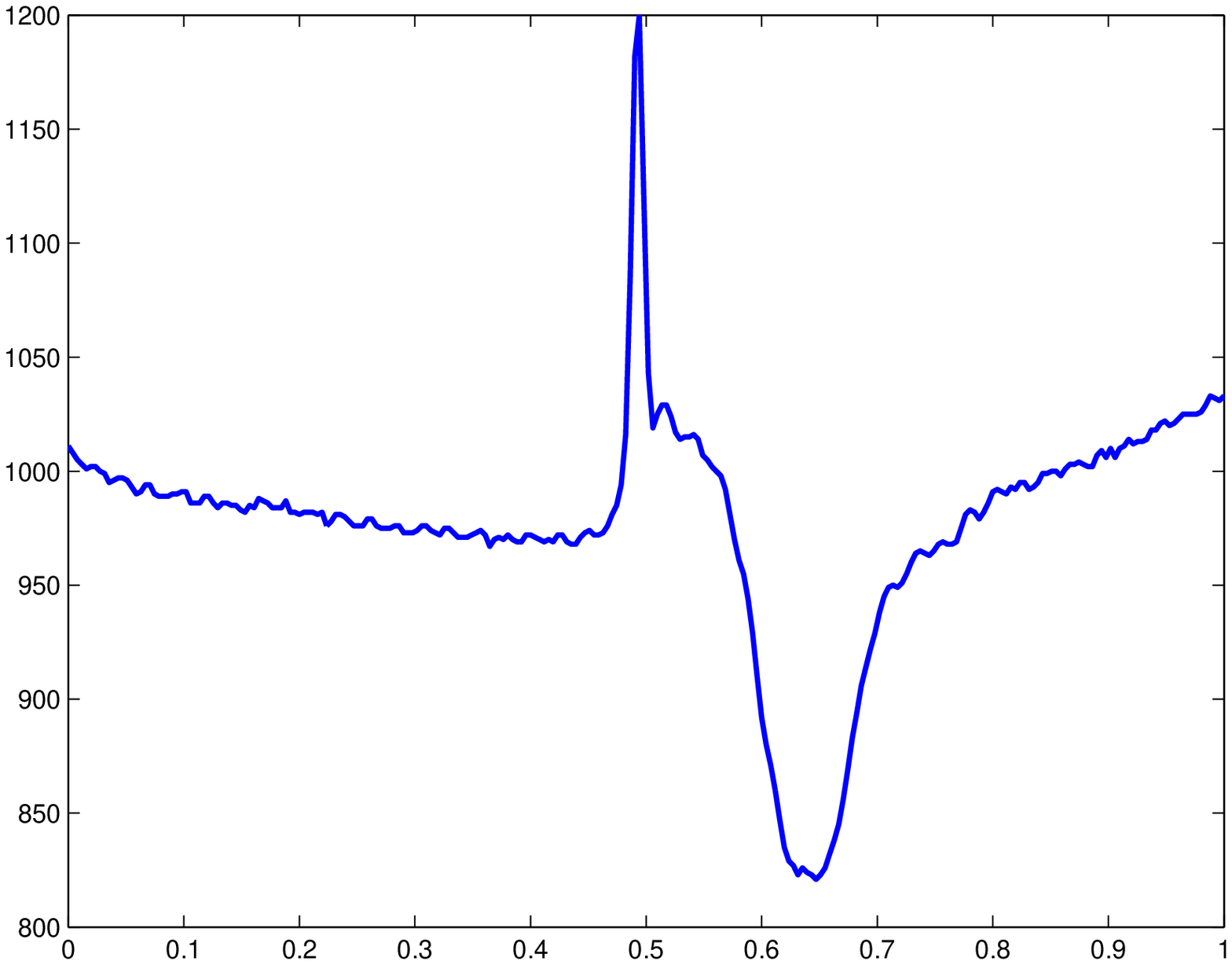} }
\subfigure[]{ \includegraphics[width=3.5cm]{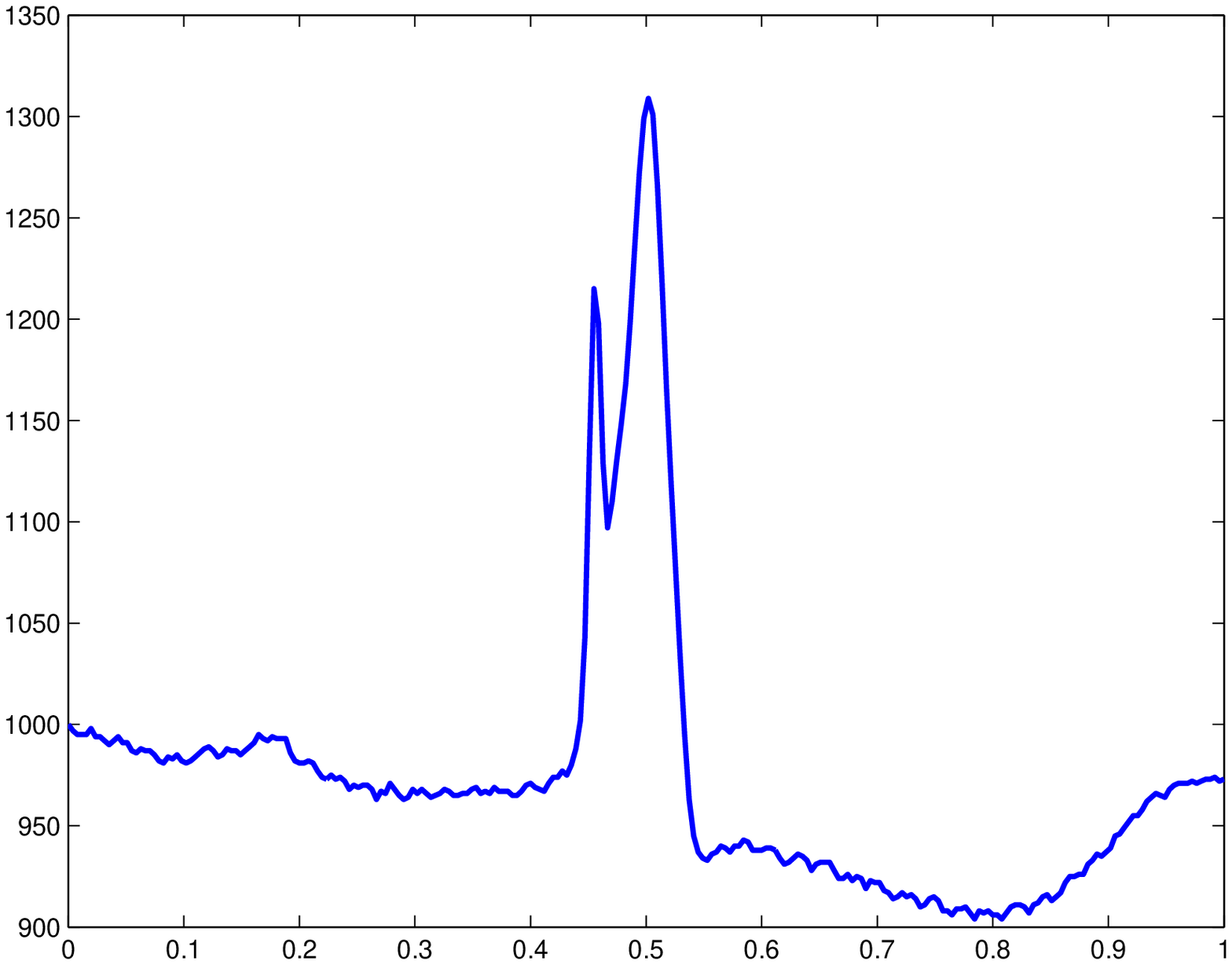} }
\subfigure[]{ \includegraphics[width=3.5cm]{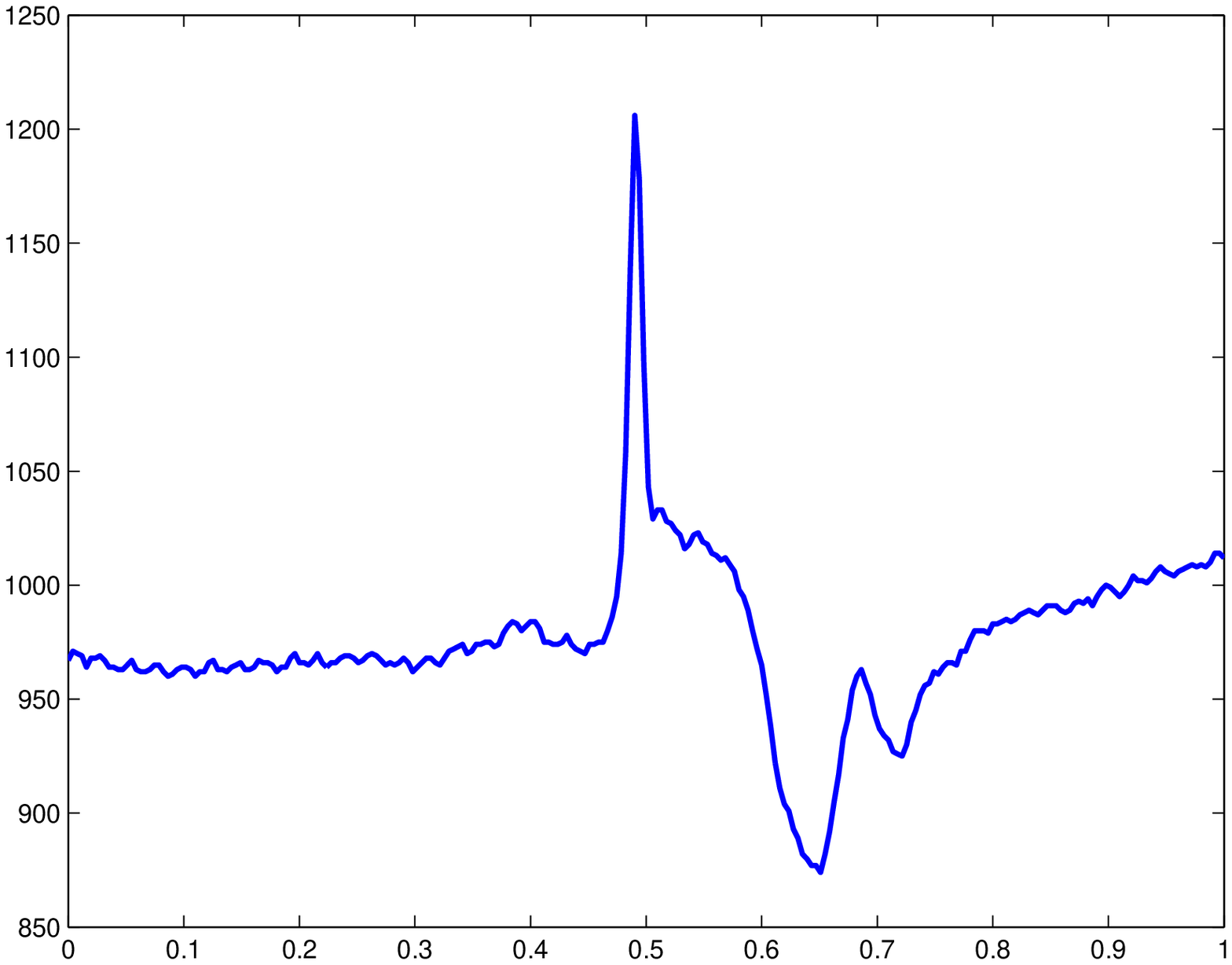} }
\caption{Case of cardiac arrhythmia: six signals containing a single QRS complex out of $J=72$ extracted from the ECG recording displayed in Figure \ref{fig:data}(b). Units on the horizontal axis are arbitrary.  } \label{fig:Arrhythmia}
\end{figure}

In the analysis of ECG data, it is generally assumed that the heart electrical activity repeats itself. Therefore, during an ECG record, one classically considers that  the heart cycle of interest remains approximately the same with every beat, and that it is embedded in a random white noise with zero expectation that is uncorrelated with the mean shape to be estimated. After an appropriate segmentation of an ECG record, one observes a set of signals of the same length such that each of them contains a single QRS complex. The preliminary segmentation step is done by taking segments (of the same length) in the ECG record that are centered around the easily detectable maxima of the QRS complex of the beats. Identification of these maxima can be done using statistical methods to identity local extrema in noisy signals \cite{MR2291263,citeulike:6080848} or by applying appropriate digital filters to identify typical parts of the QRS complex \cite{pan1985}. In this paper, we used the approach in \cite{MR2291263} to identify local maxima and to segment  the ECG record into signals of the same length containing a single QRS complex. After this preliminary segmentation, one thus observes signals with approximately the same shape. In the normal case, we obtained $J=93$ signals displayed in Figure \ref{fig:Normal}, while in the case of cardiac arrhythmia we obtained $J=72$ signals displayed in Figure \ref{fig:Arrhythmia}. 

\begin{figure}[htbp]
\centering
\subfigure[]{ \includegraphics[width=3.5cm]{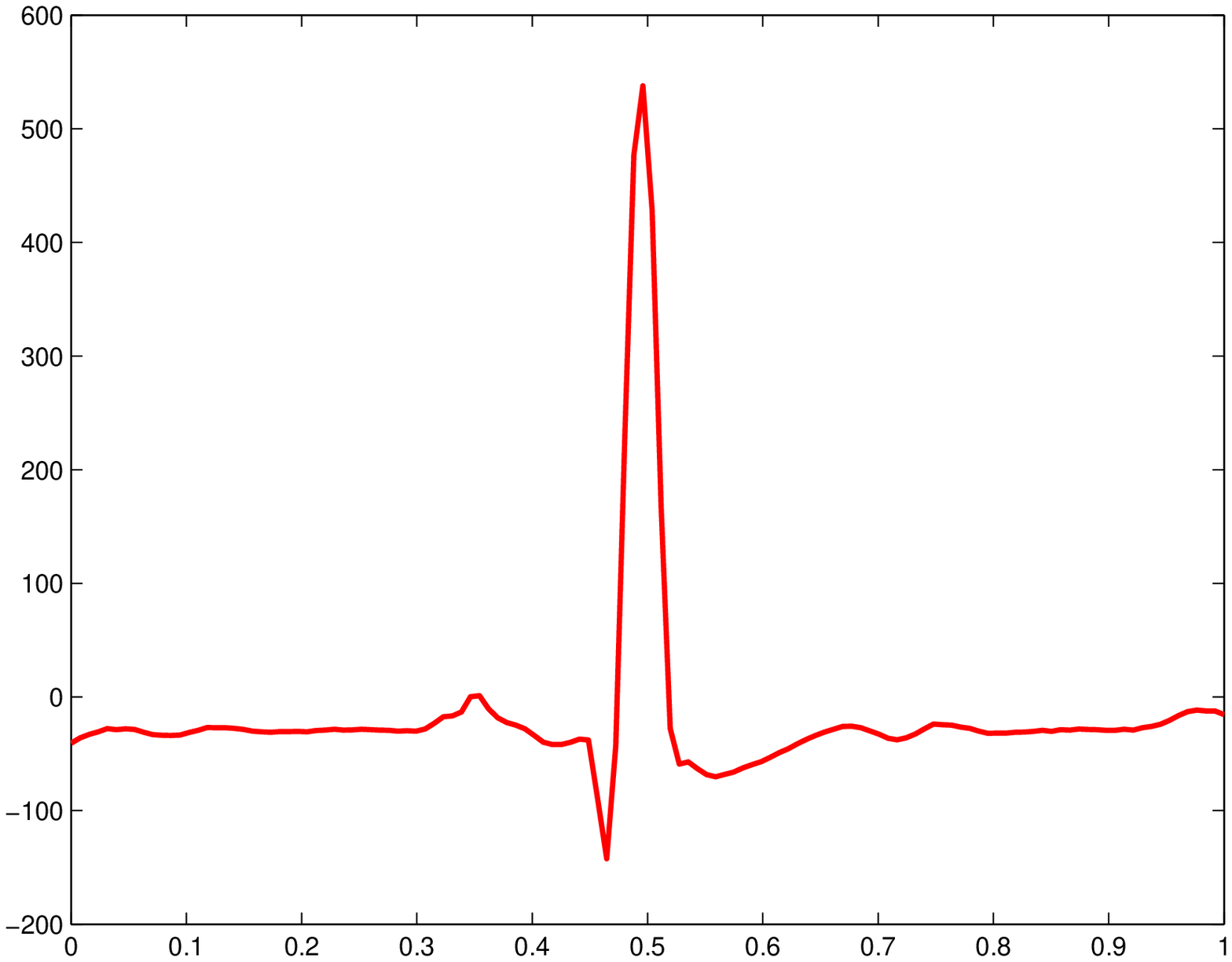} } 
\hspace{0.2cm}
\subfigure[]{ \includegraphics[width=3.5cm]{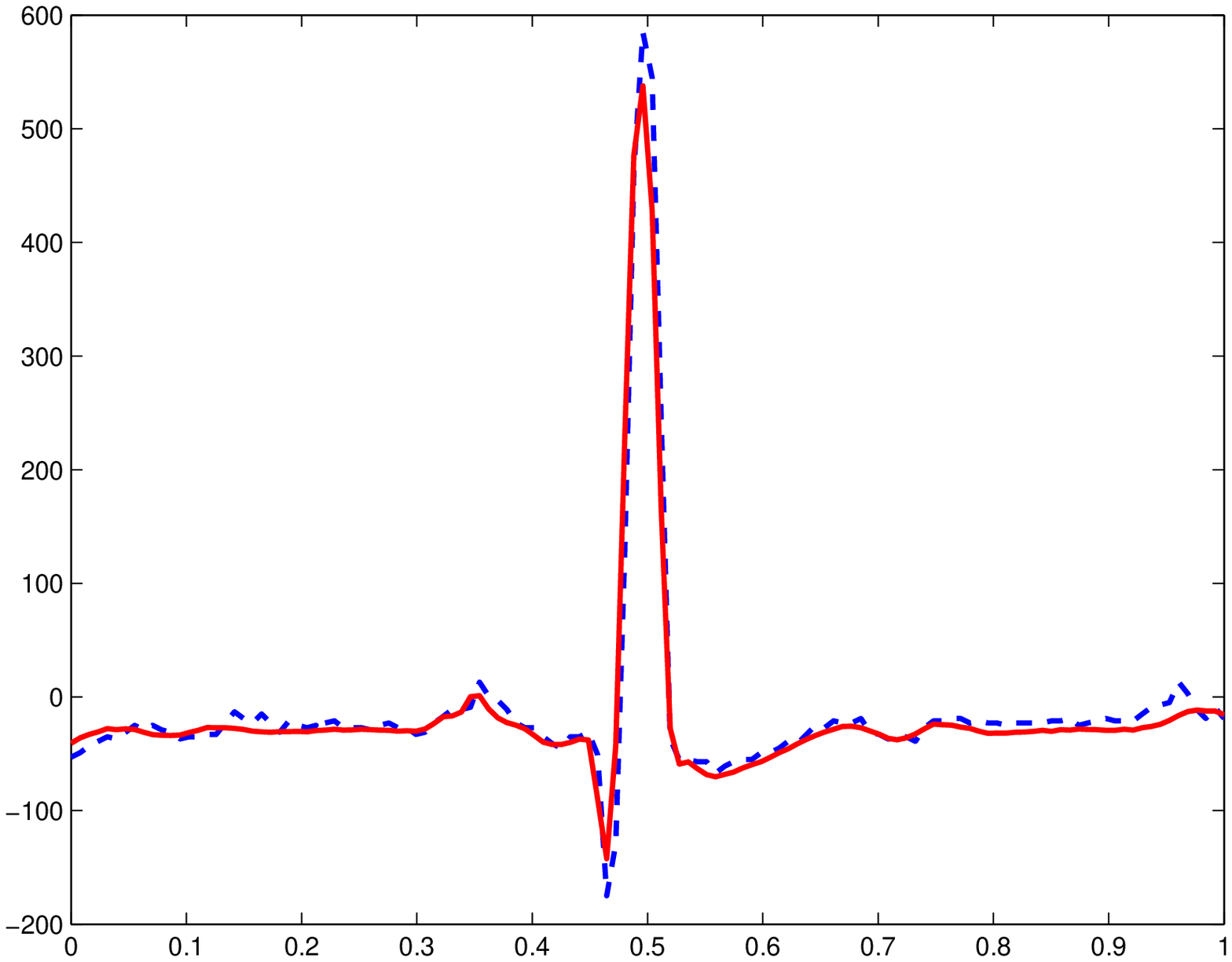} }
\subfigure[]{ \includegraphics[width=3.5cm]{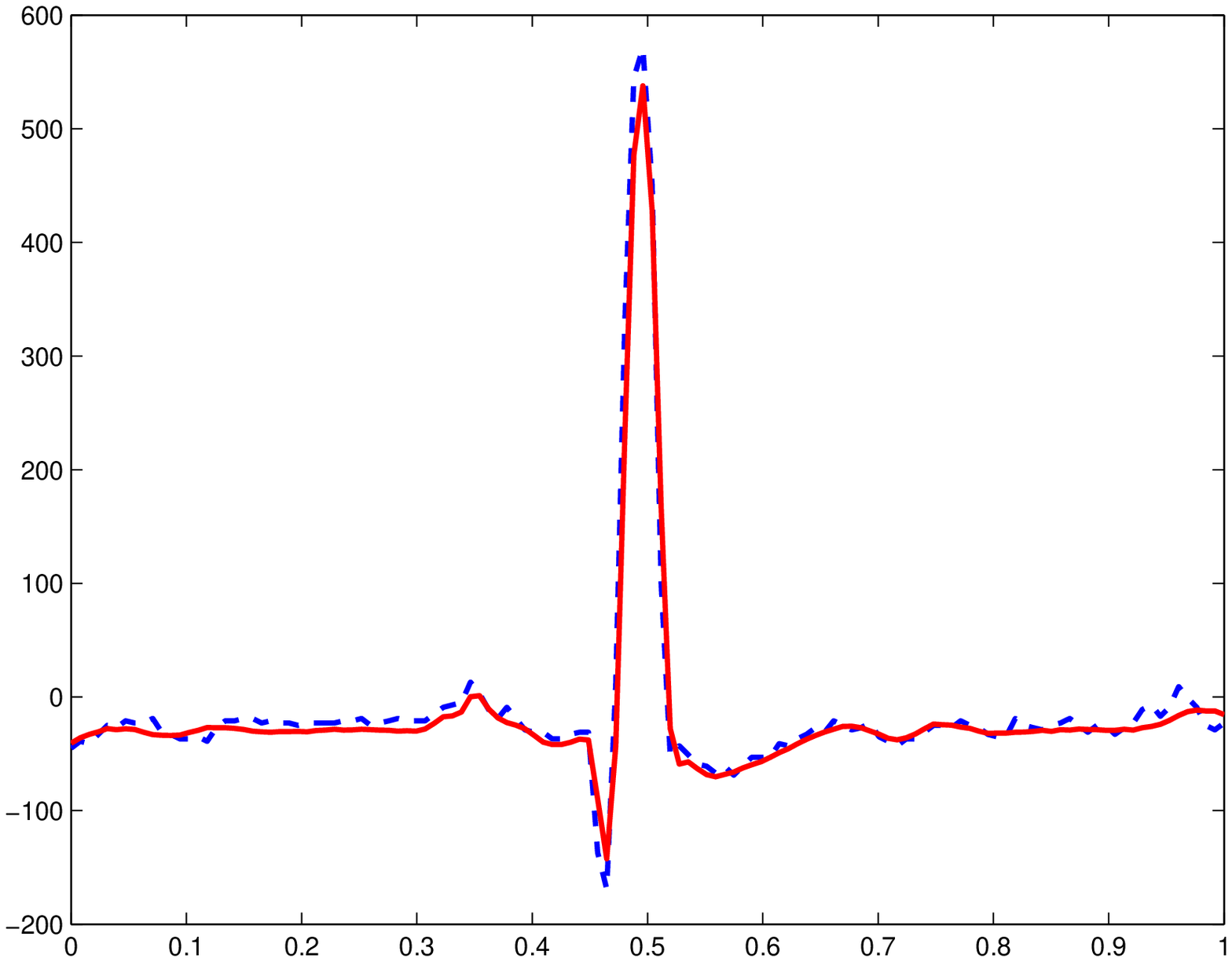} }
\subfigure[]{ \includegraphics[width=3.5cm]{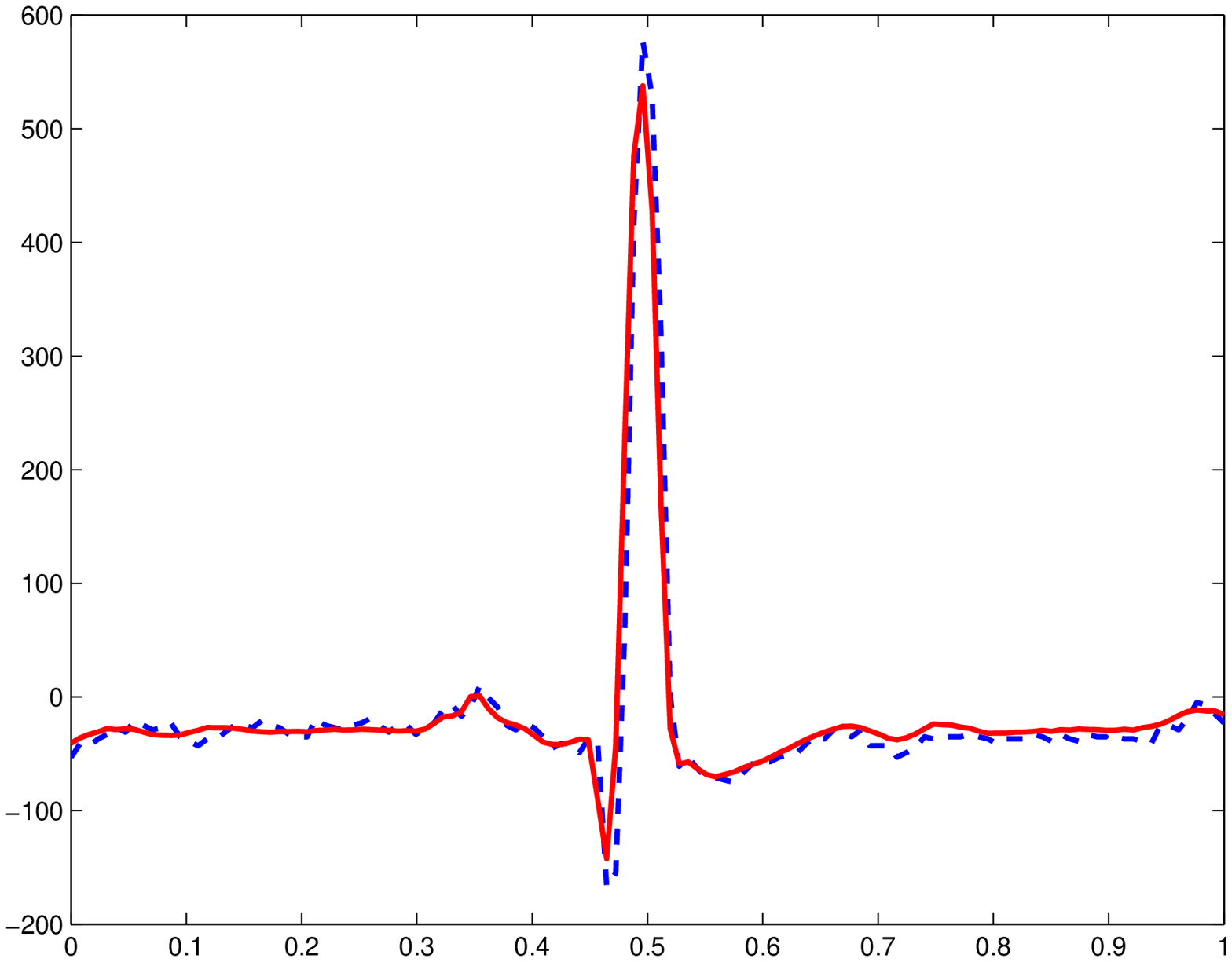} }

\hspace{4.1cm} 
\subfigure[]{ \includegraphics[width=3.5cm]{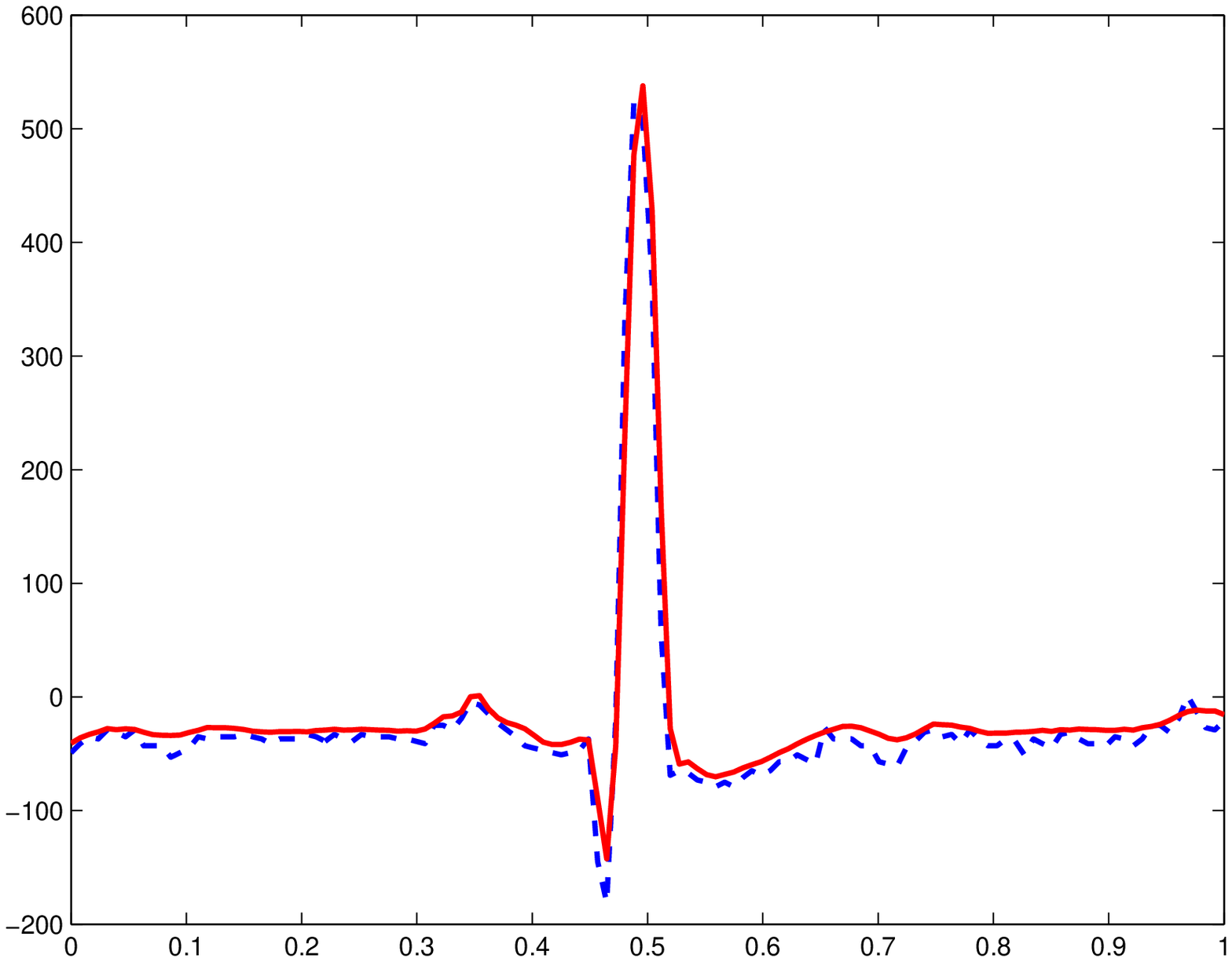} }
\subfigure[]{ \includegraphics[width=3.5cm]{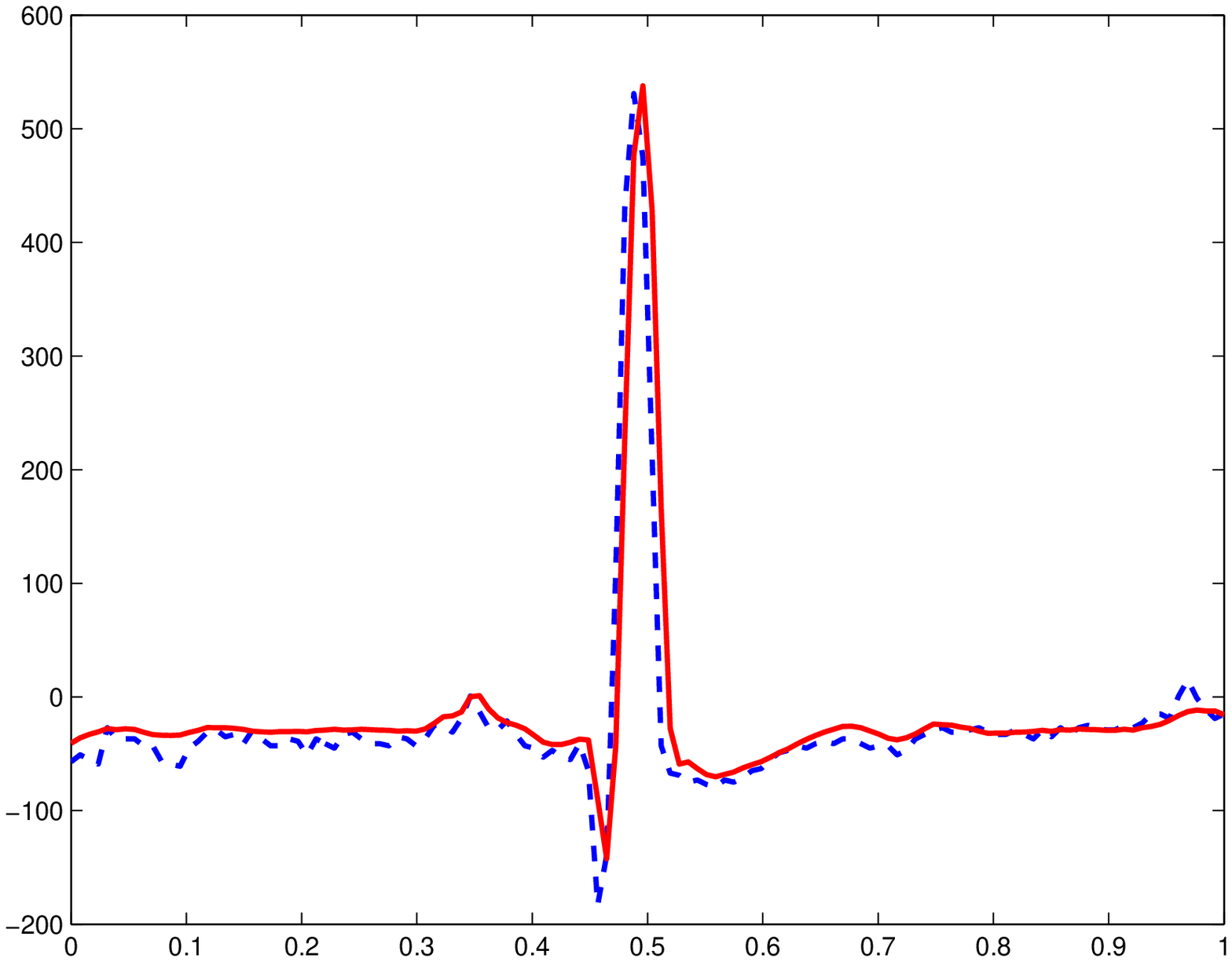} }
\subfigure[]{ \includegraphics[width=3.5cm]{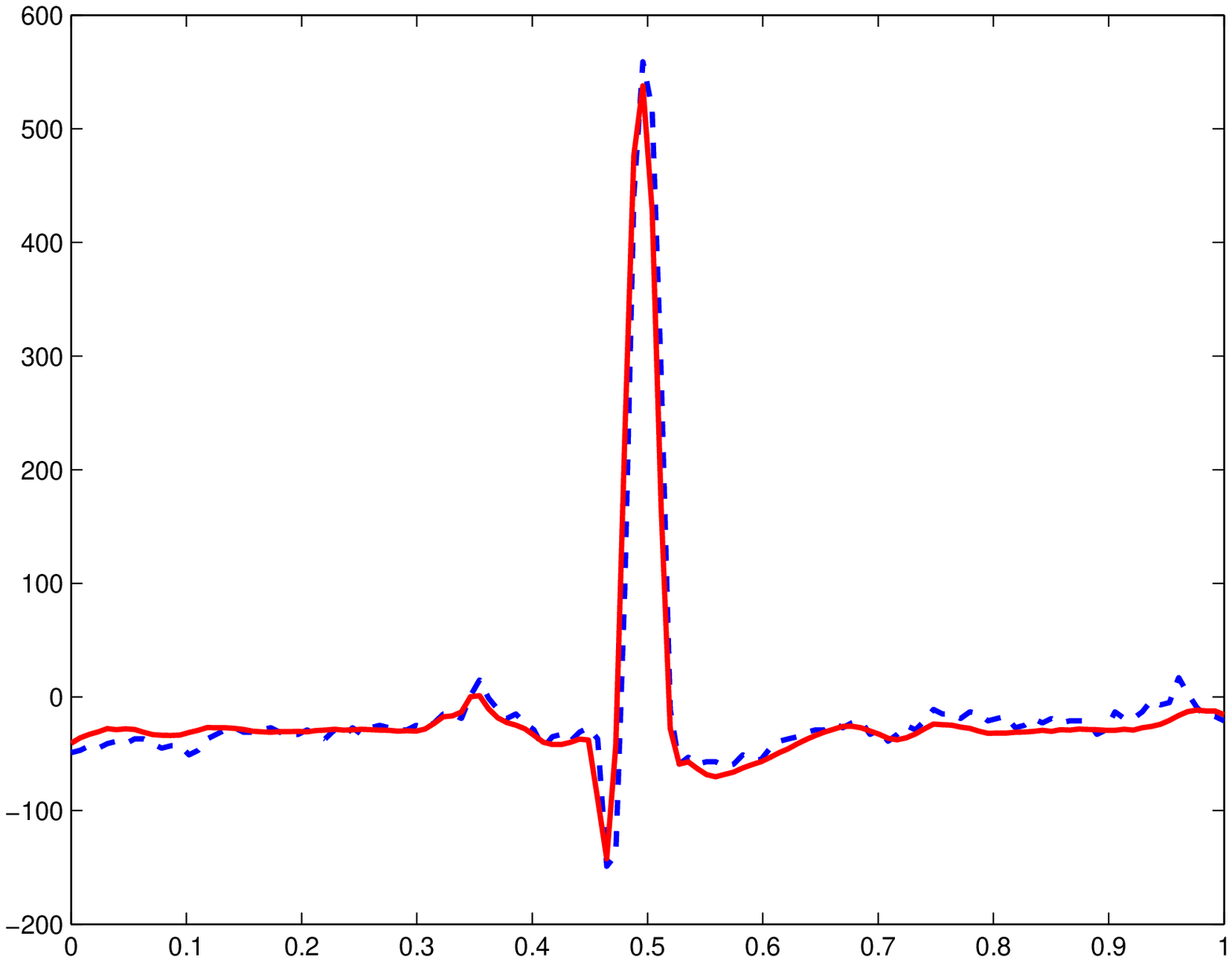} }

\caption{Normal case: (a) Euclidean mean of the $J=93$  signals after segmentation of the ECG record. (b)-(g) Superposition of six signals containing a single QRS complex (dashed curves) with the Euclidean mean (solid curve).  } \label{fig:Normal:naive}
\end{figure}

\begin{figure}[htbp]
\centering
\subfigure[]{ \includegraphics[width=3.5cm]{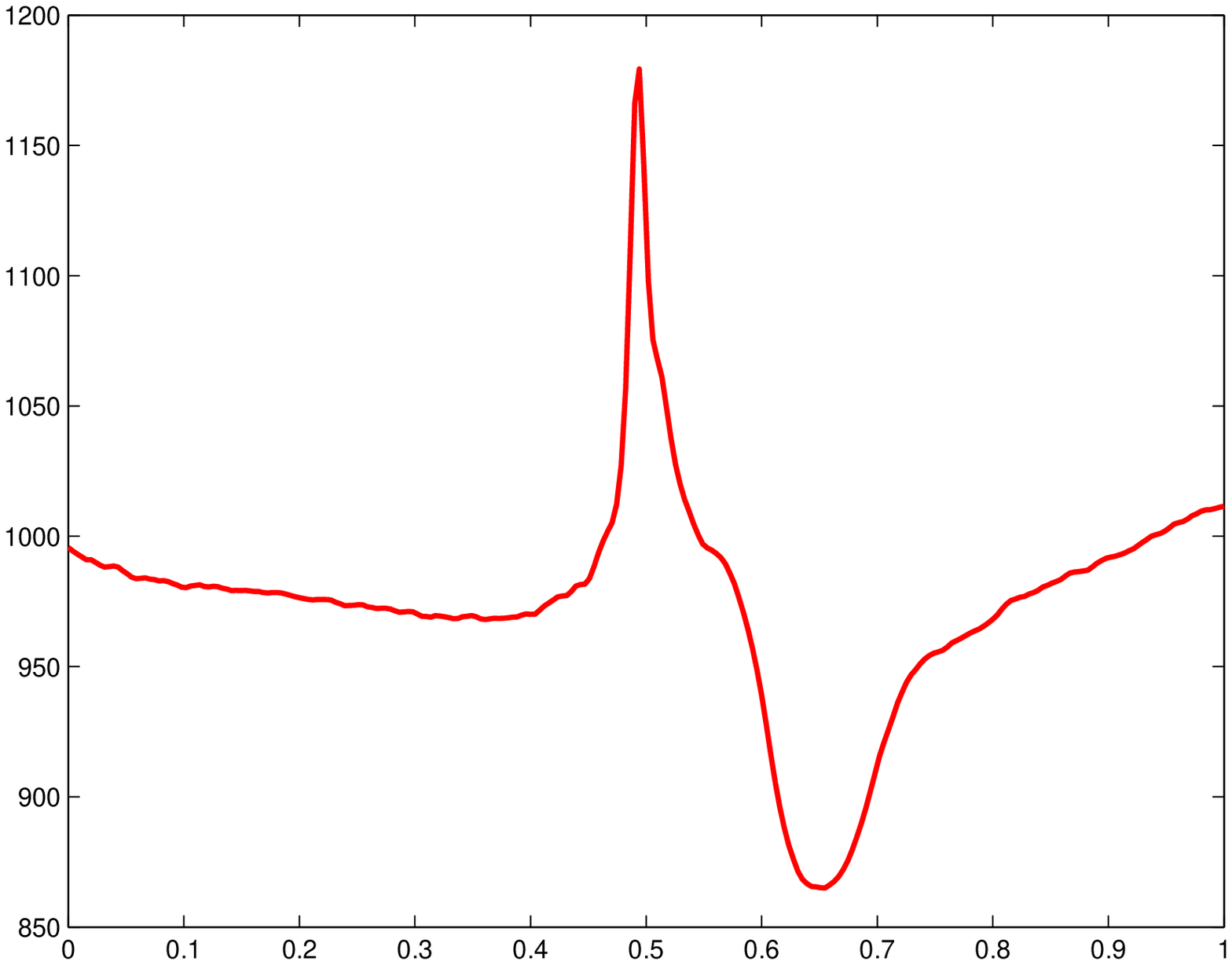} } 
\hspace{0.2cm}
\subfigure[]{ \includegraphics[width=3.5cm]{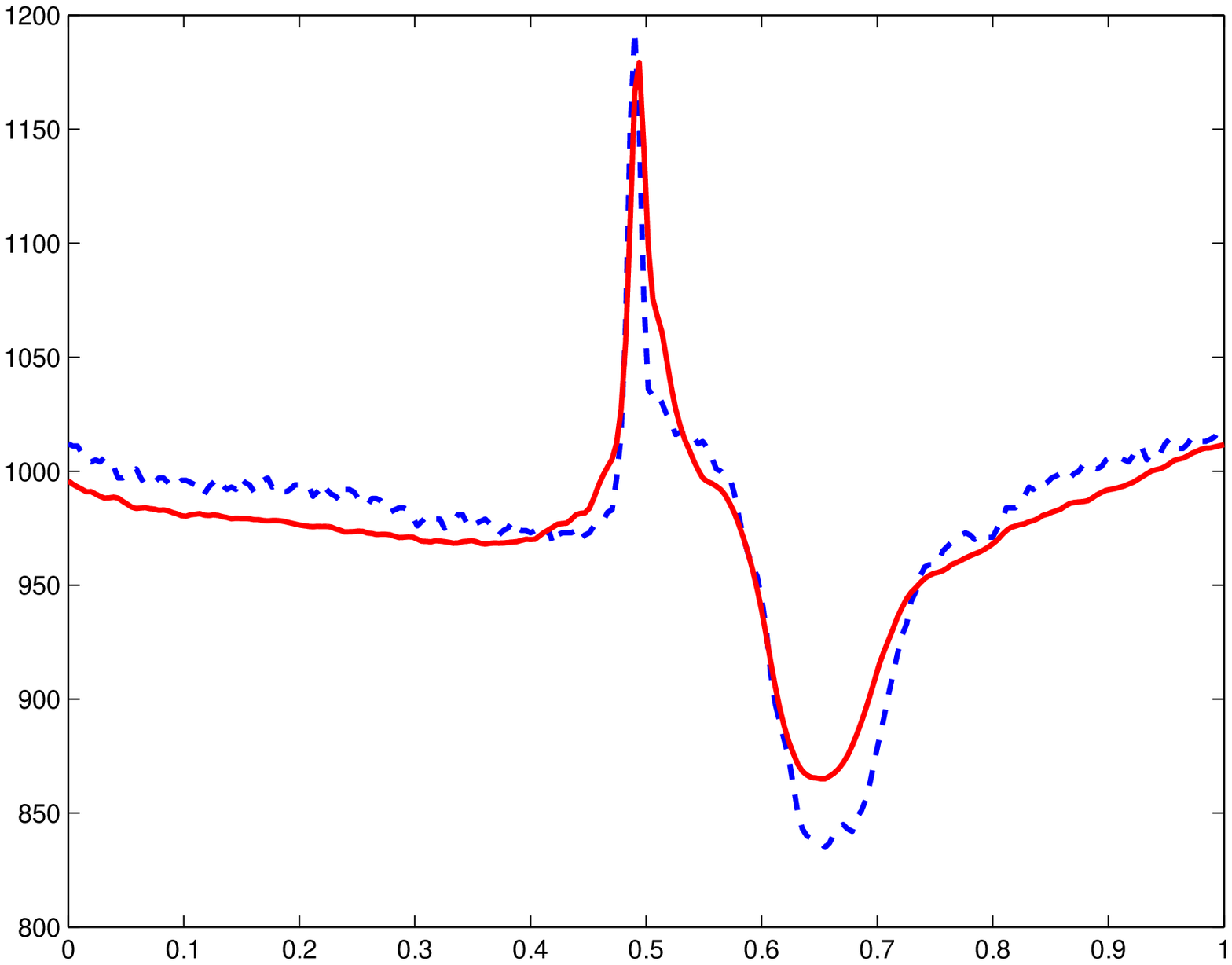} }
\subfigure[]{ \includegraphics[width=3.5cm]{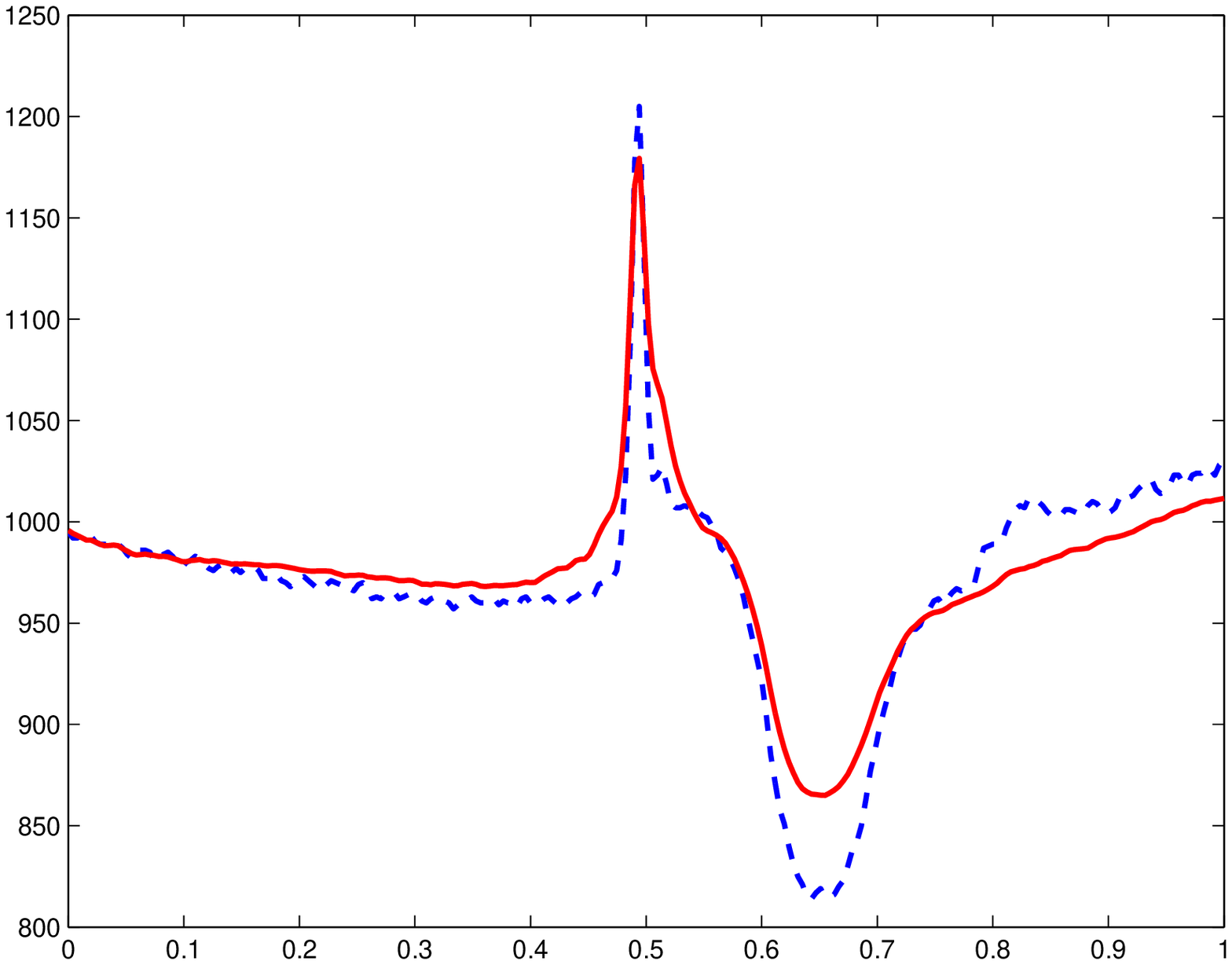} }
\subfigure[]{ \includegraphics[width=3.5cm]{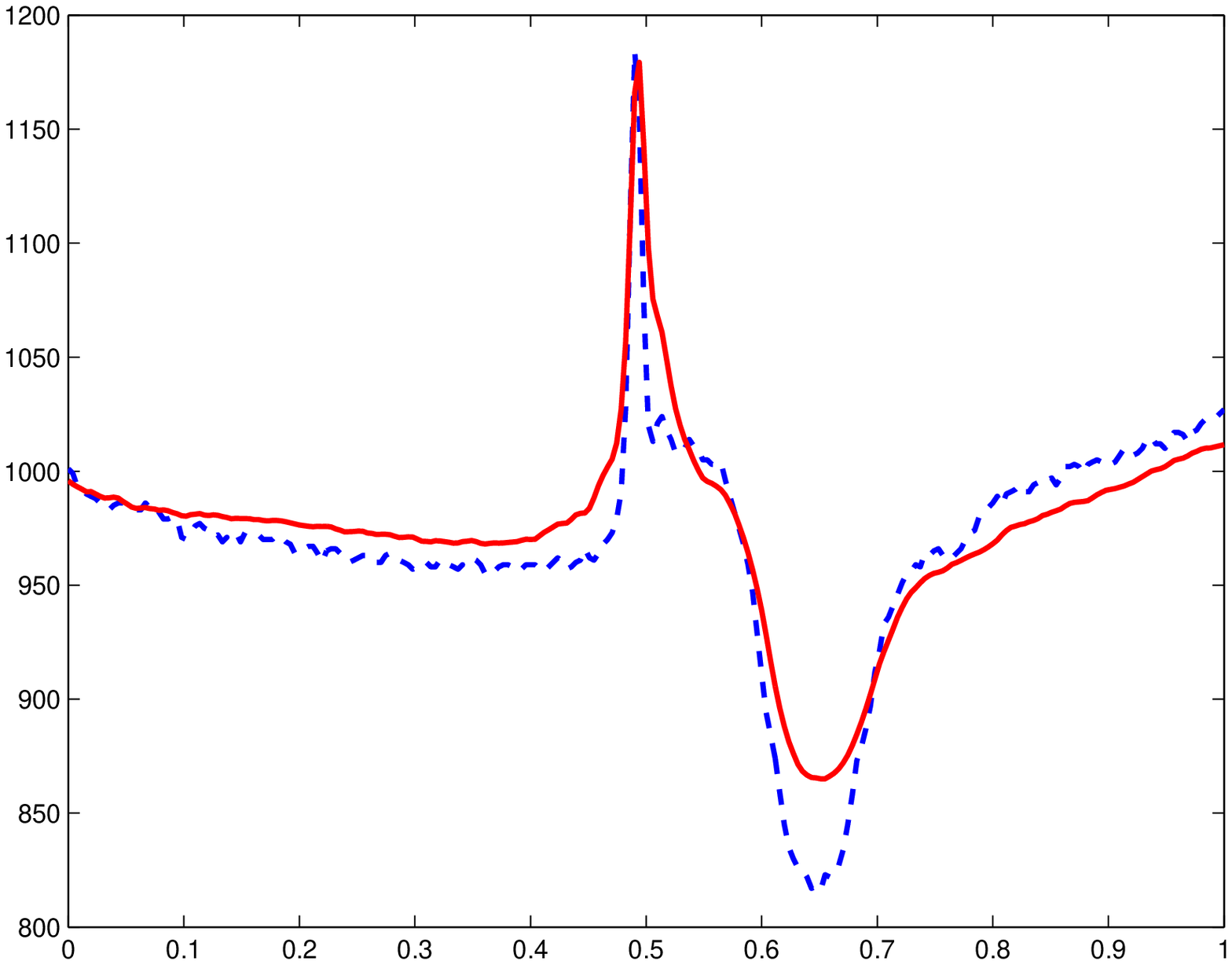} }

\hspace{4.1cm} 
\subfigure[]{ \includegraphics[width=3.5cm]{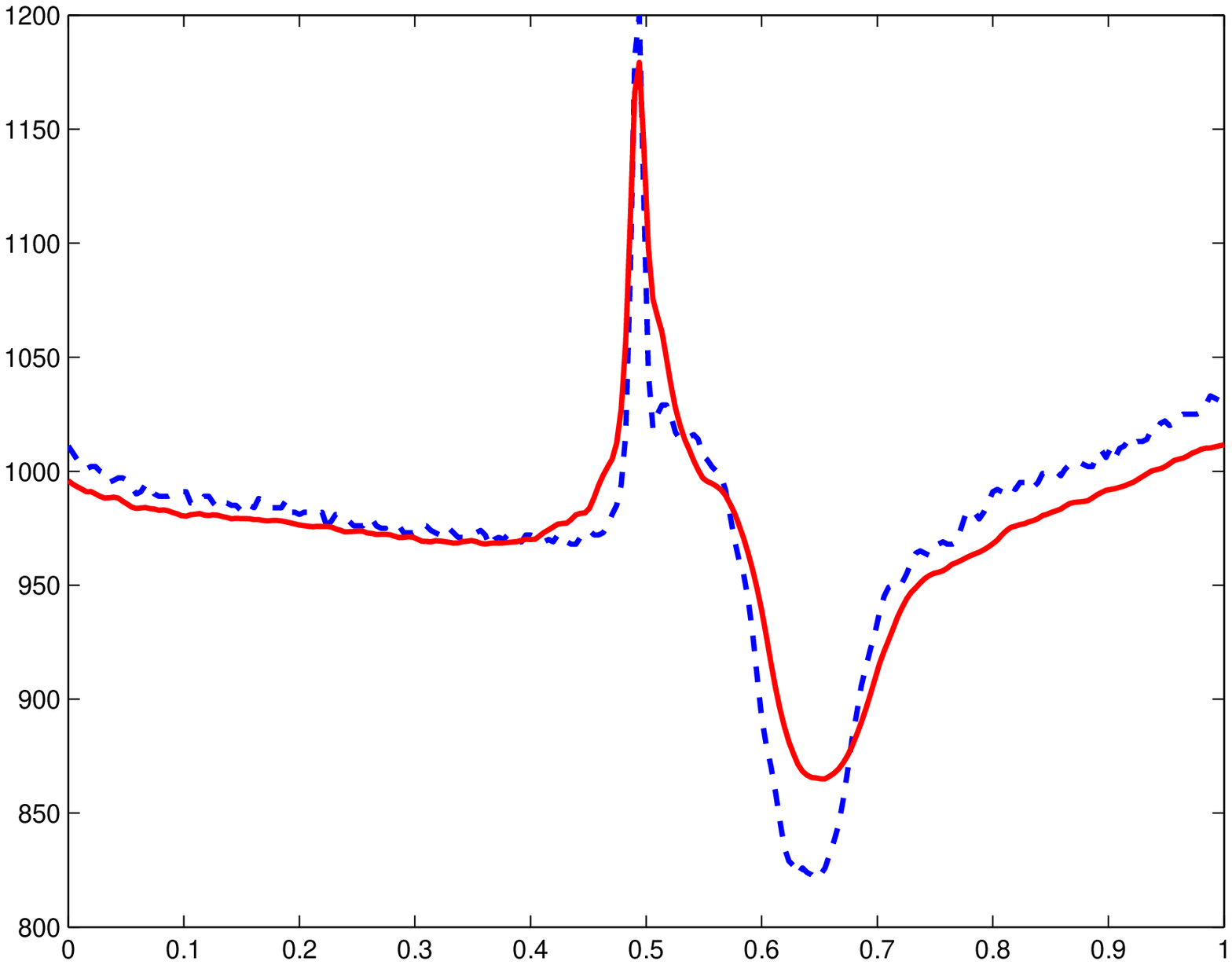} }
\subfigure[]{ \includegraphics[width=3.5cm]{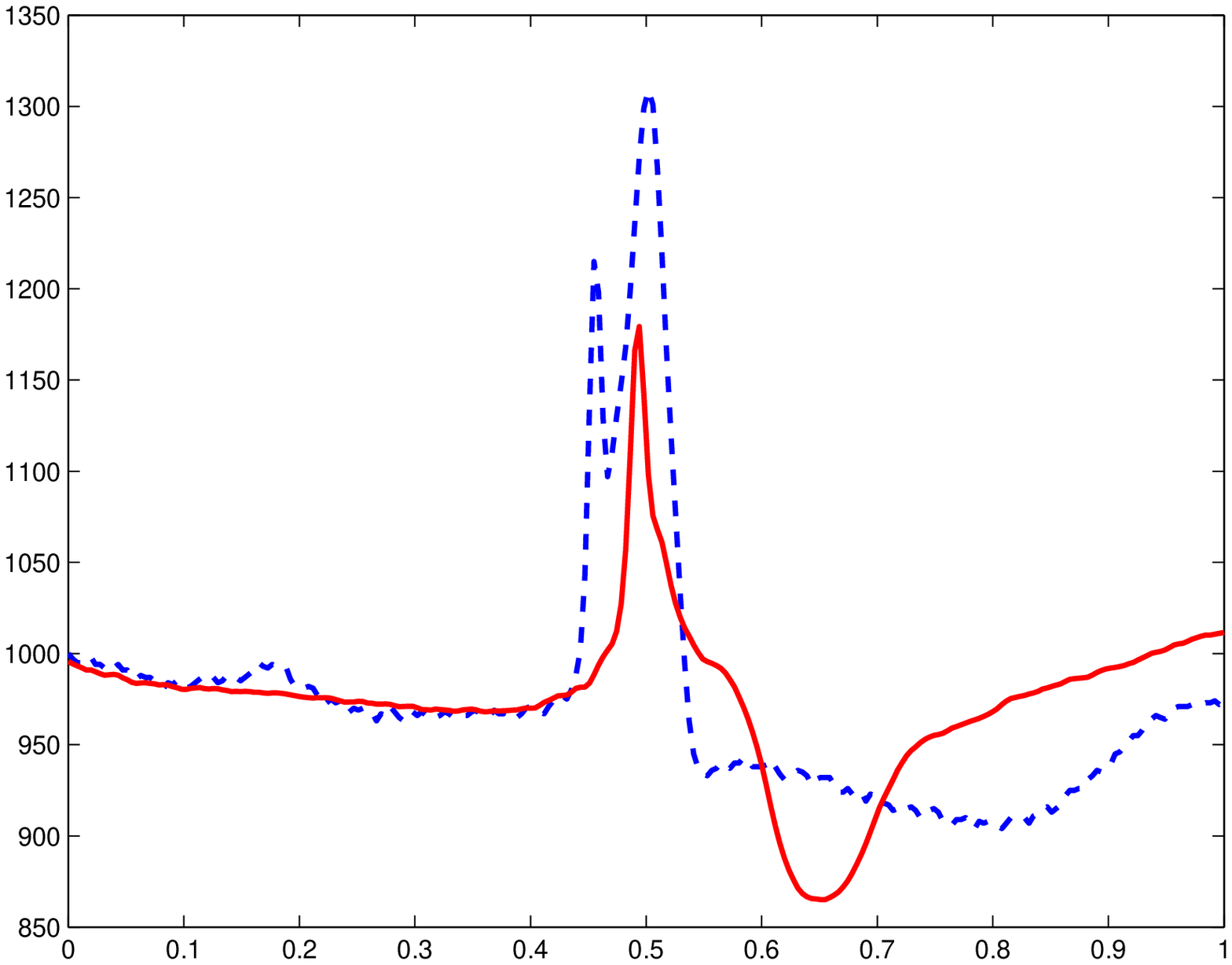} }
\subfigure[]{ \includegraphics[width=3.5cm]{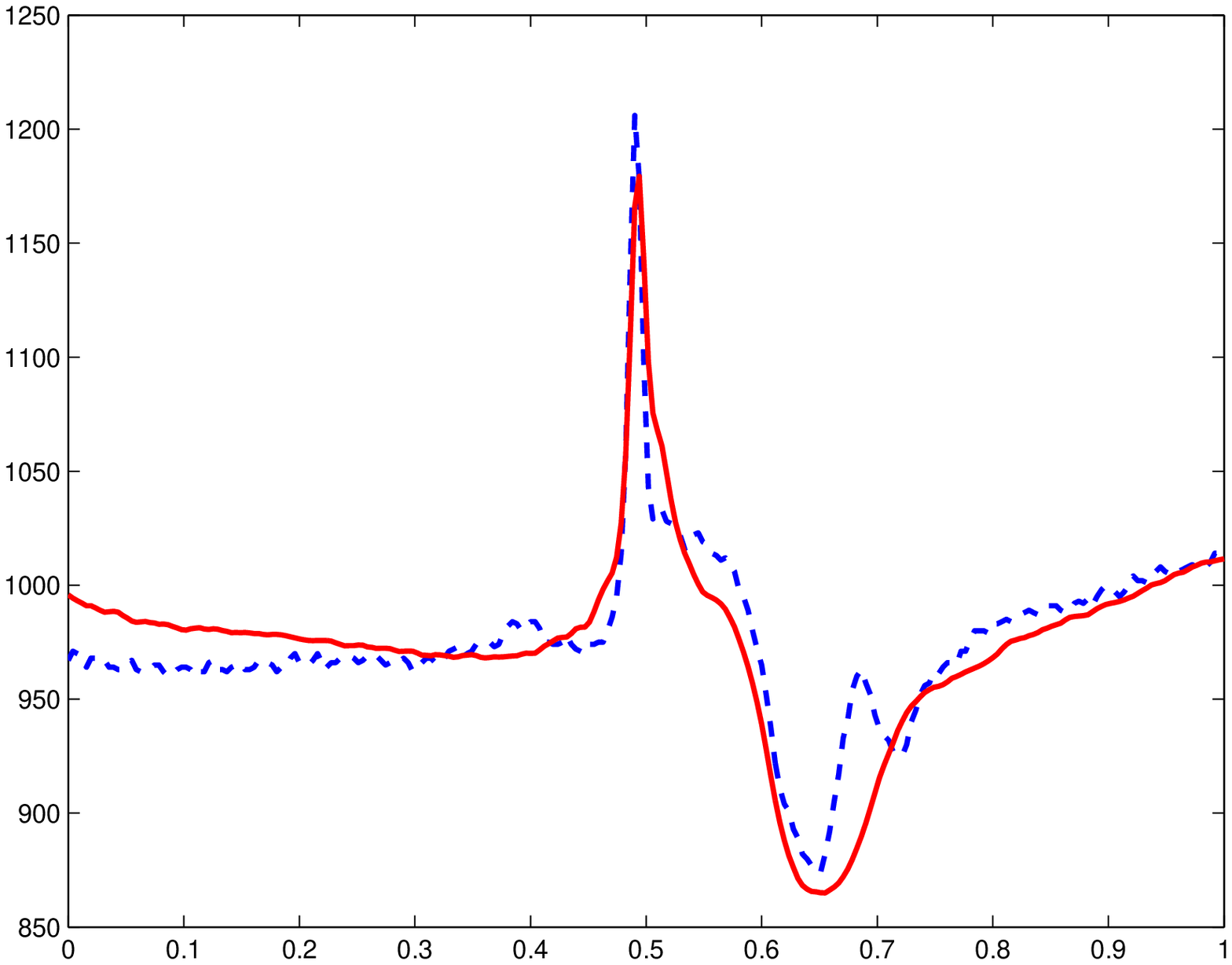} }

\caption{Case of cardiac arrhythmia: (a) Euclidean mean of the $J=72$  signals after segmentation of the ECG record. (b)-(g) Superposition of six signals containing a single QRS complex (dashed curves) with the Euclidean mean (solid curve).  } \label{fig:Arrhythmia:naive}
\end{figure}

 To estimate the typical shape of a heart cycle and to improve the signal-to-noise ratio, one may use the Euclidean mean of these signals.   In the normal case, this  leads to satisfactory results as shown in Figure \ref{fig:Normal:naive} since the  average signal displayed in Figure \ref{fig:Normal:naive}(a) clearly reflects the typical shape of the heart beats  displayed in Figure \ref{fig:Normal:naive}(b)-(g).  However, in the case of cardiac arrhythmia, the electrical activity of the heart is more irregular. This can be seen in the shape of the heart beats displayed in  Figure \ref{fig:Arrhythmia}(b)-(g) which may vary significantly from one pulse to another. Due to noise and geometric variability in the measurements, a simple averaging step may cause a low-pass filtering effect that leads to a mean  cycle that does not reflect the typical shape of the heart beats  in the ECG record \cite{3951206, 3951207,12669991}.    In the case of cardiac arrhythmia,  the Euclidean mean causes a low-pass filtering effect in the shape of the QRS complex, as shown by the results displayed in Figure \ref{fig:Arrhythmia:naive}. Indeed, it can be seen in  Figure \ref{fig:Arrhythmia} that,  around the time point $t \approx 0.45$ (which corresponds to the beginning of the QRS complex) there is rapid transition between a flat region  and the peak of the R wave. For the Euclidean mean displayed in Figure \ref{fig:Arrhythmia:naive}(a), this transition is slower, and it can be seen that the average heart cycle  in Figure \ref{fig:Arrhythmia:naive}(a) is a convolution by a smooth kernel of the signals in  Figure \ref{fig:Arrhythmia}(b)-(g) around the time point $t \approx 0.45$. To obtain better results, it is necessary to use alignment techniques for a precise synchronization of the heart beats, and then to average the aligned data.

\subsection{Deformable models}

Assume for  simplicity that the signals are observed on the time interval $[0,1]$, and  that they can be extended outside $[0,1]$ by periodicity. An alignment technique consists in finding a time synchronization of a set of signals. To be more precise, define a deformation operator $\phi_{\btheta}$ parametrized by $\btheta \in \RR^{p}$ as a smooth increasing function $\phi_{\btheta} : [0,1] \to \RR$ such that
$$
\phi_{\btheta}^{-1}(t) = \phi_{-\btheta}(t) \mbox{ for all } t \in [0,1].
$$
In the paper, we shall consider the following families of deformation operators:
\begin{description}
\item[-] {\it Translation operators}: $\phi_{\btheta}(t) = t-\btheta $ and $\phi_{\btheta}^{-1}(t) = \phi_{-\btheta}(t) = t+ \btheta $, for $\btheta \in \RR$ ($p=1$) and all $t \in [0,1]$.
\item[-] {\it Non-rigid operators}: $\phi_{\btheta} : [0,1] \to [0,1]$ is diffeomorphism of $[0,1]$ parametrized by some $\btheta \in \RR^{p}$, {\it i.e.} a smooth increasing function with $\phi_{\btheta}(0)  = 0$ and $\phi_{\btheta}(1)   = 1$ (a general method for constructing  non-rigid deformation operators is described in Section \ref{sec:diffeo}).
\end{description}

Given $f_{1},f_{2} : [0,1] \to \RR$ and a family $(\phi_{\btheta})_{\btheta \in \RR^{p}}$ of deformation operators, the problem of time synchronization of two signals is to find a $\btheta \in \RR^{p}$ such $f_{1}(\phi_{\btheta}(t)) \approx f_{2}(t)$ for all $t \in [0,1]$. In ECG data analysis, the most widely used alignment technique is time synchronization using translation operators by temporal or multiscale cross-correlation, see \cite{12669991,TriganoIR11} and references therein.

In the presence of variability in amplitude and shape in the data, it is  reasonable to suppose that the signals at hand satisfy the following deformable (or perturbation) model:
 \begin{equation} \label{eq:model}
Y_{j}^{\ell} =  f(\phi_{\btheta_{j}^{\ast}}(t_{\ell})) +  Z_{j}(\phi_{\btheta_{j}^{\ast}}(t_{\ell})) +\sigma \varepsilon_{j}^{\ell}, \quad j=1,\ldots,J,  \; \text{ and }\; \ell = 1,\ldots,n,
\end{equation}
where $Y_{j}^{\ell}$ denotes the $\ell$-th observation for the $j$-th signal and $t_{\ell} = \frac{\ell}{n}$ are equi-spaced design points in $[0,1]$.  The  function  $f : [0,1] \to \RR$ in model \eqref{eq:model} is the unknown mean shape of the signals. The  $\varepsilon_{j}^{\ell}$ are supposed to be i.i.d.\ normal variables with zero expectation and variance $1$. They represent a standard white noise in the measurements.  The $Z_j$'s are supposed to be independent of the $\varepsilon_{j}^{\ell}$'s and are i.i.d.\ realizations of  a smooth random process $Z : [0,1] \to \R$ with zero expectation.  Therefore, the $Z_{j}$'s represent  smooth  variations in amplitude of the signals around $f$. Finally, the $\btheta_{j}^{\ast}$'s are assumed to be i.i.d.\ random variables in $\RR^{p}$ with zero expectation.  Hence, the random deformation operators  $\phi_{\btheta_{j}^{\ast}}$ represent geometric variability in time in the data.

Note that the deformable model \eqref{eq:model} is well adapted to ECG data processing. The main type of perturbations related to the analysis of ECG data (see {\it e.g.}\ \cite{12669991,TriganoIR11}) are the baseline wandering effect (a low-frequency signal), electromyographic (EMG) noise, and power-line interference which is an amplitude and frequency varying sinusoid. This source of additive noise is modeled in \eqref{eq:model} by the zero-mean processes $Z_{j}$. The physiological nature of the electrocardiographic signal also alters the recording from heartbeat to heartbeat in lag and duration. In the ECG signal, there are therefore variations in time of the heart cycle from one beat to another. This makes the heartbeats look shorter or longer in duration. After the segmentation of an ECG record into signals containing a single QRS complex,  this local variability in time is modeled in \eqref{eq:model} by the  non-rigid deformation operators  $\phi_{\btheta_{j}^{\ast}}$. Aligning heart beats using non-rigid deformation operators is an alternative to the cross-correlation method which is classical used in ECG data analysis. One of the goals of this paper is to show that using non-rigid operators to align signals obtained by the segmentation of an  ECG record  may significantly improve the  shape of the mean heart cycle computed by averaging the data after a standard alignment using cross-validation.

\subsection{Fr残het means of curves}

The problem of estimating $f$ and the deformation parameters $\btheta_{j}^{\ast}$ in model \eqref{eq:model} has been studied in \cite{BC11}  using the following procedure. First, for each $j=1,\ldots,J$, smooth the data $(Y_{j}^{\ell})_{\ell=1}^{n}$ to construct an estimator $\hat{f}_{j} : [0,1] \to \RR$ of $f \circ \phi_{\btheta_{j}^{\ast}} $. In this paper, this smoothing step is  done either by low-pass Fourier filtering or by wavelet thresholding. In a second step, estimate simultaneously the deformation parameters $\btheta_{j}^{\ast},j=1,\ldots,J$ by minimizing  the following criterion 
\begin{equation} \label{eq:crittheta}
(\hat{\btheta}_{1},\ldots,\hat{\btheta}_{J}) = \argmin_{ (\btheta_{1},\ldots,\btheta_{J}) \in  \bTheta_{0}} M(\btheta_{1},\ldots,\btheta_{J}),
\end{equation}
where
\begin{equation} \label{eq:Mcrittheta}
M(\btheta_{1},\ldots,\btheta_{J}) = \frac{1}{J} \sum_{j=1}^{J}   \int_{0}^{1} \bigg(   \hat f_{j} (\phi_{-\btheta_{j} }(t)) - \frac{1}{J} \sum_{j'=1}^{J}  \hat f_{j'}(\phi_{-\btheta_{j'} }(t))  \bigg)^{2} dt,
\end{equation}
and
$$
\bTheta_0  = \{(\btheta_1,\ldots,\btheta_J)\in (\RR^{p})^J, \ \btheta_1+\ldots+\btheta_J = 0  \}.
$$
Finally, in a third  step take 
\begin{equation} \label{eq:frechetmean}
\hat{f}(t) =  \frac{1}{J} \sum_{j=1}^{J} \hat f_{j}(\phi_{-\hat{\btheta}_{j} }(t)),
\end{equation}
as an estimator of the mean shape $f$. 

As explained in  \cite{BC11},  the estimator $\hat{f}$ can be interpreted as a smoothed Fr残het mean of the observed signals. The Fr残het mean  \cite{fre} is an extension of the usual Euclidean mean to   spaces endowed with non-Euclidean metrics. We refer to \cite{Afsari} and \cite{HuckemannAOS} for  recent overviews of this notion and its application to the analysis of random variables taking their values in non-linear manifolds.  Note that   this procedure does not require the use of a reference template to compute the estimators $\hat{\btheta}_{1},\ldots,\hat{\btheta}_{J}$ of the deformation parameters. Instead, one can interpret $\frac{1}{J} \sum_{j'=1}^{J}  \hat f_{j'}(\phi_{-\hat{\btheta}_{j'} }(t))$ as a template that is automatically computed during the minimization of \eqref{eq:Mcrittheta}, and onto which one searches to align all the curves $\hat f_{j}$.  Therefore, this approach avoids the problem of taking one of the observed signals as a reference template which can lead to poor alignment results  for data with a low signal-to-noise ratio. It should be also remarked that the constrained set $\bTheta_0$ (onto which the minimization of $M(\btheta_{1},\ldots,\btheta_{J})$ is done) reflects the assumption that  the deformation parameters $\btheta_{j}^{\ast}$ in model \eqref{eq:model} have zero expectation. The choice of this constraint    is also related to identifiability issues in model \eqref{eq:model}, and we refer to  \cite{BC11} for a detailed discussion on that point.

In \cite{BC11}, the statistical properties of $\hat{f}$ and the $\hat{\btheta}_{j}$'s have been studied in details in the asymptotic setting $n \to + \infty$ and/or $J \to + \infty$. However, the numerical performances of Fr残het means of curves have not been much studied in \cite{BC11}. The goal of this paper is then two folds. First, we describe new algorithms to compute $\hat{f}$ and the $\hat{\btheta}_{j}$'s in the case of translation and non-rigid operators. We also discuss the choice of the smoothing estimators $\hat{f}_{j}$. Secondly,  we show the benefits of the methodology proposed in \cite{BC11} for signal averaging of ECG data.

\subsection{Previous work on signal averaging}

In statistics, the problem of estimating the mean shape of a set of curves that differ by a time transformation is usually referred to as the curve registration problem. It has received a lot of attention in the statistical literature over the last two decades. Among the various methods that have been proposed, one can distinguish between landmark-based approaches see e.g.\ \cite{kneipgas}, \cite{MR2291263}, and  time synchronization to align a set of curves see e.g.\ \cite{ramli}, \cite{wanggas}, \cite{liumuller}, \cite{kg},  \cite{TriganoIR11}.

All these approaches are based either on the use of one the observed signals as a reference template (see {\it e.g.}\  \cite{TriganoIR11} for curve alignment with translation operators), or on the Procrustean mean which is a standard algorithm to estimate a mean shape without a reference template.  The Procrustean mean is  based on an alternative scheme between estimation of  deformation operators and averaging of back-transformed curves given estimated values of the deformation parameters. To be more precise, assume that $Y_{j} : [0,1] \to \RR$ denotes a linear interpolation of the data $(Y_{j}^{\ell})_{\ell=1}^{n}$. Using our notations, Procrustean mean consists of an initialization step $\hat{f}_{0} = \frac{1}{J} \sum_{j=1}^{J} Y_{j}$ which is the Euclidean mean of the raw data taken as a first reference template.  Then, at iteration $1 \leq i \leq i_{\max}$, it computes for all $1 \leq j \leq J$ the estimators $\hat{\btheta}_{j,i}$ of the $j$-th deformation parameter as $$\hat{\btheta}_{j,i} =  \arg \min_{\btheta \in \RR^{p}} \int_{0}^{1} |Y_{j}( \phi_{\btheta_{j} }(t) ) - \hat{f}_{i-1}(t) |^{2} dt $$ and then takes $ \hat{f}_{i}(t) =  \frac{1}{J} \sum_{j=1}^{J} Y_{j}( \phi_{\hat{\btheta}_{j,i}}(t) ) ) $  as a new reference template. This procedure is repeated until the estimated reference template does not change. Usually the algorithm converges in a few steps.

However, in the case of a low signal-to-noise ratio, it is preferable to first smooth the data before alignment. In this paper, to minimize the criterion \eqref{eq:Mcrittheta}, we  propose an alternative algorithm to  Procrustean mean. We also show that the obtained smoothed Fr残het mean \eqref{eq:frechetmean}  yields better results. In the statistical literature, the criterion \eqref{eq:crittheta} has been first proposed by \cite{mazaloubgam}  in the case of translation operators and then further studied by  \cite{vimond} and \cite{BG10}. Its generalization to other deformation operators and the connection between minimizing  \eqref{eq:Mcrittheta} and the computation of Fr残het means of curves has been investigated in \cite{BC11}.

\subsection{Organization of the paper}

In Section \ref{sec:algo} we describe a gradient descent algorithm to compute the $\hat{\btheta}_{j}$'s. We detail this procedure in the case of translation and non-rigid operators. We  also compare our approach with  the Procrustean mean of curves. Numerical experiments on signal averaging for ECG data are presented in Section \ref{sec:numerics}.  We conclude the paper by a brief discussion on the numerical results and the benefits of our approach.

\section{Methodology for mean pattern estimation} \label{sec:algo}

\subsection{A gradient descent algorithm with an adaptive step}

Denote by
$$
\nabla M (\btheta) = \left( \frac{\partial}{\partial \btheta_{1}}  M (\btheta), \dots, \frac{\partial}{\partial \btheta_{J}}  M (\btheta)  \right)  \in (\RR^{p})^J,
$$
the gradient of the criterion   \eqref{eq:Mcrittheta}  at $\btheta = (\btheta_{1},\ldots, \btheta_{J}) \in (\RR^{p})^J$. A closed form expression of this gradient is given in Section \ref{sec:shifts}  for translation operators, and in Section \ref{sec:diffeo} for non-rigid operators. Computation of the deformation parameters can then be done by minimizing the criterion $M(\btheta)$ using the following gradient descent algorithm with the constraint that $\sum_{j=1}^{J} \btheta_{j} = 0$:\\

\noindent {\bf Initialization : } let $\btheta^{0}  = 0$, $\delta_{0} = \frac{1}{\| \nabla M (\btheta^{0}) \|}$, $C(0) =M(\btheta^{0})$, and set $i = 0$.  \vspace{0.2cm}

\noindent {\bf Step 2 :} let $\tilde{\btheta} = \btheta^{i} - \delta_{i} \nabla M(\btheta^{i})$ and let $\btheta^{new} = \left(\tilde{\btheta}_{1} - \frac{1}{J} \sum_{j=1}^{J} \tilde{\btheta}_{j} ,\ldots, \tilde{\btheta}_{J} - \frac{1}{J} \sum_{j=1}^{J} \tilde{\btheta}_{j} \right) $.

\noindent Let $C(i+1) =M(\btheta^{new})$. 

\noindent {\bf While $C(i+1) > C(i)$ do}
$$
\delta_{i} = \delta_{i} / \kappa ,   \quad \tilde{\btheta} = \btheta^{i} - \delta_{i} \nabla M(\btheta^{i}),  \quad \btheta^{new} = \left(\tilde{\btheta}_{1} - \frac{1}{J} \sum_{j=1}^{J} \tilde{\btheta}_{j} ,\ldots, \tilde{\btheta}_{J} - \frac{1}{J} \sum_{j=1}^{J} \tilde{\btheta}_{j} \right) ,
$$
and set  $C(i+1) =M(\btheta^{new})$.\\
{\bf End while}

\noindent  Take $\btheta^{i+1} = \btheta^{new}$. \vspace{0.2cm}

\noindent {\bf Step 3 :} if $C(i) - C(i+1) \geq \rho(C(1) - C(i+1))$ then  set $i = i + 1$ and return to {\bf Step 2}, else stop the iterations, and take $\hat{\btheta} = \btheta^{i+1}$.  \\

In the above algorithm, $\rho > 0$ is a  stopping parameter and $\kappa > 1$ is a parameter to control the choice of the adaptive step $\delta_{i}$. 

\subsection{Choice of  the regularization parameter in the smoothing step} \label{sec:smoothing}

For the smoothing step, we present numerical results for
\begin{description}
\item[-] {\it low-pass Fourier filtering}: for $t \in [0,1] $
$$
\hat{f}_{j}(t) = \sum_{|k| \leq \hat{\lambda}_{j}} c_{k}^{(j)} e^{i 2 \pi k t}, 
$$
with $c_{k}^{(j)} = \frac{1}{n} \sum_{\ell = 1}^{n} Y_{j,\ell}  e^{-i 2 \pi k \frac{\ell}{n}} $, and where $\hat{\lambda}_{j} \in \N$ is a regularization parameter (cut-off frequency). A possible data-based choice for $\hat{\lambda}_{j} $ is to use generalized cross validation (GCV), see e.g.\ \cite{Craven}. 
\item[-] {\it wavelet smoothing by hard thresholding:} for $t \in [0,1] $
$$
\hat{f}_{j}(t) = \sum_{k = 0}^{2^{m_{0}}} \alpha_{m_{0},k}^{(j)} \phi_{m_{0},k}(t) + \sum_{m=m_{0}}^{m_{1}}  \sum_{k = 0}^{2^{m}}  \beta_{m,k}^{(j)} \1_{\left\{ \left| \beta_{m,k}^{(j)} \right| \geq  \hat{\sigma}_{j} \sqrt{2  \log(n) } \right\}} \psi_{m,k}(t),
$$
where $\phi_{m_{0},k}(t) = 2^{\frac{m_{0}}{2}} \phi(2^{m_{0}} t-  k)$ and $\psi_{m,k}(t) = 2^{\frac{m}{2}} \psi(2^{m} t-  k)$ are the usual scaling and wavelet basis functions at resolution levels $0 \leq m_{0} \leq m \leq m_{1}$ and location $k$, $\alpha_{m_{0},k}^{(j)} , \beta_{m,k}^{(j)}$ are respectively the empirical scaling and wavelet coefficients computed from the data $(Y_{j}^{\ell})_{\ell=1}^{n}$ (see {\it e.g.}\ \cite{ABS01} for further details on wavelet thresholding). The universal threshold $\hat{\sigma}_{j} \sqrt{2  \log(n) }$ depends on the estimation $\hat{\sigma}_{j}$ of the level of noise in the $j$-th signal. It is given by the  median absolute deviation (MAD) of the empirical wavelet coefficients at the highest level of resolution $m_{1}$ (see {\it e.g.}\ \cite{ABS01}).
\end{description}


\subsection{The case of translation operators} \label{sec:shifts}

When $\phi_{\theta}(t) = t-\theta$, it follows by simple computations that for $j=1,\ldots,J$
$$
\frac{\partial}{\partial \btheta_{j}} M (\btheta) = -\frac{2}{J} \int_{0}^{1} \frac{\partial}{\partial t} \hat f_{j}(t+\btheta_{j}) \left( \frac{1}{J} \sum_{j'=1}^{J}  \hat f_{j'}(t + \btheta_{j'}) \right) dt + \frac{2}{J} \left( |\hat f_{j}(1+\btheta_{j})|^2 - |\hat f_{j}(\btheta_{j})|^2\right).
$$
Note that in the case of low-pass Fourier filtering, $\hat f_{j}(t+1) = \hat f_{j}(t)$ for any $t \in \RR$, and the gradient of $M$ can be computed directly in the Fourier domain thanks to Parseval's relation
$$
\frac{\partial}{\partial \btheta_{j}} M (\btheta) =    -\frac{2}{J}   \sum_{|k| \leq \lambda } \Re \left[ 2 i \pi k  c_{k}^{(j)}e^{2 i k \pi \btheta_{j}}  \left( \overline{  \frac{1}{J} \sum_{j'=1}^{J} c_{k}^{(j')} e^{2 i k \pi  \btheta_{j'} } } \right) \right],
$$
where $\Re(c)$ denotes the real part of a complex number $c$.

\subsection{The case of non-rigid operators} \label{sec:diffeo}

 To build a family  $(\phi_{\btheta})_{\btheta \in \RR^{p}}$ of parametric diffeomorphisms of $[0,1]$, we adapt to one-dimensional curves the approach proposed in  \cite{BGL09} to compute the mean pattern of a set of two-dimensional images. Let $v : [0,1] \to \R$ be a smooth parametric vector field given by a linear combination of $p$ basis functions
$\{h_k : [0,1] \to \RR,\ k=1,\dots, p\}$, such that
$$
v(t) =  \sum_{k=1}^p \theta_k  h_k(t)    \mbox{ for } t \in [0,1],
$$
where  $\btheta = (\theta_1,\ldots,\theta_p) \in \RR^{p} $ is a set of  reals coefficients. The function $v$ is thus parametrized by the set of coefficients $\btheta$, and we write $v = v_{\btheta}$ to stress this dependency. In what follows, it will be assumed that the basis functions are continuously differentiable on $[0,1]$ and such that $h_{k}$ and $\frac{\partial}{\partial t} h_{k}$ vanish at $t=0$ and $t=1$.  For the $h_{k}$'s we took in our numerical experiments a set of $p=10$ B-spline functions of degree $3$ using equally-spaced knots on $[0,1]$. Then, let $t \in [0,1]$ and for $u \in [0,1]$  consider the following ordinary differential equation (ODE)
\begin{equation}
\frac{\partial}{\partial u} \psi(u,t) = v_{\btheta}( \psi(u,t) ) \label{eq:ODE}
\end{equation}
with initial condition $\psi(0,t) = t$. Then, it can be shown (see e.g.\ \cite{MR2656312}) that for any $u \in [0,1]$ the solution of the above ODE is unique and such that
$$
t \mapsto \psi_{\btheta}(u,t) = t + \int_{0}^{u} v_{\btheta}( \psi_{\btheta}(s,t) ) ds
$$
 is a diffeomorphism of $[0,1]$ i.e.\ a smooth increasing function with   $\psi_{\btheta}(u,0) = 0$ and $\psi_{\btheta}(u,1) = 0$. Then, we denote by $$\phi_{\btheta}(t) = \psi_{\btheta}(1,t)$$ the solution at $u=1$ of the ODE \eqref{eq:ODE}. In this way, we finally obtain a diffeomorphism $\phi_{\btheta}$ that is parametrized by the set of coefficients $\btheta \in \RR^{p}$, and that is such that $\phi^{-1}_{\btheta}(t) = \phi_{-\btheta}(t)$. Then, it can be also shown that the mapping $\btheta \mapsto \phi_{\btheta}$ is differentiable, and thanks to Lemma 2.1 in \cite{beg} it follows that
$$
\frac{\partial}{\partial \theta_k }  \phi_{\btheta}(t) = \frac{\partial}{\partial t } \phi_{\btheta}(t) \int_{0}^{1} \left(  \frac{\partial}{\partial t } \psi_{\btheta}(u,t)  \right)^{-1} h_k( \psi_{\btheta}(u,t)) du. 
$$

Now, one obtains that if $\phi_{\theta}$ is a diffeomorphism  generated by the ODE \eqref{eq:ODE}, then  for $\btheta_{j} = (\theta_{j}^{(1)},\ldots,\theta_{j}^{(p)}) \in \RR^{p}$
$$
\frac{\partial}{\partial \btheta_{j}} M (\btheta) = \left(\frac{\partial}{\partial \theta_{j}^{(1)}} M (\btheta), \ldots, \frac{\partial}{\partial \theta_{j}^{(p)}} M (\btheta)  \right),
$$
with for $q = 1,\ldots,p$
$$
\frac{\partial}{\partial \theta_{j}^{(q)}} M (\btheta) = -\frac{2}{J} \int_{0}^{1} \frac{\partial}{\partial \theta_{j}^{(q)}}  \phi_{-\btheta_{j}}(t) \frac{\partial}{\partial t} \hat f_{j}(\phi_{-\btheta_{j}}(t)) \left( \frac{1}{J} \sum_{j'=1}^{J}  \hat f_{j'}(\phi_{-\btheta_{j'}}(t)) -  \hat f_{j}(\phi_{-\btheta_{j}}(t)) \right) dt  
$$
and
$$
 \frac{\partial}{\partial \theta_{j}^{(q)}}  \phi_{-\btheta_{j}}(t) = \frac{\partial}{\partial t } \phi_{-\btheta_{j}}(t) \int_{0}^{1} \left(  \frac{\partial}{\partial t } \psi_{-\btheta_{j}}(u,t)  \right)^{-1} h_q( \psi_{-\btheta_{j}}(u,t)) du.
$$

\subsection{Comparison between the smoothed Fr残het mean and the Procrustes mean}

\begin{figure}[htbp]
\centering
\subfigure[]{ \includegraphics[width=3.5cm]{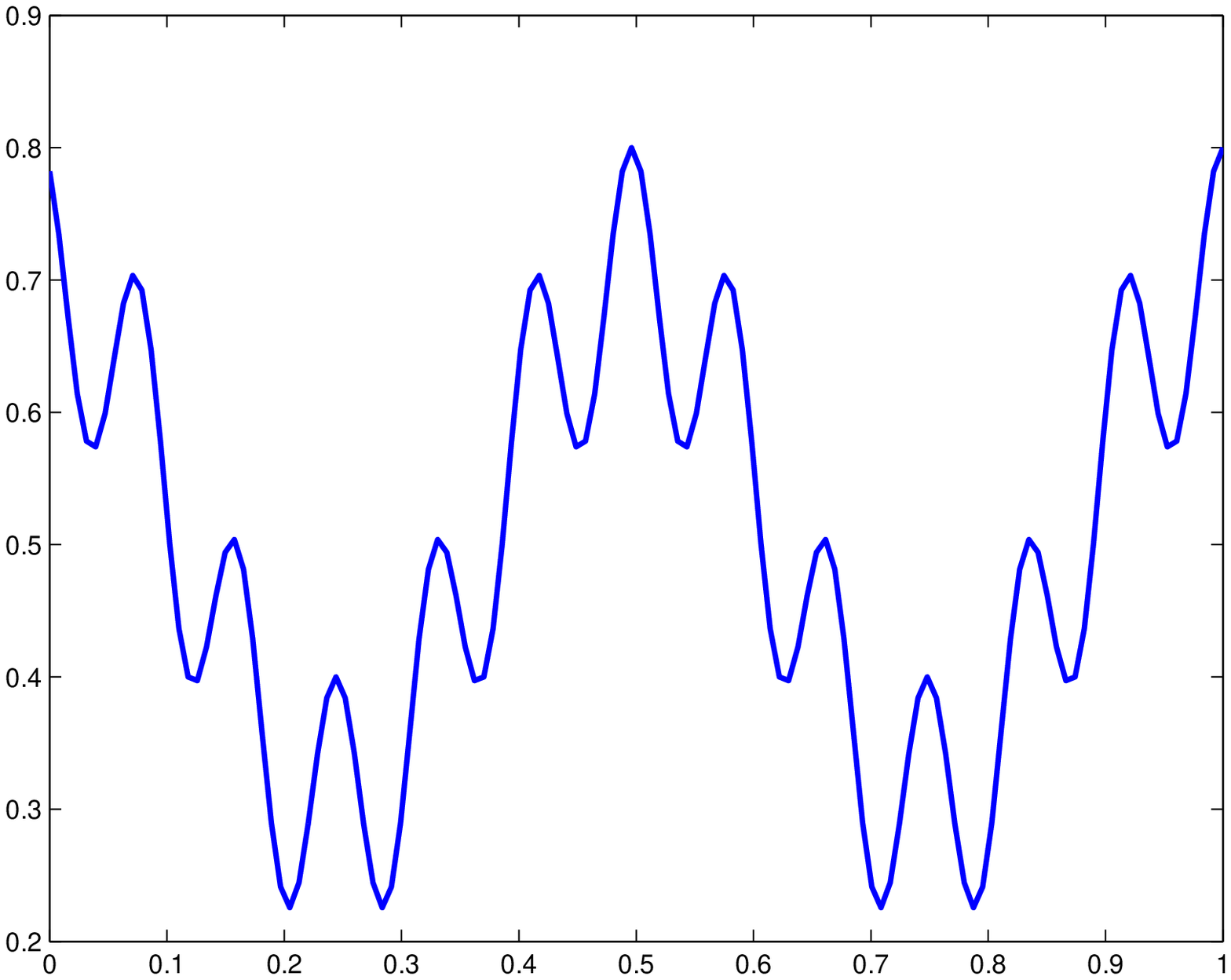} }
\hspace{0.2cm}
\subfigure[]{ \includegraphics[width=3.5cm]{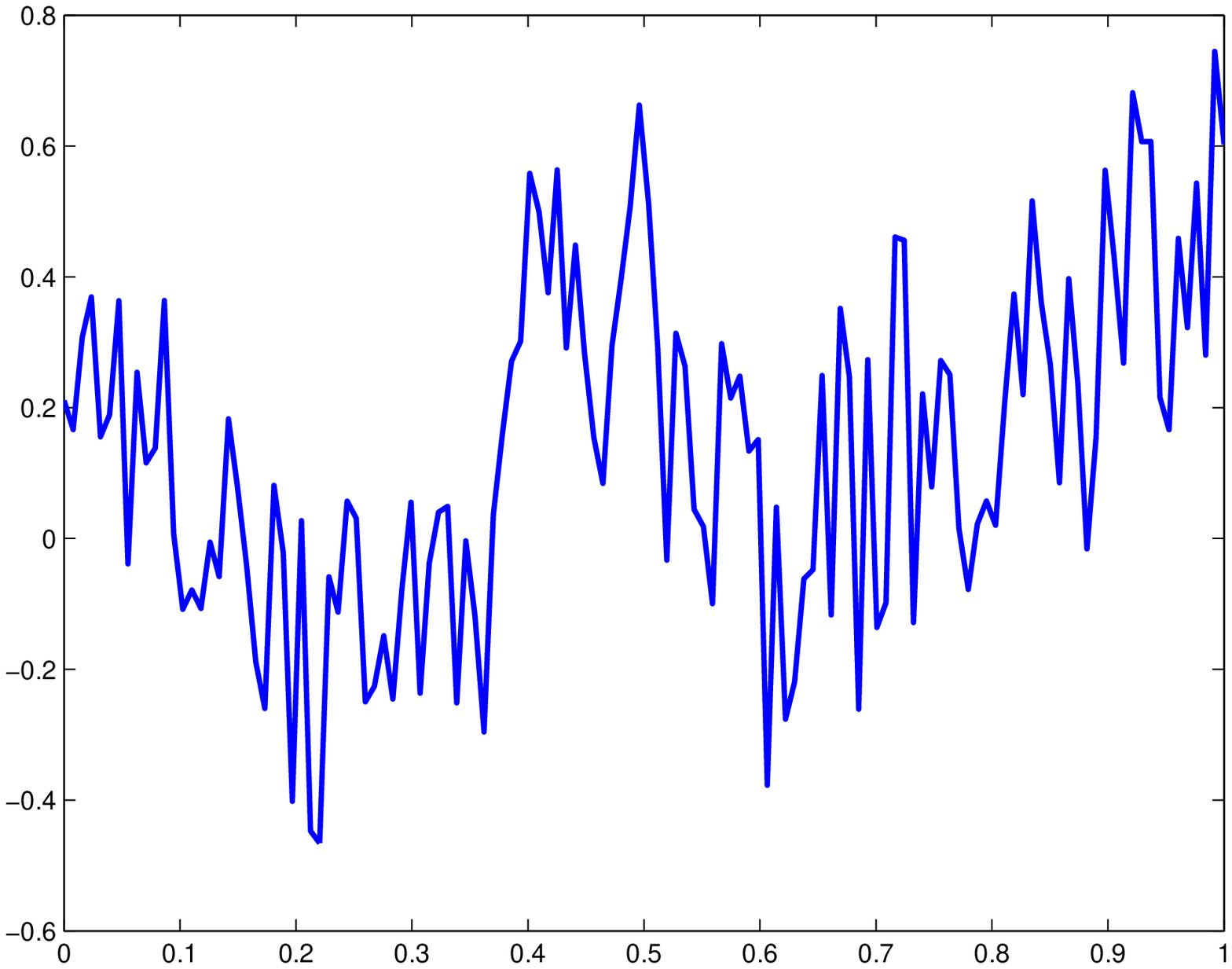} }
\subfigure[]{ \includegraphics[width=3.5cm]{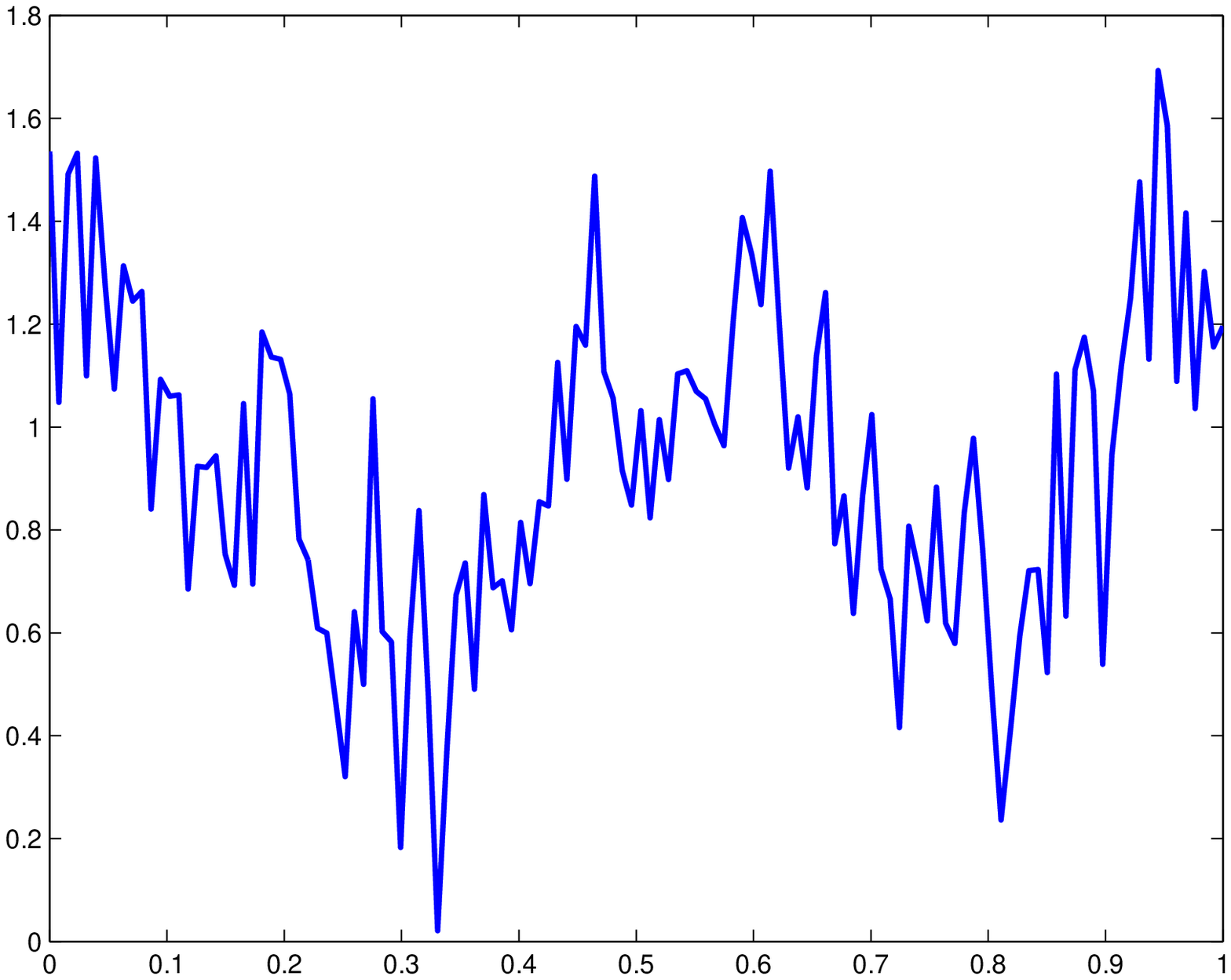} }
\subfigure[]{ \includegraphics[width=3.5cm]{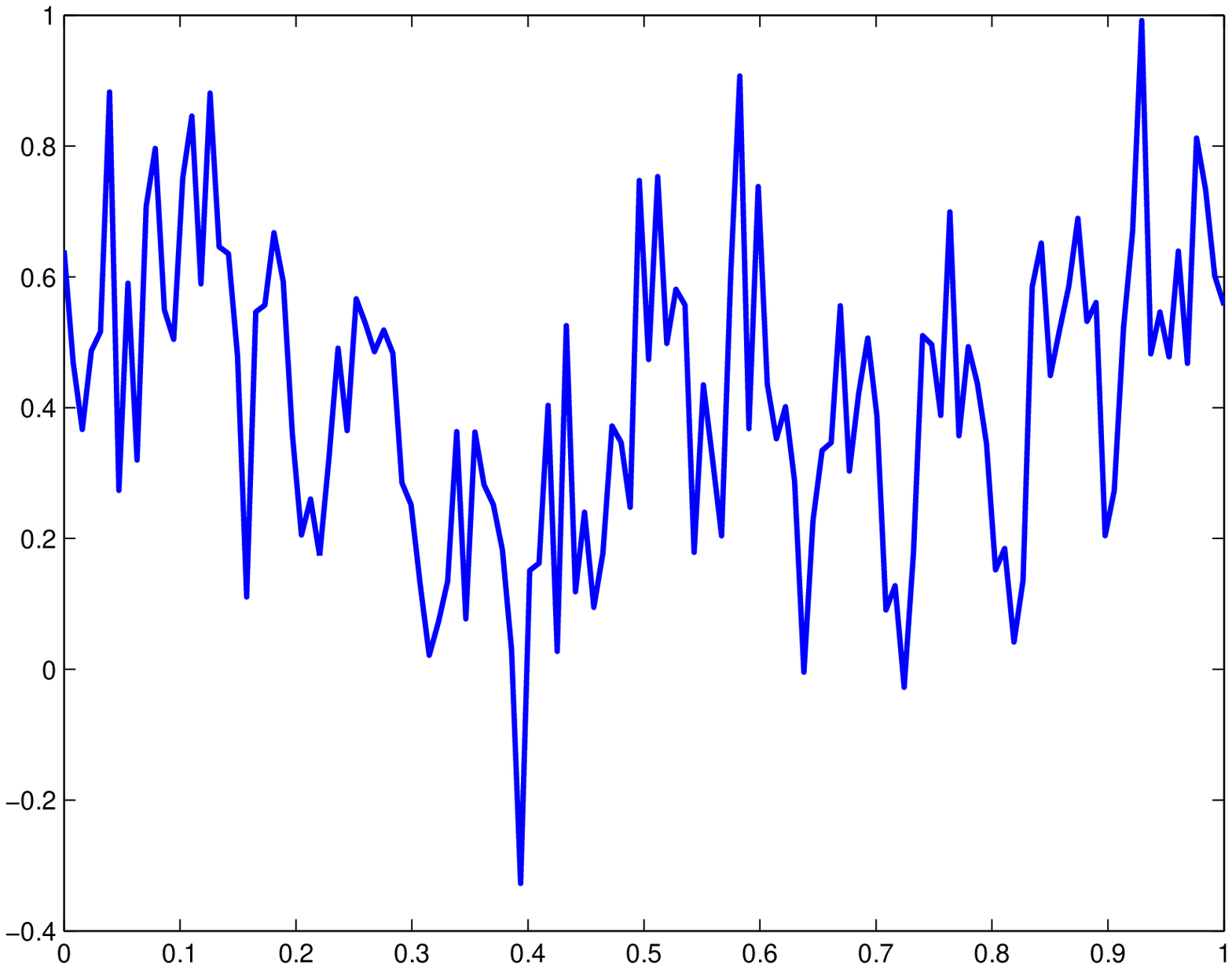} }

\hspace{0.3cm}
\subfigure[]{ \includegraphics[width=3.5cm]{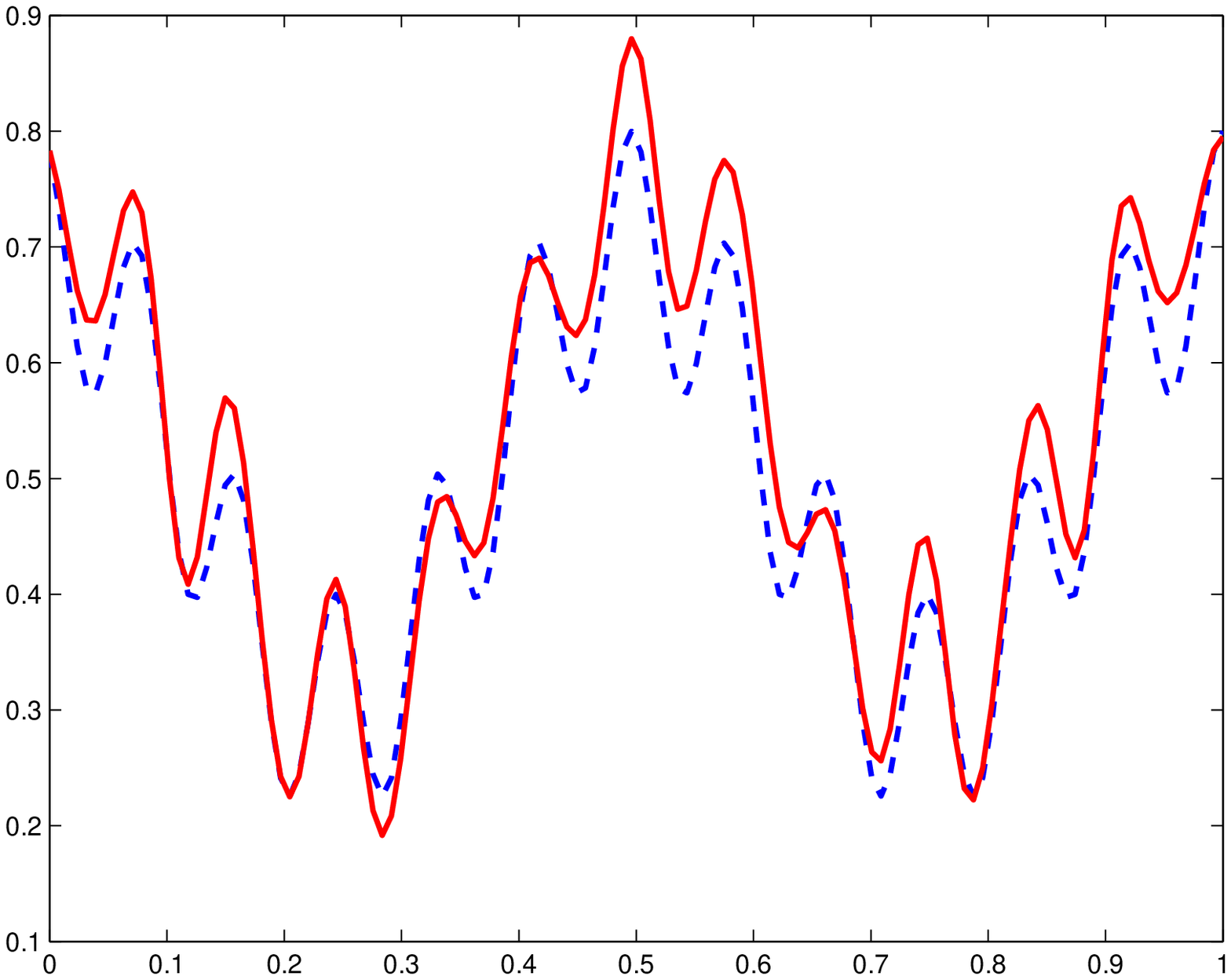} }
\subfigure[]{ \includegraphics[width=3.5cm]{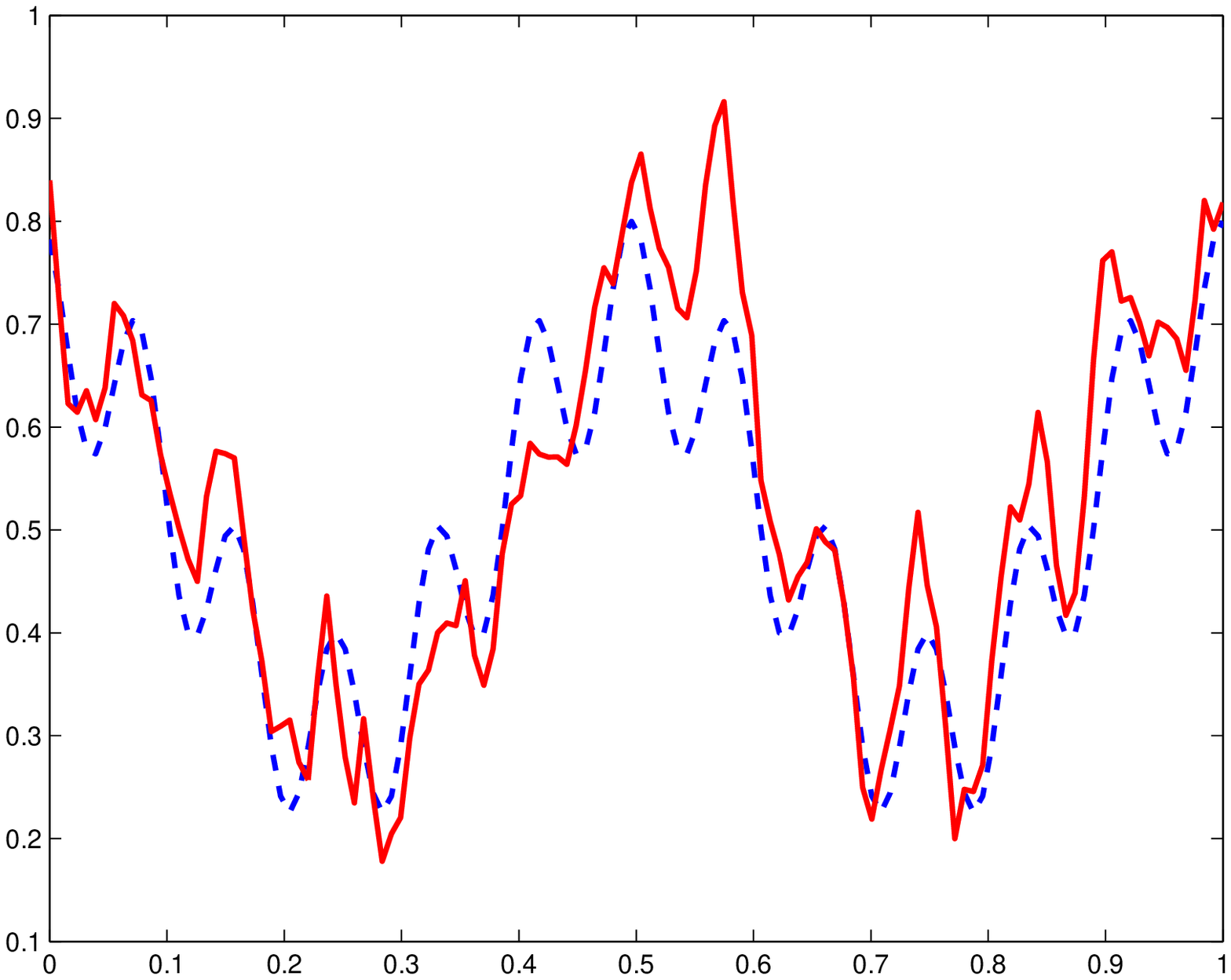} }

\caption{(a) Unknown mean shape $f$, (b)-(d) Three noisy signals out of $J=15$ sampled from model \eqref{eq:model:shift},  (e) Estimation of $f$ by smoothed Fr残het mean (solid curve) using low-pass Fourier filtering with $\hat{\lambda}_{j}$ chosen by GCV, (f) Estimation of $f$ by  Procrustes mean  without smoothing  (solid curve). The dashed curve in (e)  and (f) is the mean shape $f$.} \label{fig:wave}
\end{figure}

To illustrate the advantages of the smoothed Fr残het mean over the Procrustes mean of the raw data (without any smoothing), let us consider a set of $J=15$ signals generated from the following deformable model using translation operators:
 \begin{equation} \label{eq:model:shift}
Y_{j}^{\ell} =  f(t_{\ell} -\btheta_{j}^{\ast} ) +  Z_{j}(t_{\ell} -\btheta_{j}^{\ast} ) +\sigma \varepsilon_{j}^{\ell}, \quad j=1,\ldots,J,  \; \text{ and }\; \ell = 1,\ldots,n,
\end{equation}
where $n=128$, the $\btheta_{j}^{\ast}$'s are i.i.d.\ normal variables with zero mean and variance $\mu^{2} = 0.004$, the $\varepsilon_{j}^{\ell}$'s are i.i.d.\ normal variables with zero mean and variance $\sigma^2$, $f$ is the signal displayed in Figure \ref{fig:wave}(a), and the $Z_{j}$'s are  i.i.d.\ copies of the  Gaussian process $Z$ defined by
$$
Z(t) = a_{0} + \sum_{k=1}^{+ \infty}  k^{-3/2} (a_{k} c_{k}(t) + b_{k} s_{k}(t)), \; t  \in [0,1],
$$
where $a_{0}$ and $(a_{k},b_{k})_{k \geq 1}$ are i.i.d.\ normal variables with zero mean and variance $\sigma^{2}$, $c_{k}(t) = \sqrt{2} \cos(2 \pi k t)$ and $s_{k}(t) = \sqrt{2} \sin(2 \pi k t)$. A sample of three signals out of $J=15$ generated from model \eqref{eq:model:shift} is displayed in Figure \ref{fig:wave}(b)-(d). It can be seen that $\sigma$ has been chosen to have a relatively low signal-to-noise ratio. The estimation of $f$ by smoothed Fr残het mean   using low-pass Fourier filtering with $\hat{\lambda}_{j}$ chosen by GCV for each $j=1,\ldots,J$ is displayed in Figure \ref{fig:wave}(e), while the estimation obtained by Procrustes mean of the  data without  smoothing is shown in  Figure \ref{fig:wave}(f). Clearly, the result obtained  by the smoothed Fr残het mean  is much more satisfactory. 

\begin{figure}[htbp]
\centering
\includegraphics[width=9cm]{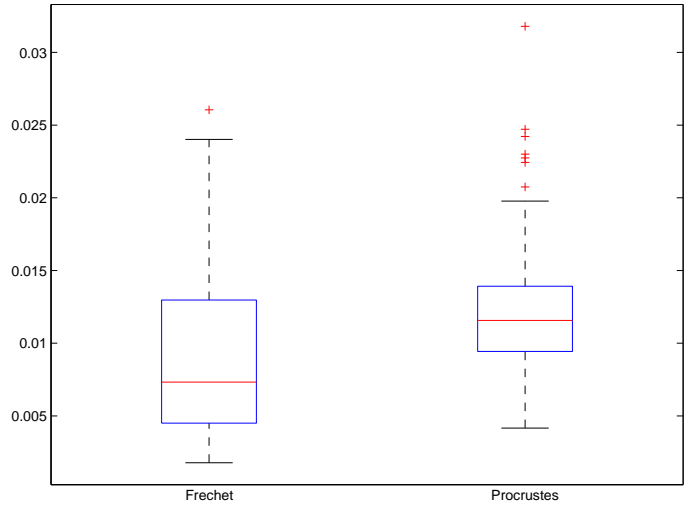}

\caption{Boxplots of $MSE( \hat{f}^{(m)}_{Frechet} )$ (left-hand boxplot)  and $MSE( \hat{f}^{(m)}_{Procrustes} )$ (right-hand boxplot) for $m=1,\ldots,M$.} \label{fig:boxplot}
\end{figure}

An important choice is thus the calibration of the regularization parameters $\hat{\lambda}_{j}$. To illustrate the benefits of using  the smoothed Fr残het mean with an automatic choice of the $\hat{\lambda}_{j}$'s by GCV, we used $M=100$ replications of model \eqref{eq:model:shift} with $J=15$. For each repetition $m=1,\ldots,M$, we have computed an estimator $\hat{f}^{(m)}_{Procrustes}$ of $f$ using the Procrustes mean  without smoothing, and an estimator $\hat{f}^{(m)}_{Frechet}$ of $f$ using a smoothed Fr残het mean where each $\hat{\lambda}_{j}$ is chosen by denoising the $j$-th signal by GCV. In Figure \ref{fig:boxplot}, we display boxplots (for $m=1,\ldots,M$) of the  mean squared errors
$$
MSE( \hat{f}^{(m)}_{Frechet} ) := \frac{1}{n} \sum_{\ell =1}^{n} \left| \hat{f}^{(m)}_{Frechet}(t_{\ell}) - f(t_{\ell}) \right|^2  \mbox{ and } MSE( \hat{f}^{(m)}_{Procrustes} ) := \frac{1}{n} \sum_{\ell =1}^{n} \left| \hat{f}^{(m)}_{Procrustes}(t_{\ell}) - f(t_{\ell}) \right|^2.
$$
It can be seen that, over $M=100$ replications, the mean squared error of the smoothed Fr残het mean is clearly much lower than the one obtained by Procrustes mean.

\section{Application to ECG data analysis} \label{sec:numerics}

Now, let us return to the analysis of the  ECG records displayed in Figure \ref{fig:data}. We propose to compare the results obtained by the smoothed Fr残het mean using either translation operators or non-rigid operators given by the family of parametric diffeomorphisms of $[0,1]$  described in Section \ref{sec:diffeo}. The smoothing of the $J$ signals obtained after segmentation of the ECG records is done by wavelet thresholding with a data-based choice of the regularization parameters $\hat{\sigma}_{j}$ as explained in Section \ref{sec:smoothing}.

\begin{figure}[htbp]
\centering
\subfigure[]{ \includegraphics[width=3.5cm]{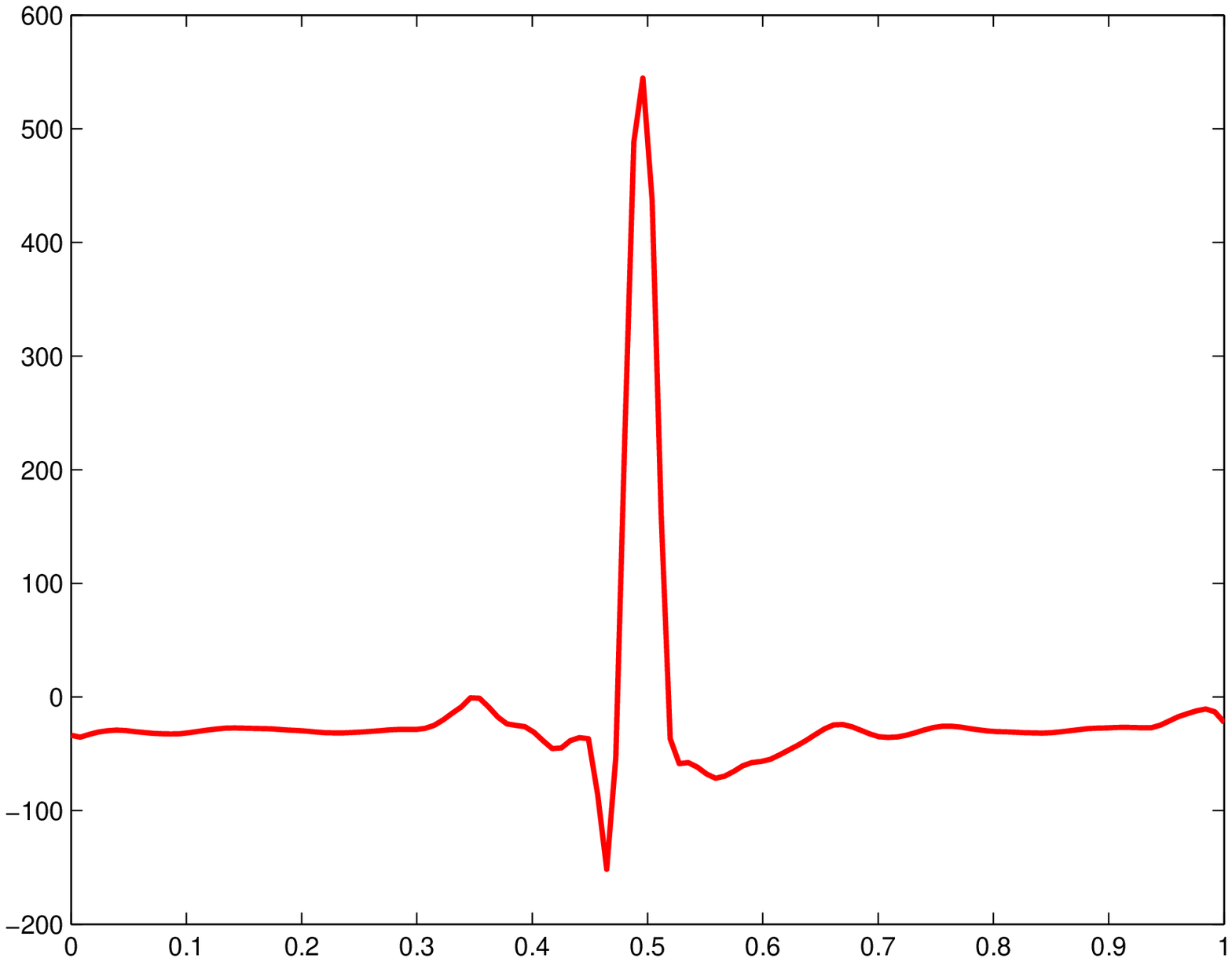} }
\hspace{0.2cm}
\subfigure[]{ \includegraphics[width=3.5cm]{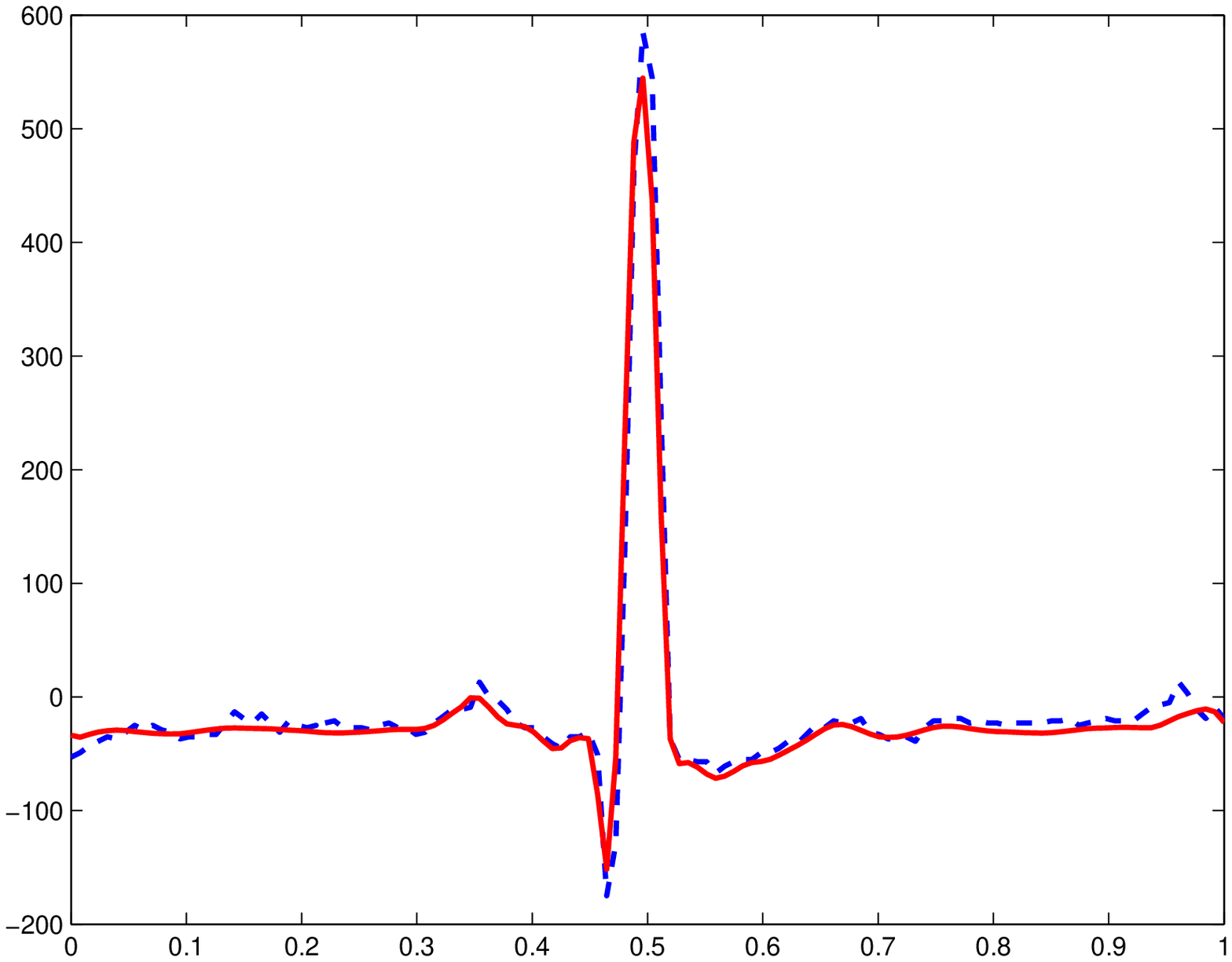} }
\subfigure[]{ \includegraphics[width=3.5cm]{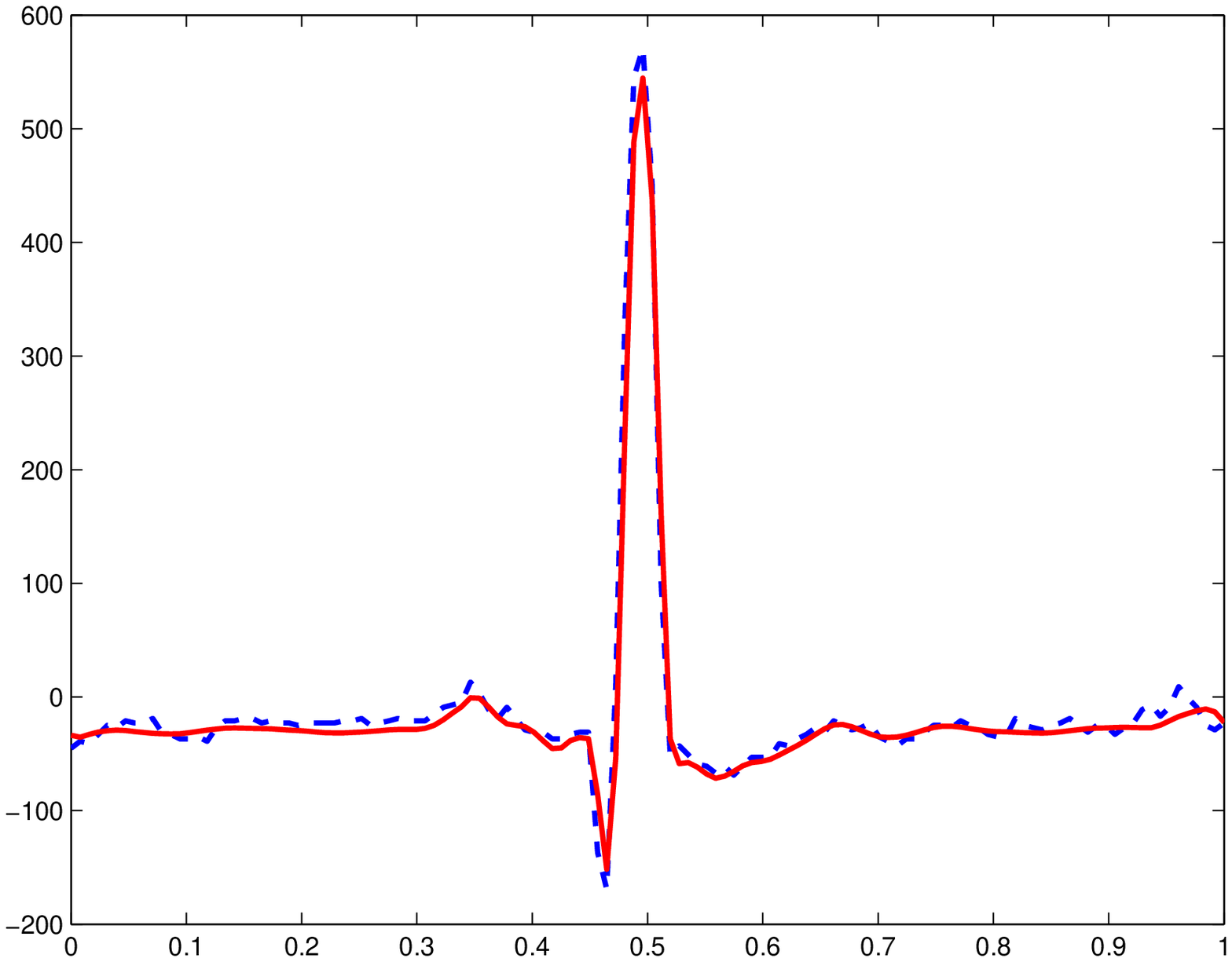} }
\subfigure[]{ \includegraphics[width=3.5cm]{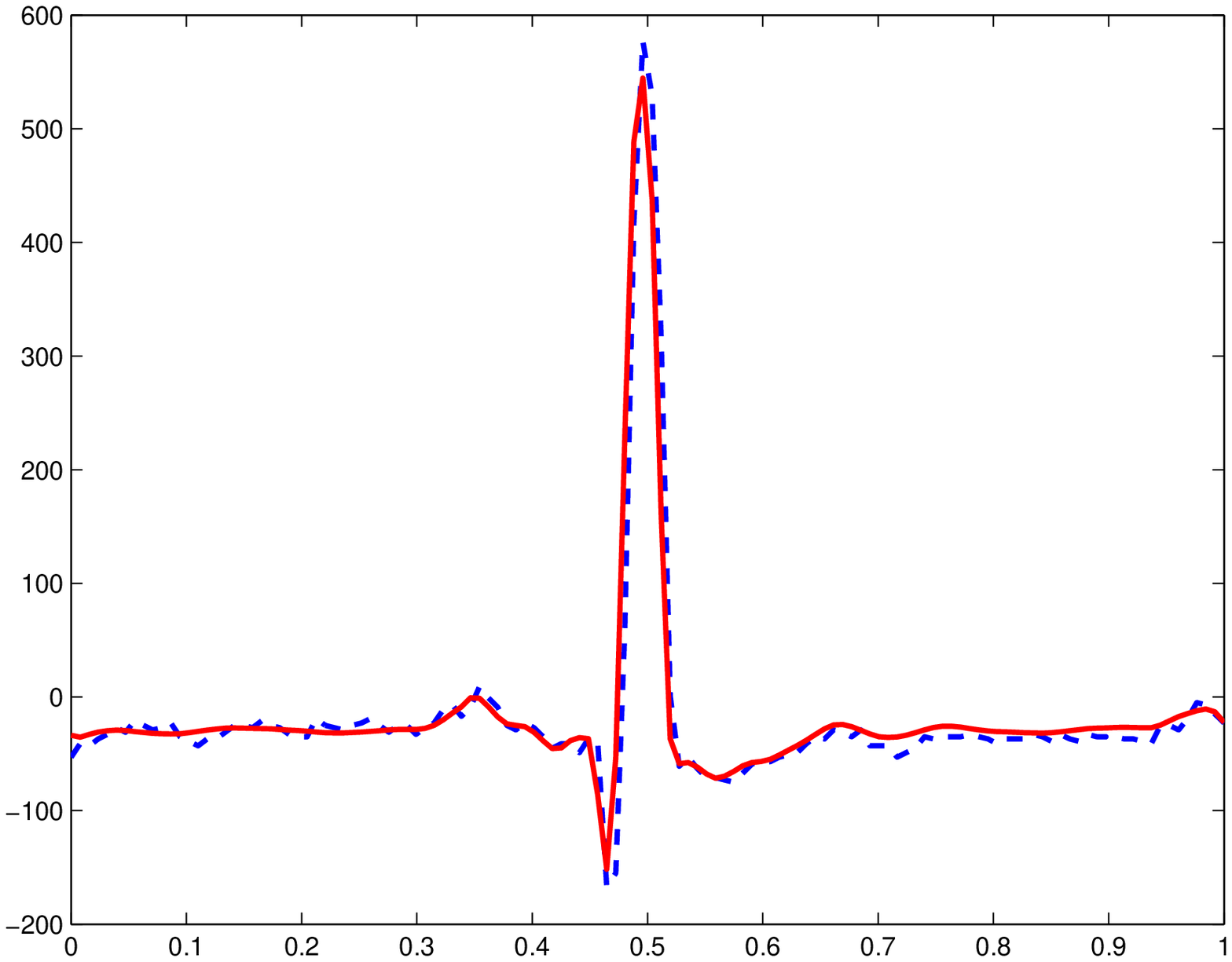} }

\hspace{4.1cm}
\subfigure[]{ \includegraphics[width=3.5cm]{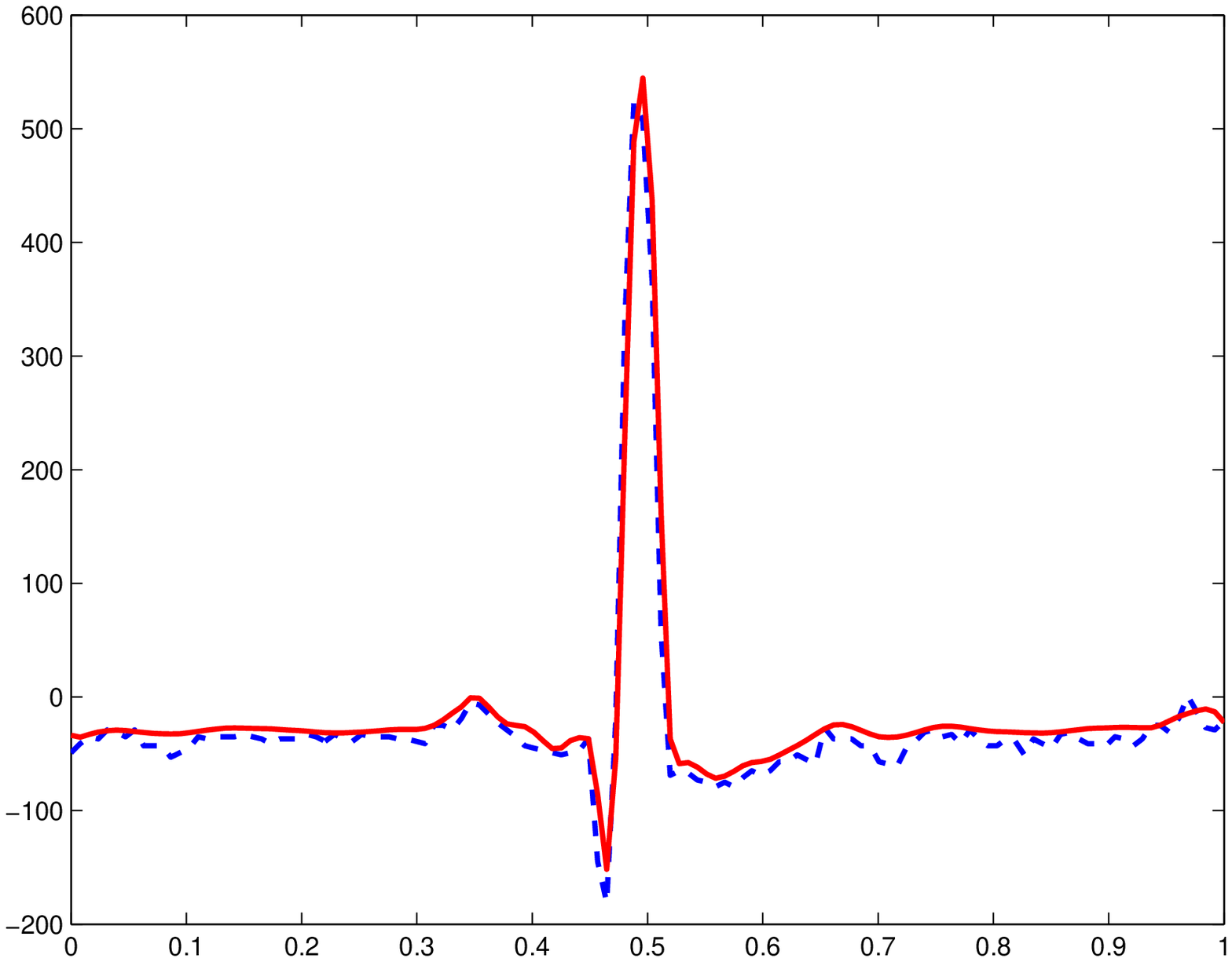} }
\subfigure[]{ \includegraphics[width=3.5cm]{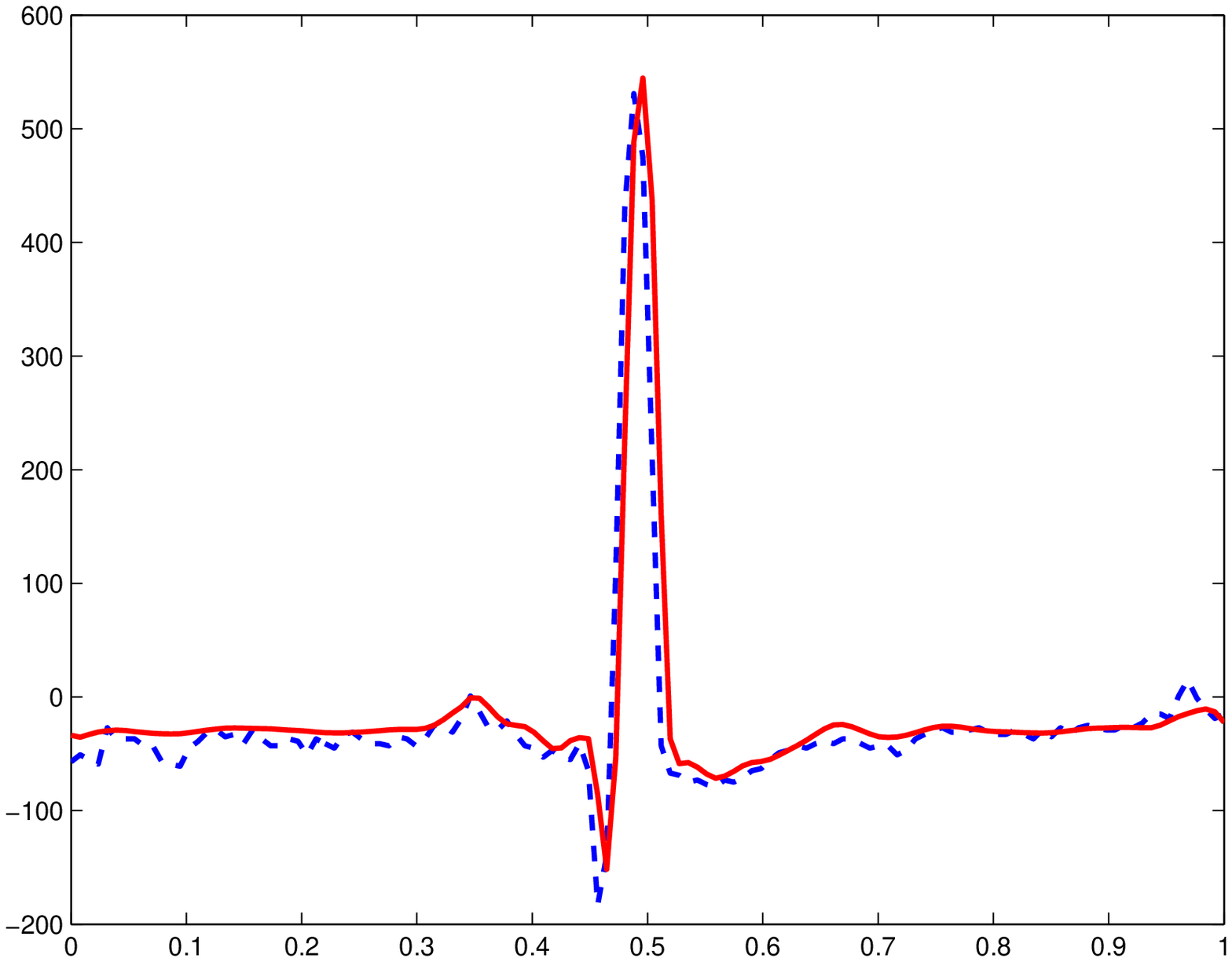} }
\subfigure[]{ \includegraphics[width=3.5cm]{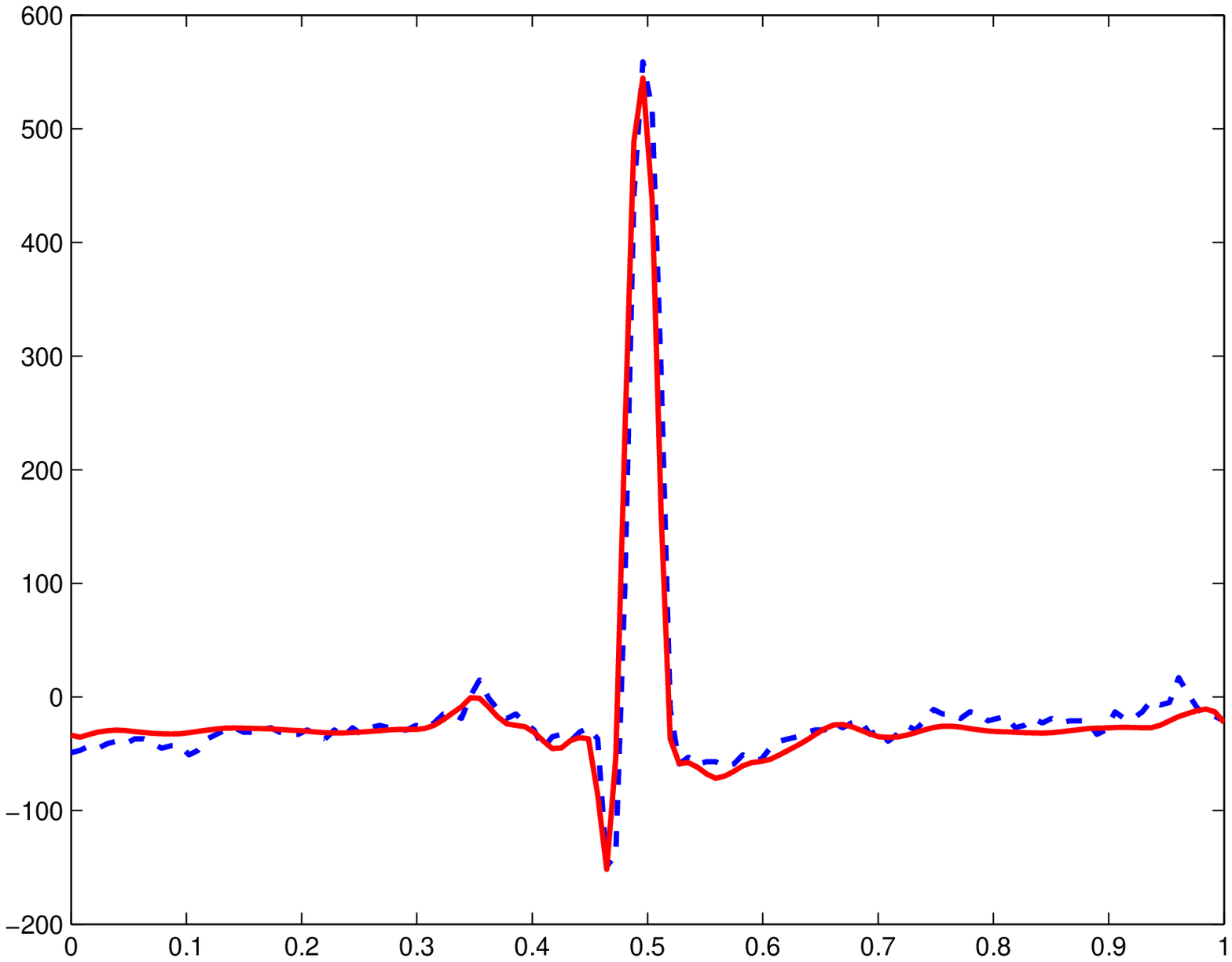} }
\caption{Normal case: (a) Fr残het mean using translation operators of the $J=93$  signals after segmentation of the ECG record. (b)-(g) Superposition of six signals containing a single QRS complex (dashed curves) with the Fr残het mean using translation operators (solid curve).  } \label{fig:Normal:shift}
\end{figure}

\begin{figure}[htbp]
\centering
\subfigure[]{ \includegraphics[width=3.5cm]{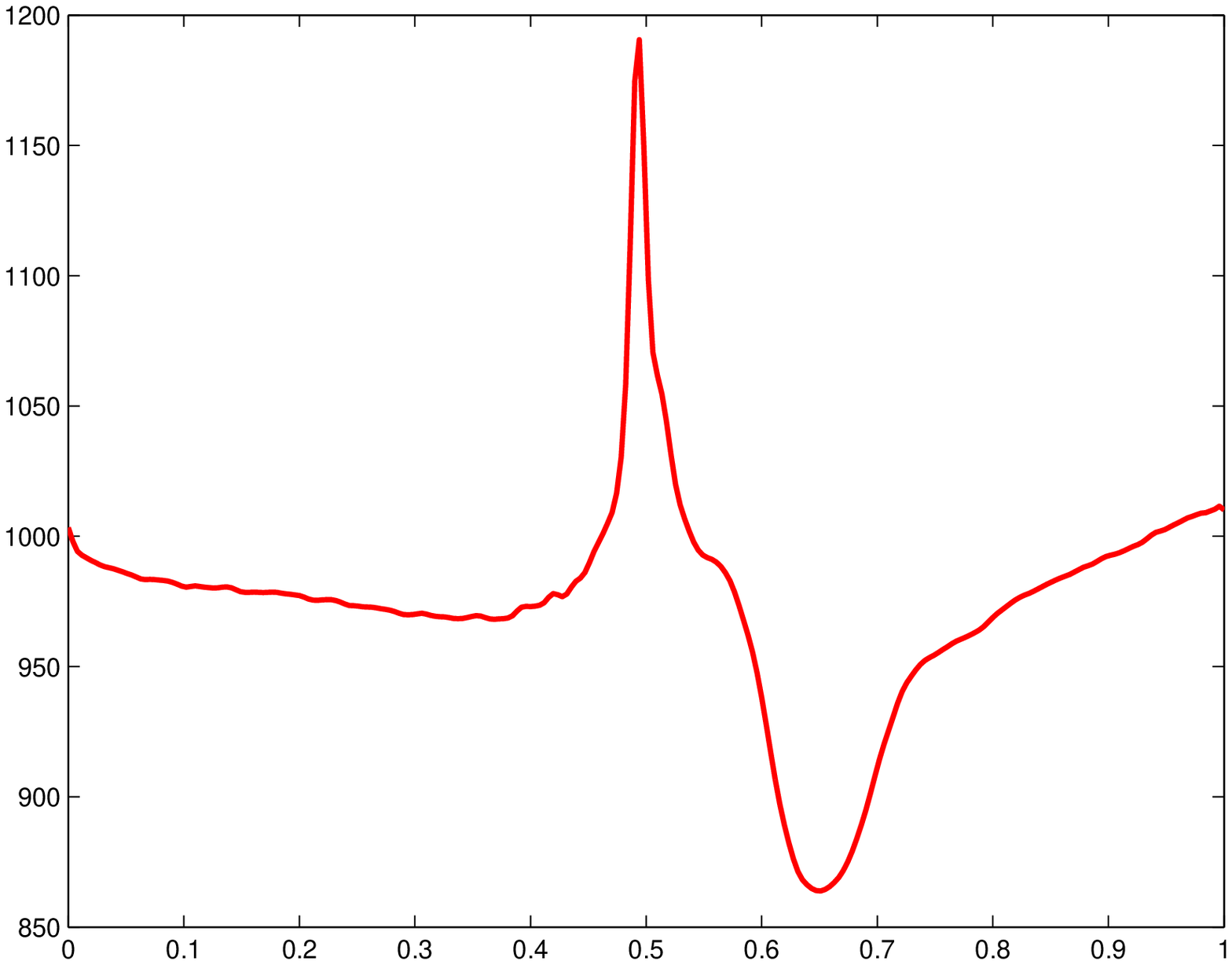} }
\hspace{0.2cm}
\subfigure[]{ \includegraphics[width=3.5cm]{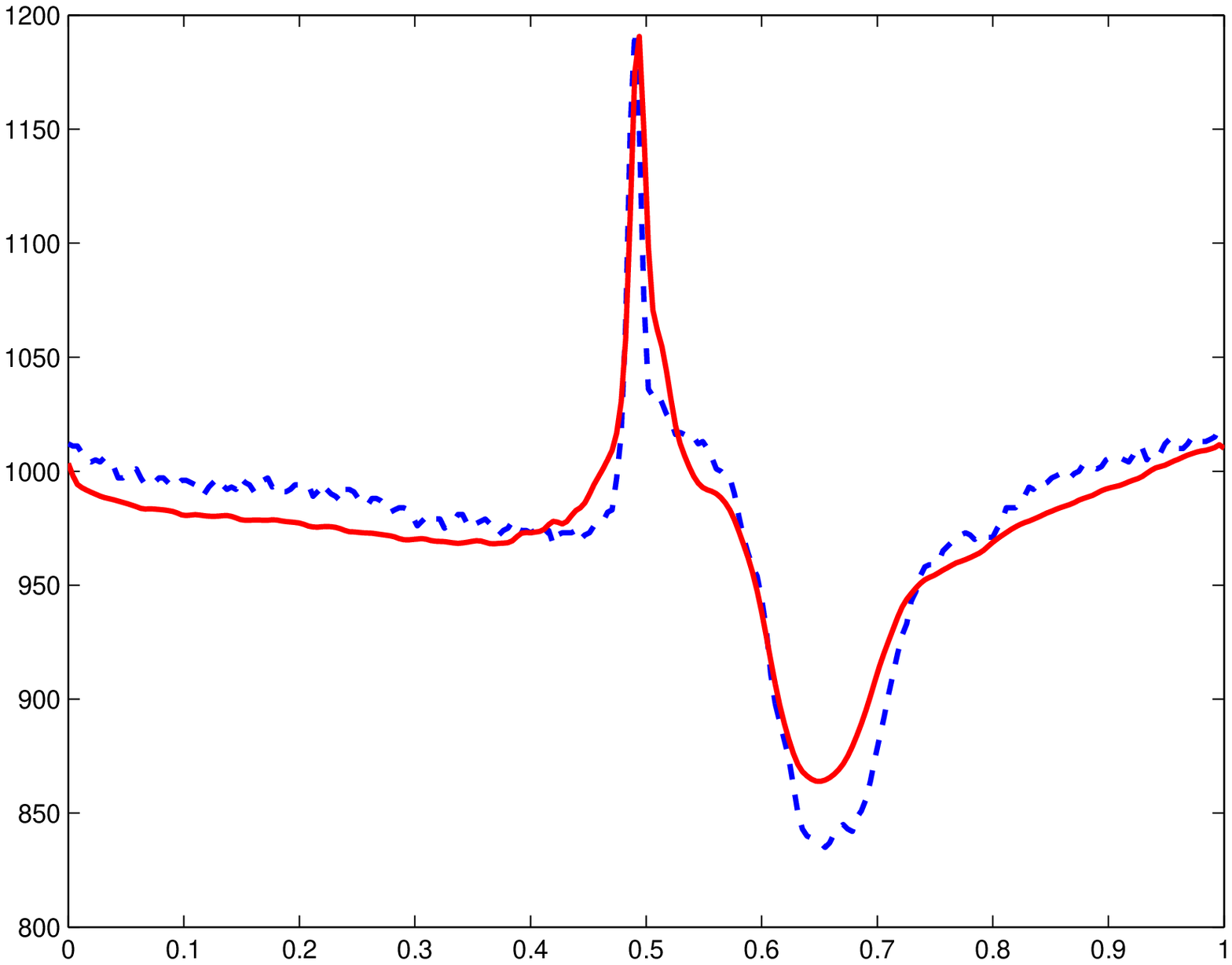} }
\subfigure[]{ \includegraphics[width=3.5cm]{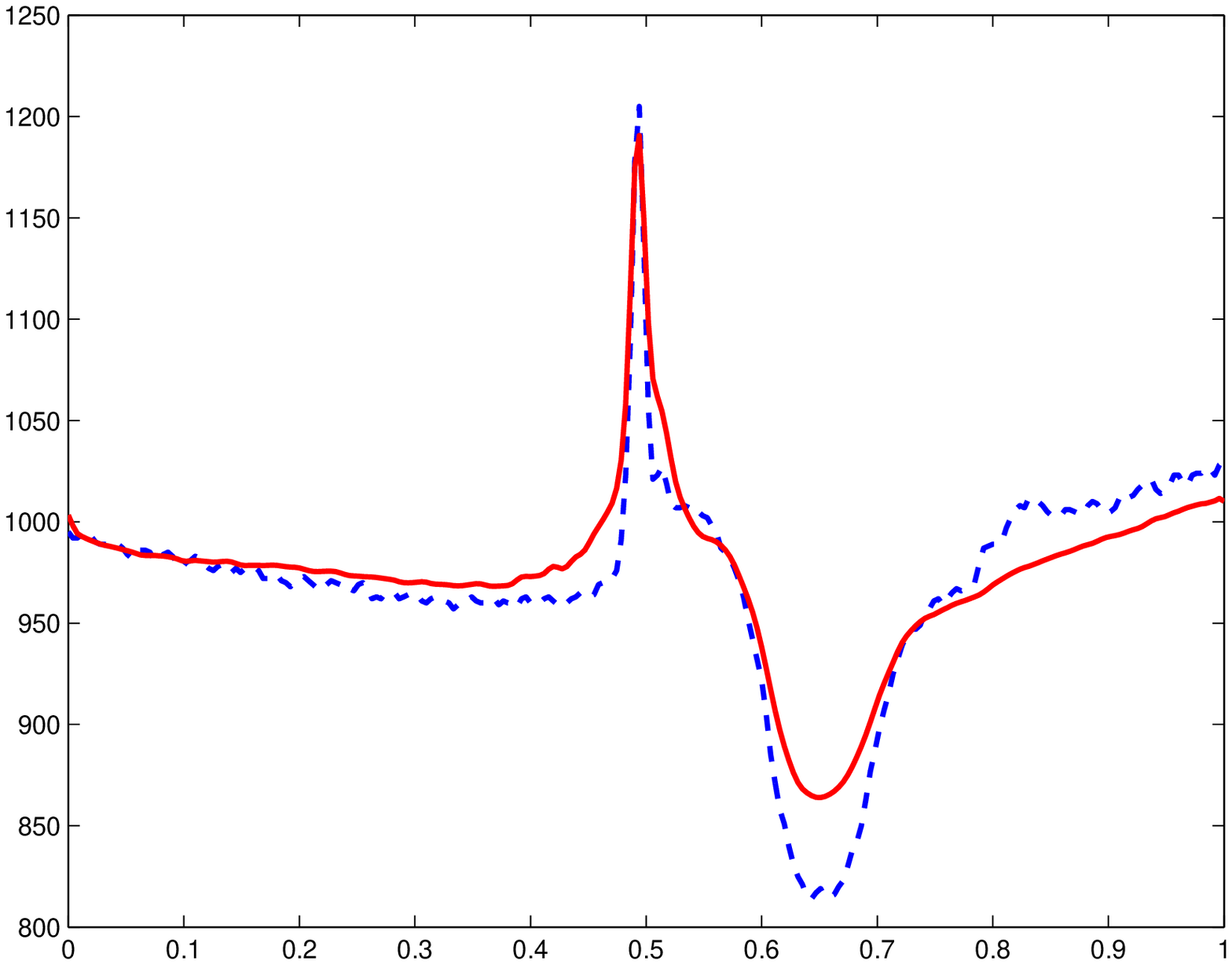} }
\subfigure[]{ \includegraphics[width=3.5cm]{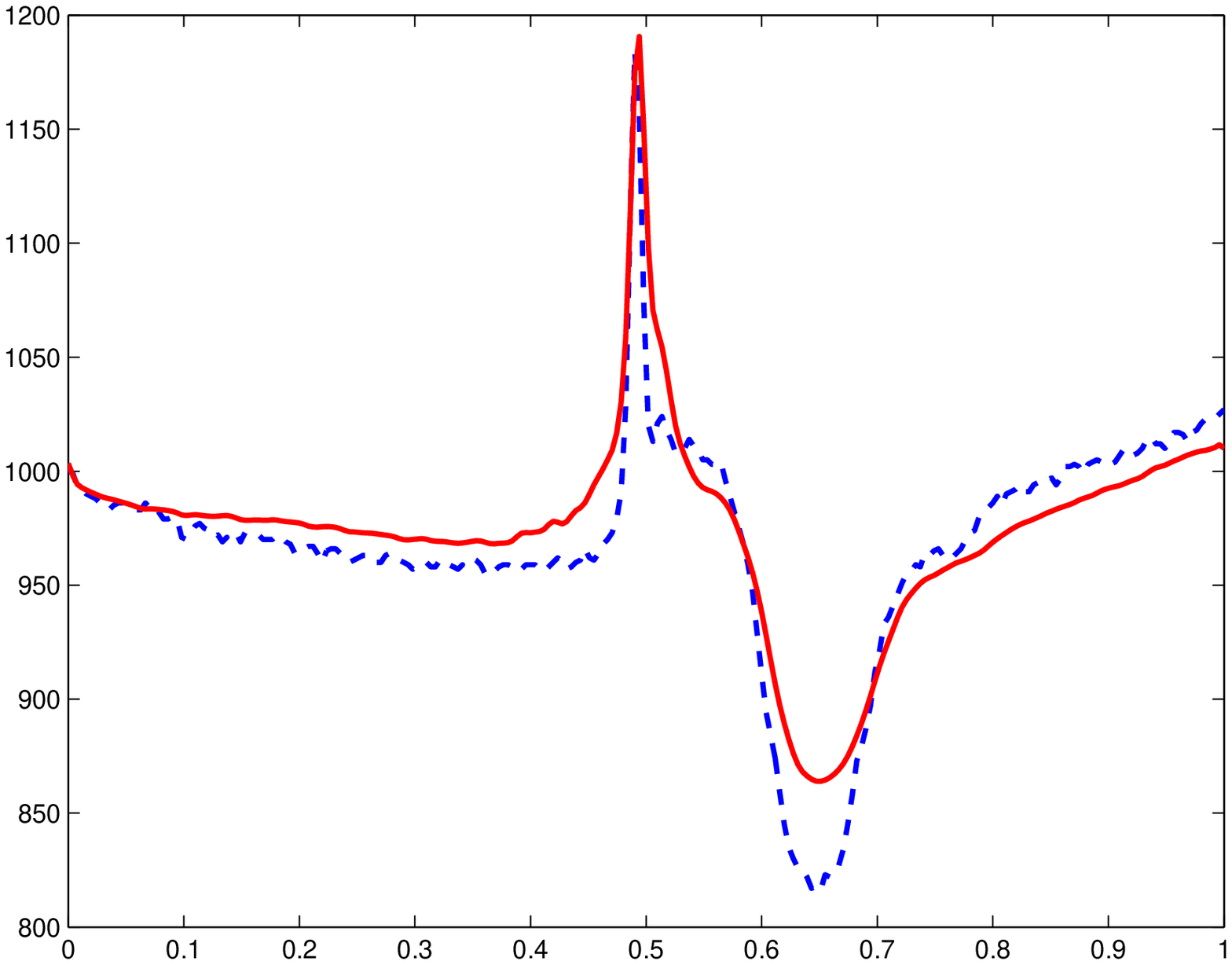} }

\hspace{4.1cm}
\subfigure[]{ \includegraphics[width=3.5cm]{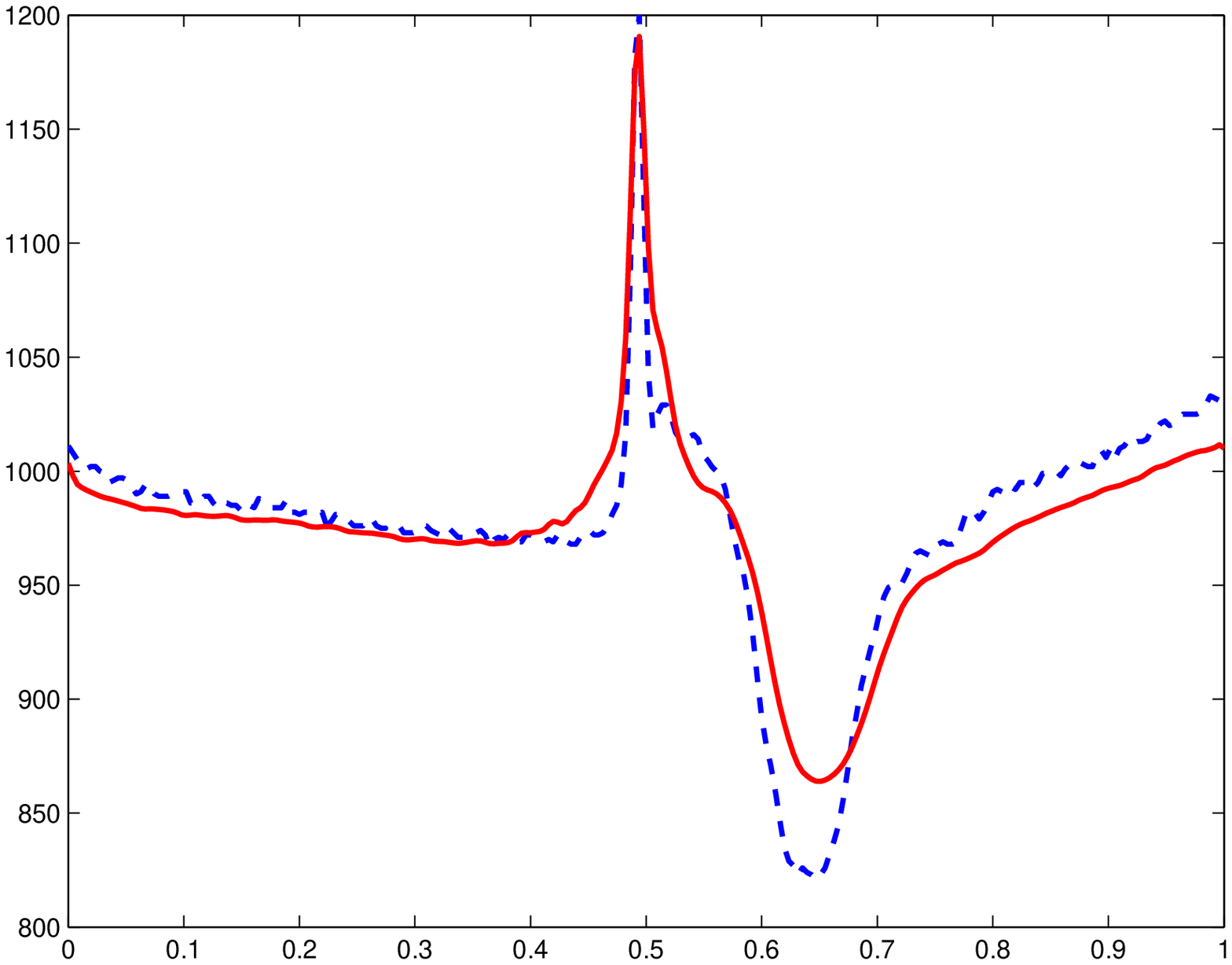} }
\subfigure[]{ \includegraphics[width=3.5cm]{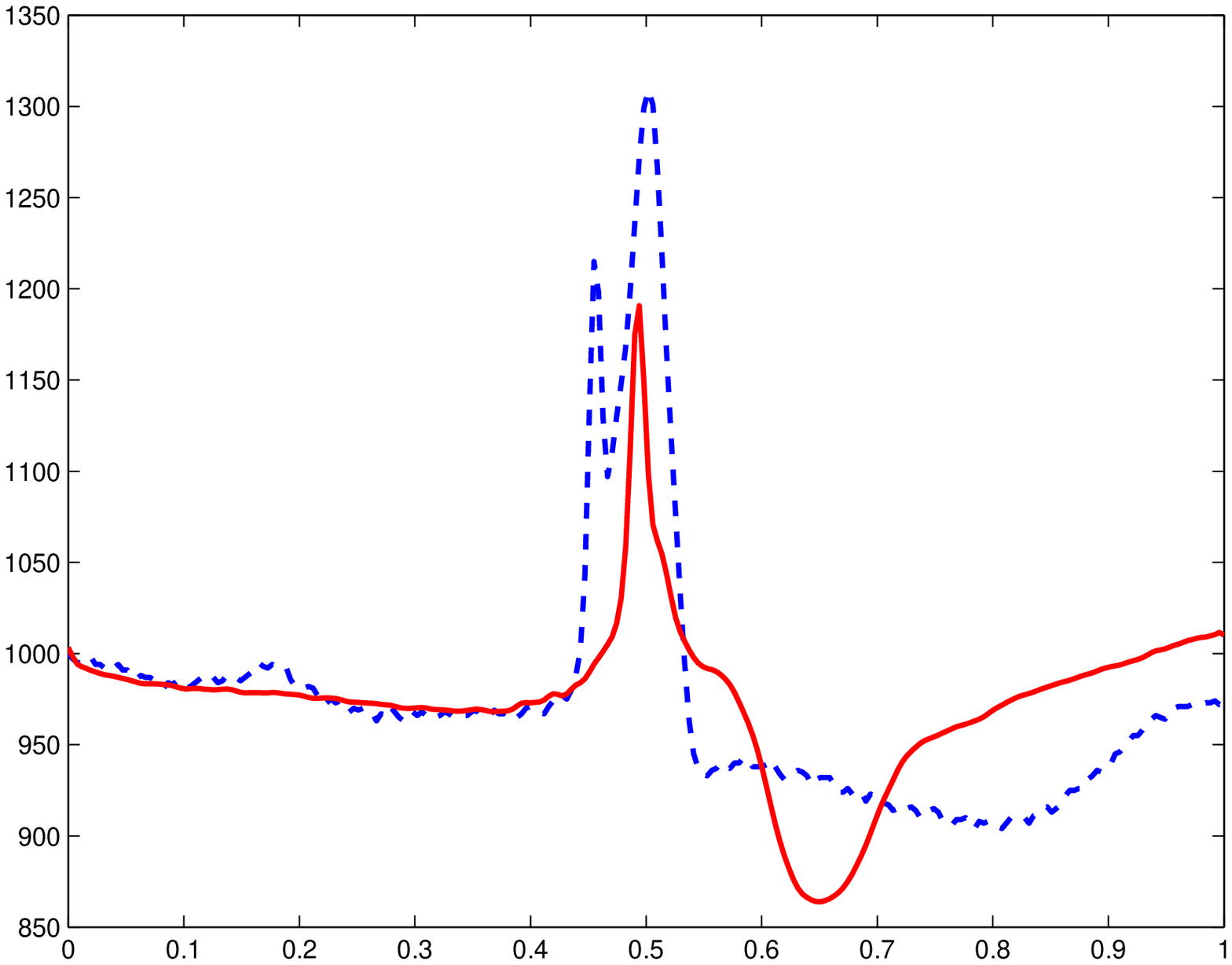} }
\subfigure[]{ \includegraphics[width=3.5cm]{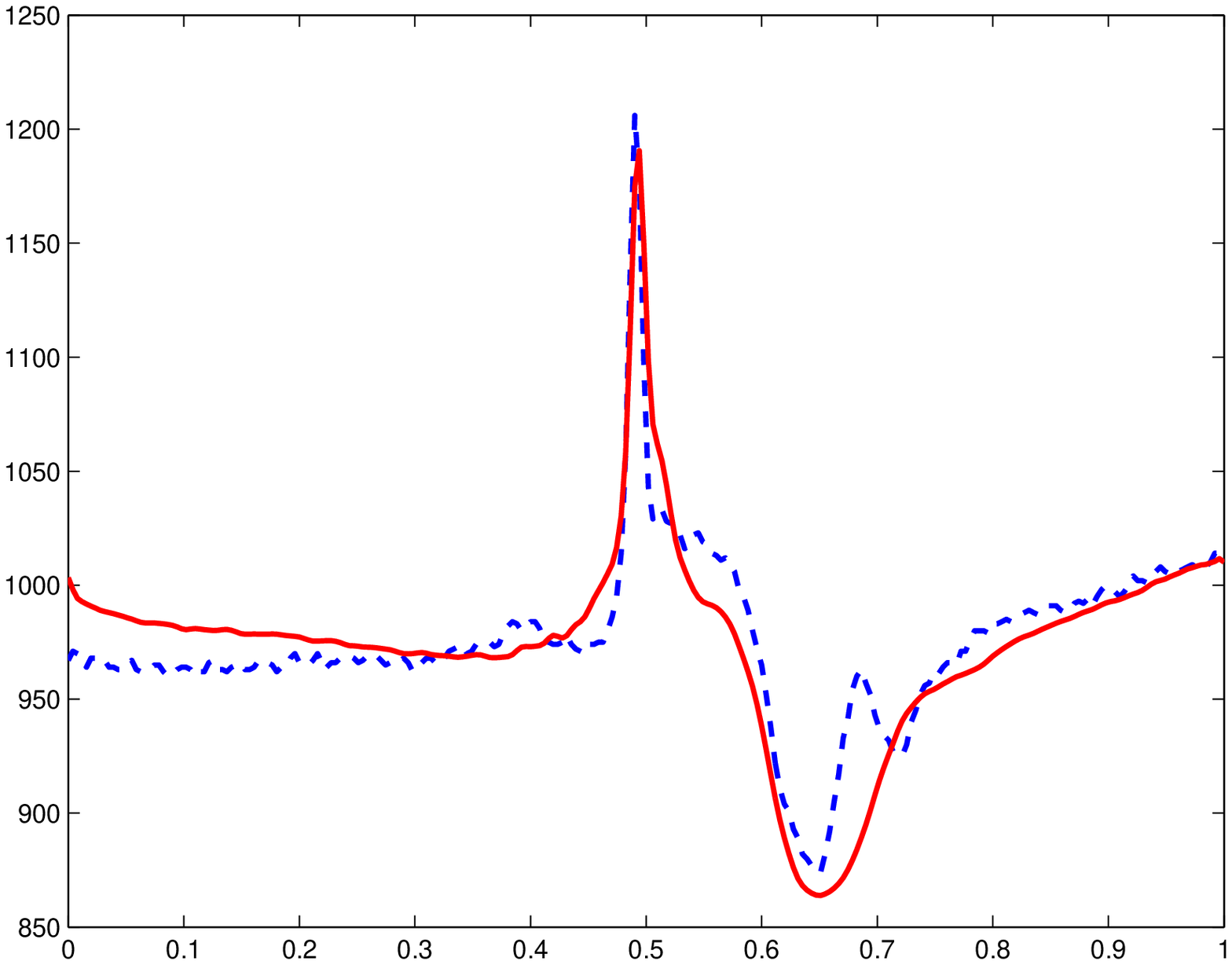} }

\caption{Case of cardiac arrhythmia: (a) Fr残het mean using translation operators of the $J=72$  signals after segmentation of the ECG record. (b)-(g) Superposition of six signals containing a single QRS complex (dashed curves) with the Fr残het mean using translation operators (solid curve).  } \label{fig:Arrhythmia:shift}
\end{figure}

The computation of smoothed Fr残het means using  translation operators in the normal and cardiac arrhythmia cases are displayed in Figure \ref{fig:Normal:shift} and Figure \ref{fig:Arrhythmia:shift}. The results obtained using the Procrustes mean are very similar, and they are thus not reported. Note that using Fr残het or Procrustes means  with translation operators is very similar to the alignment of signals using cross-correlation which is the widely used technique for signal averaging in ECG data analysis.

One can see that, for the normal case, the results are very satisfactory as the mean heart cycle displayed in Figure \ref{fig:Normal:shift}(a) is a good estimation of the typical shape of the signals displayed in Figure \ref{fig:Normal:shift}(b)-(g). For the case of cardiac arrhythmia, the results are not very satisfactory since there is still a low-pass filtering effect on the shape of the QRS complex (around the time point $t \approx 0.45$) in the mean heart cycle displayed in  Figure \ref{fig:Arrhythmia:shift}(a). In this case, the Fr残het mean does not represent very well the shape of the signals in Figure \ref{fig:Arrhythmia:shift}(b)-(g). This  low-pass filtering effect is due to a much larger variability in shape of the signals (after segmentation) in the case of cardiac arrhythmia than in the normal case. Since the activity of the heart can be very irregular in  cases of cardiac arrhythmia, modeling shape variability using only translation operators is not flexible enough. A more precise alignment to take into account a local variability in lag and duration of the heart beats is thus needed.

\begin{figure}[htbp]
\centering
\subfigure[]{ \includegraphics[width=3.5cm]{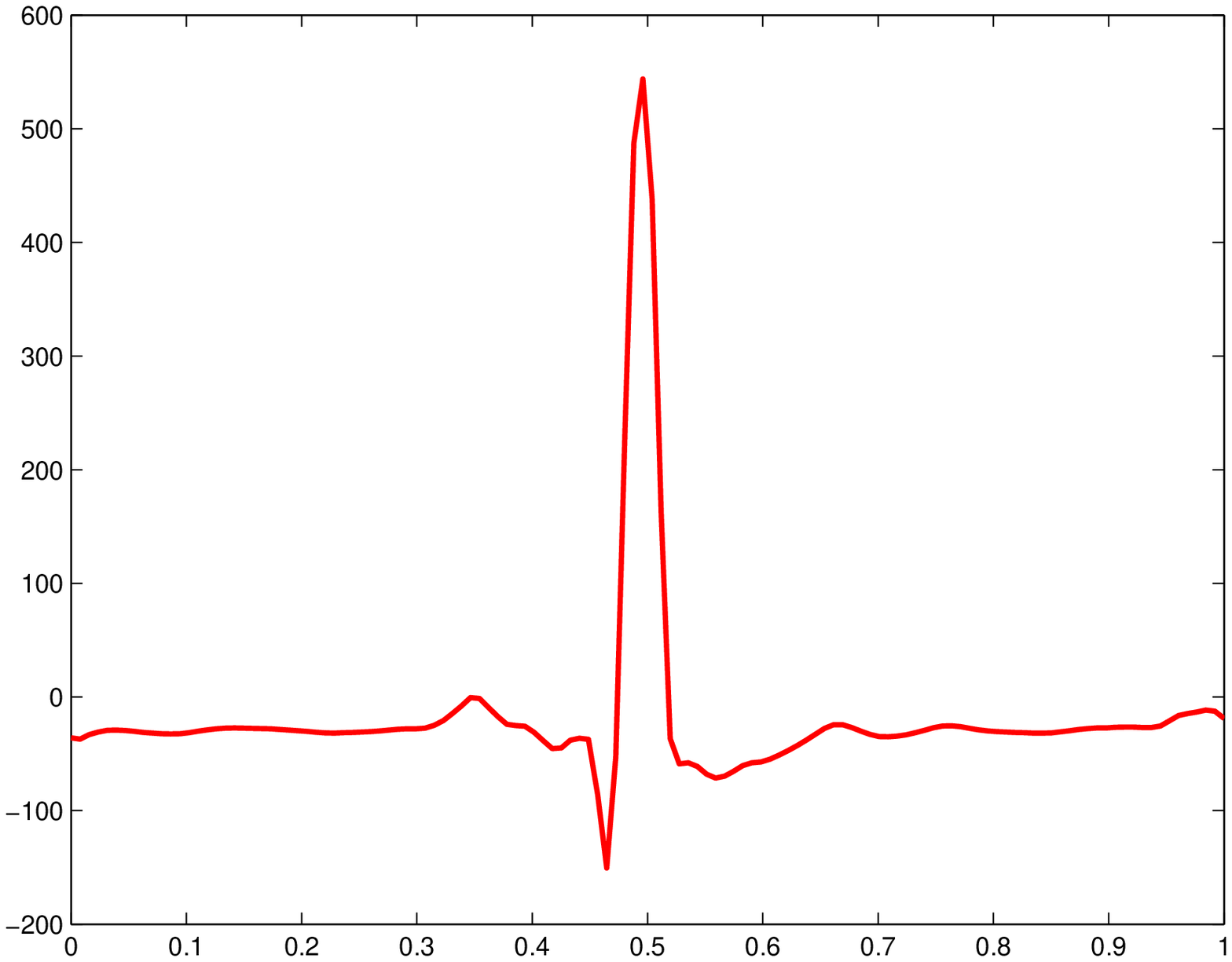} }
\hspace{0.2cm}
\subfigure[]{ \includegraphics[width=3.5cm]{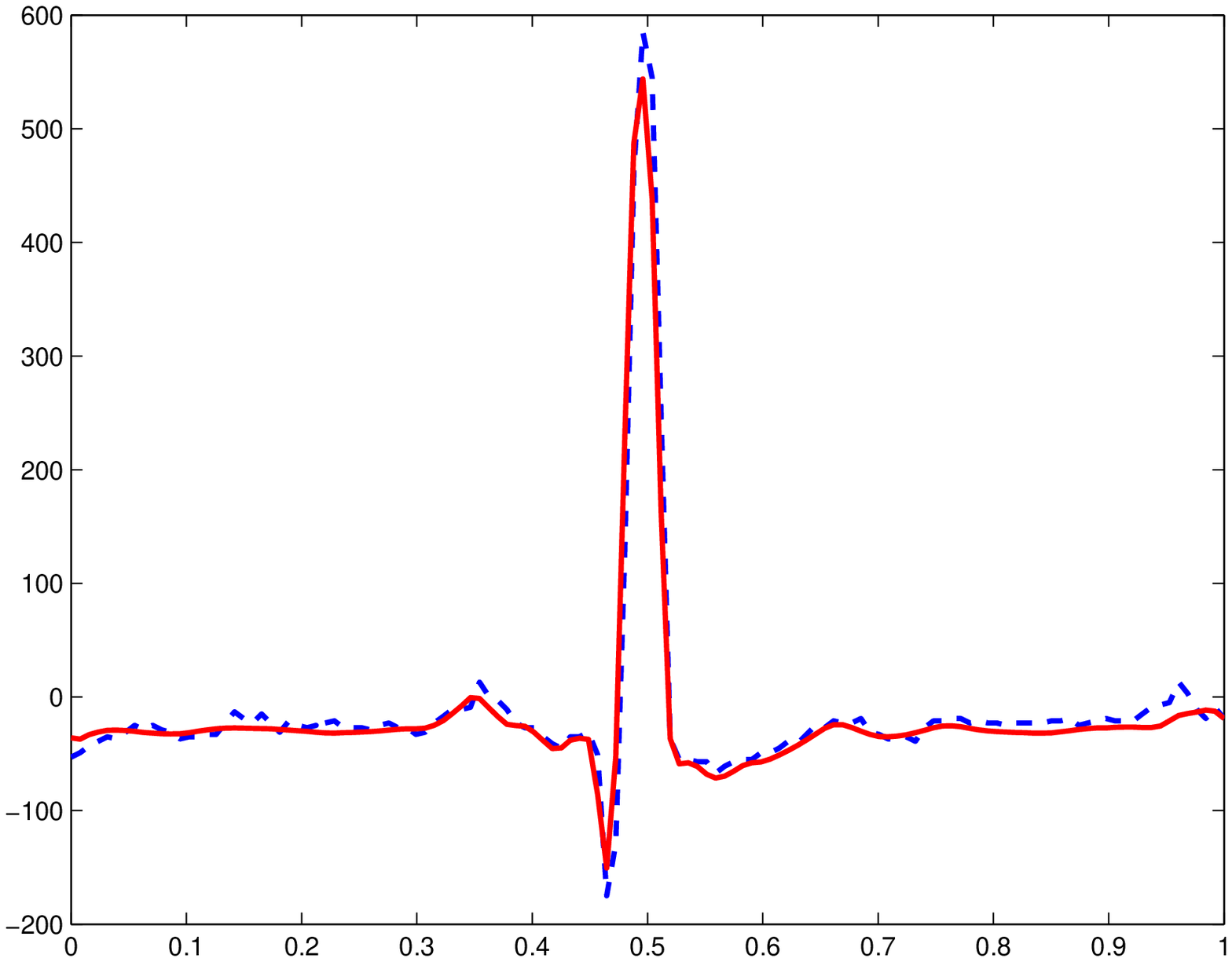} }
\subfigure[]{ \includegraphics[width=3.5cm]{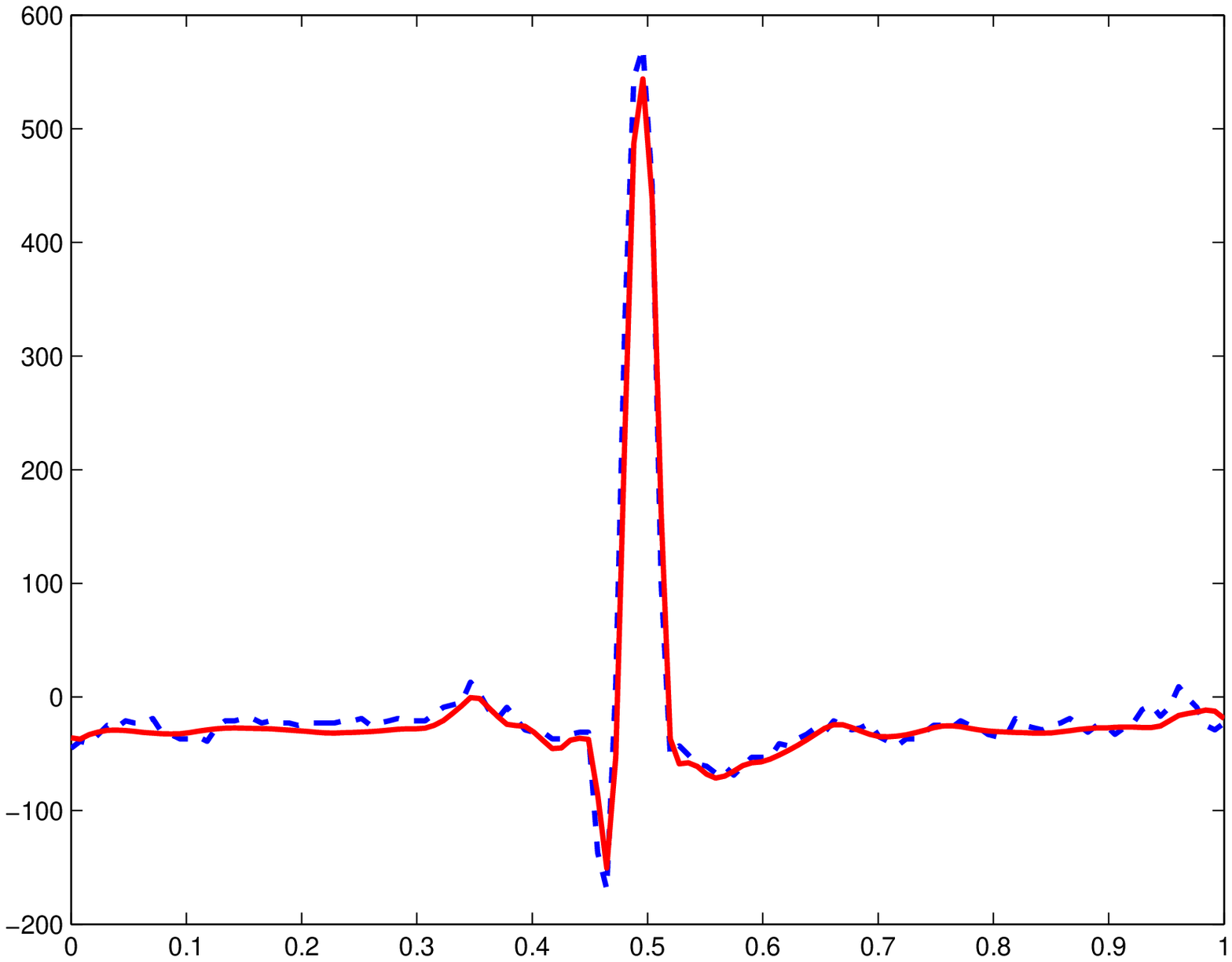} }
\subfigure[]{ \includegraphics[width=3.5cm]{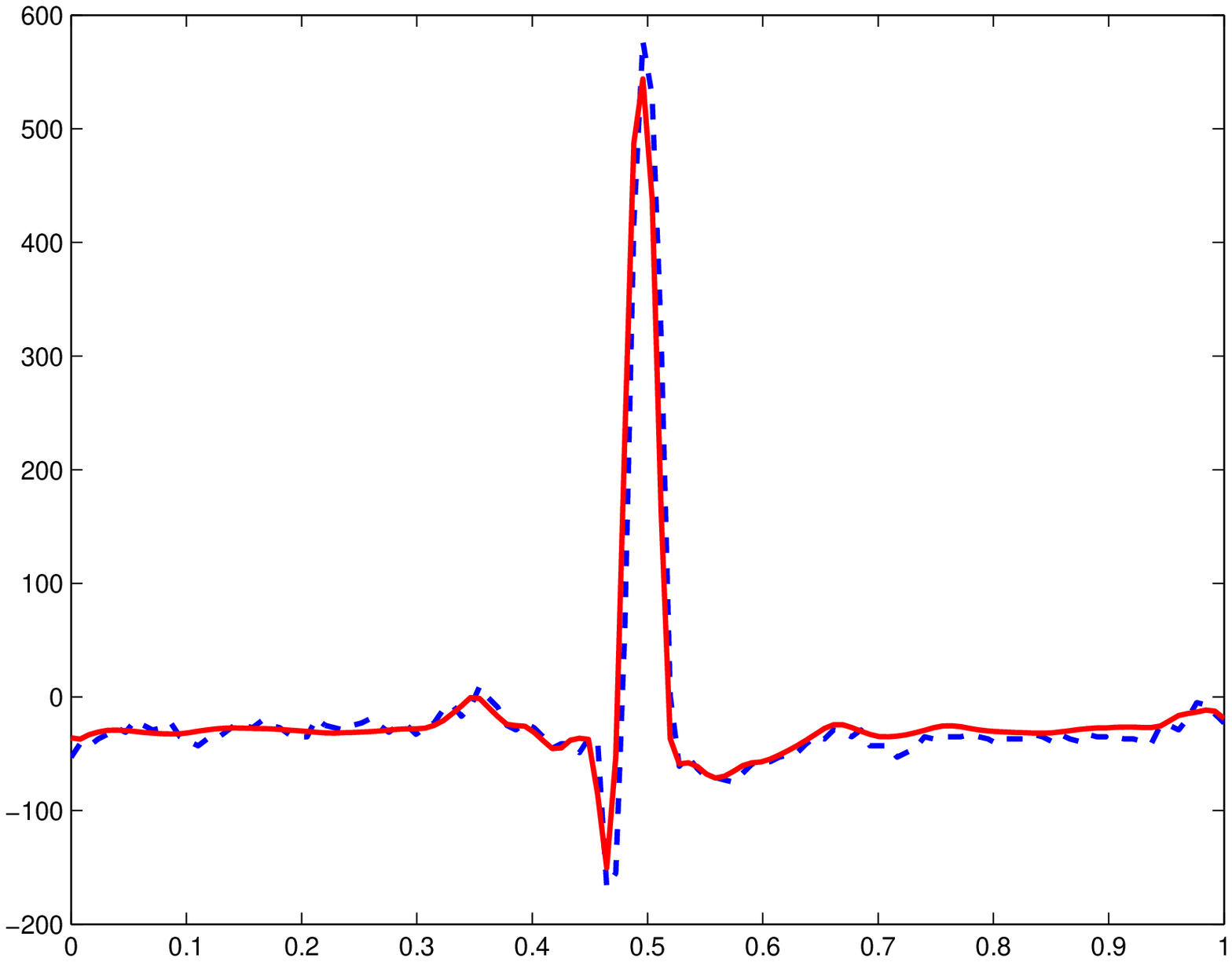} }

\hspace{4.1cm}
\subfigure[]{ \includegraphics[width=3.5cm]{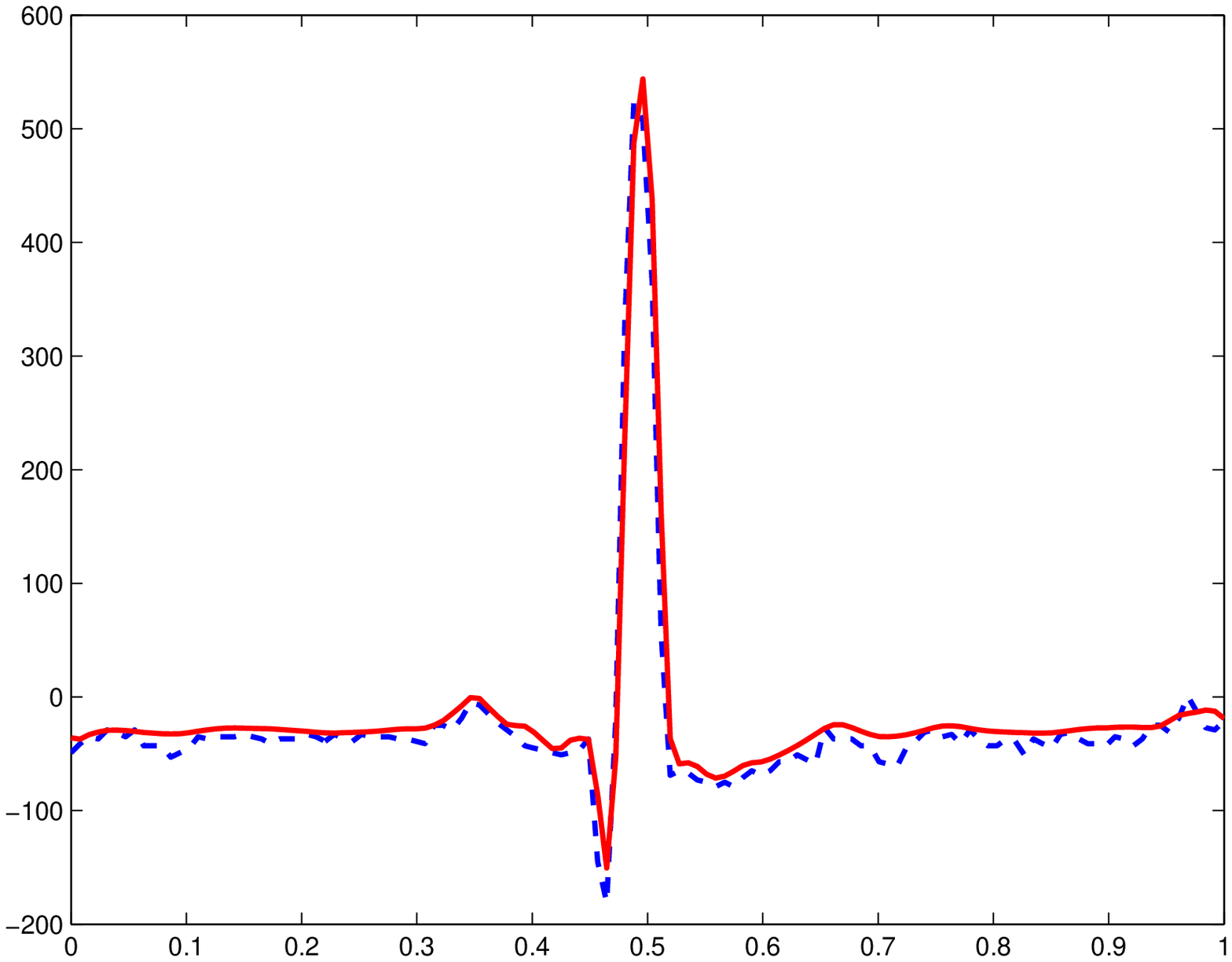} }
\subfigure[]{ \includegraphics[width=3.5cm]{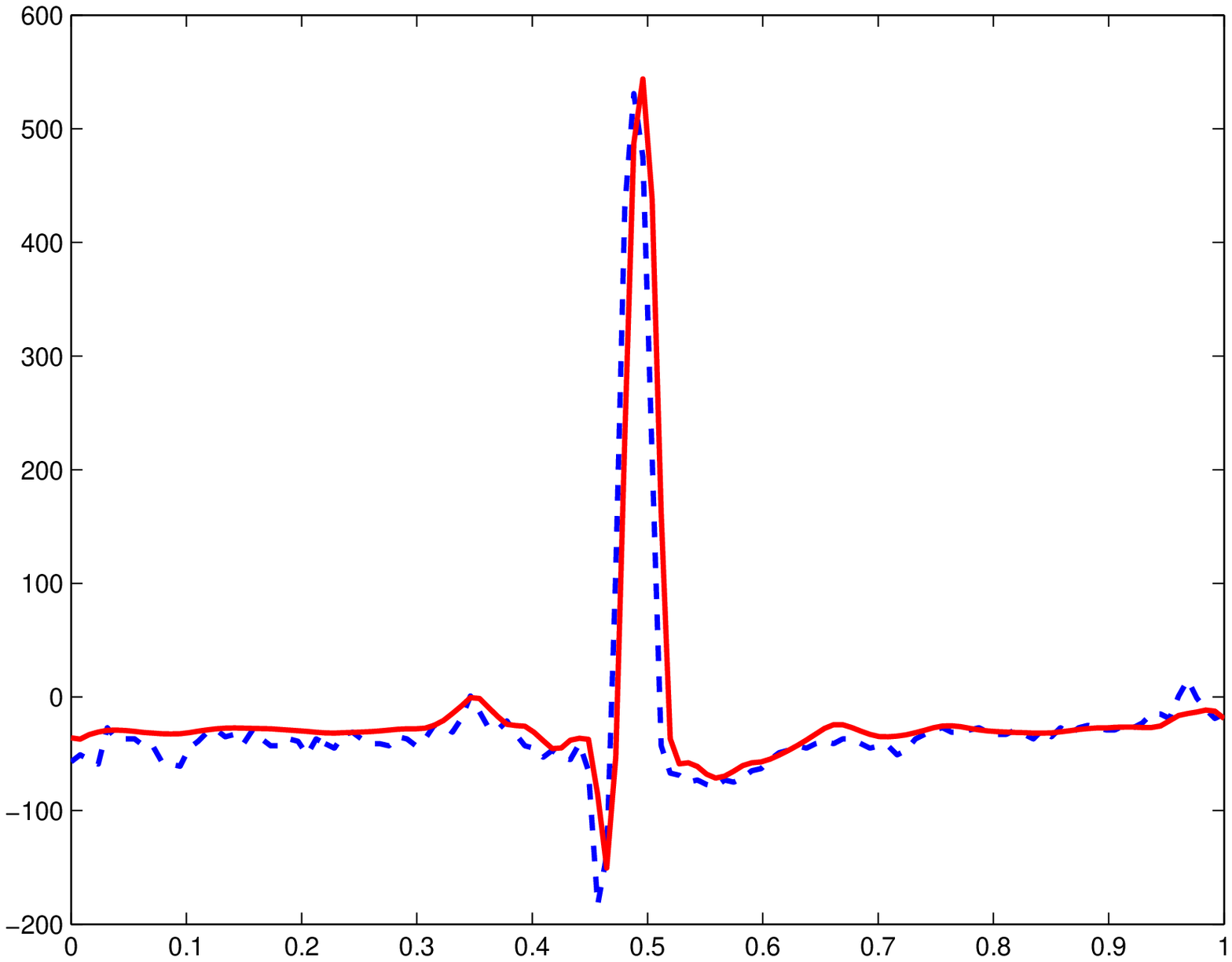} }
\subfigure[]{ \includegraphics[width=3.5cm]{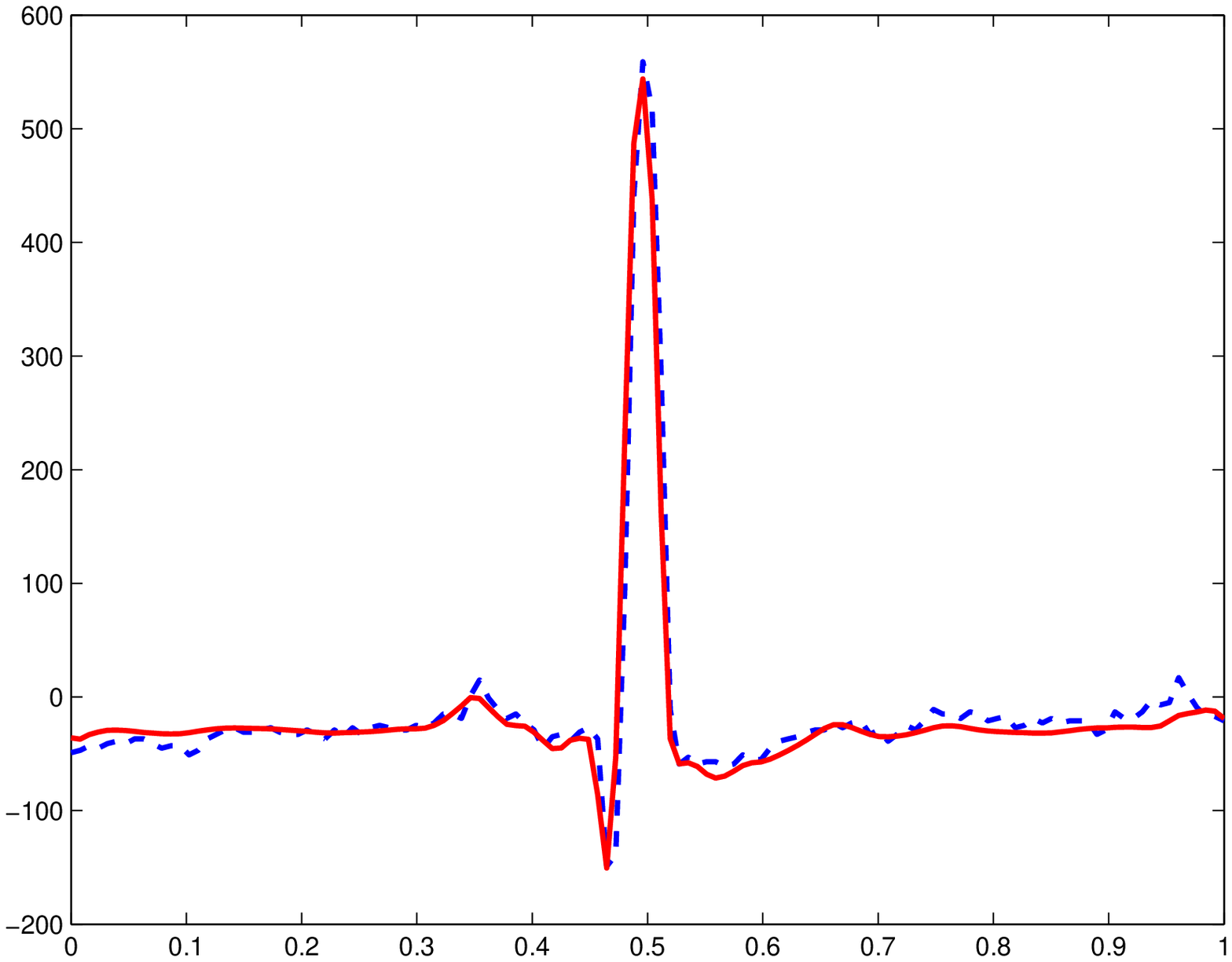} }

\caption{Normal case: (a) Fr残het mean using non-rigid operators of the $J=93$  signals after segmentation of the ECG record. (b)-(g) Superposition of six signals containing a single QRS complex (dashed curves) with the Fr残het mean using non-rigid operators (solid curve).  } \label{fig:Normal:diffeo}
\end{figure}

\begin{figure}[htbp]
\centering
\subfigure[]{ \includegraphics[width=3.5cm]{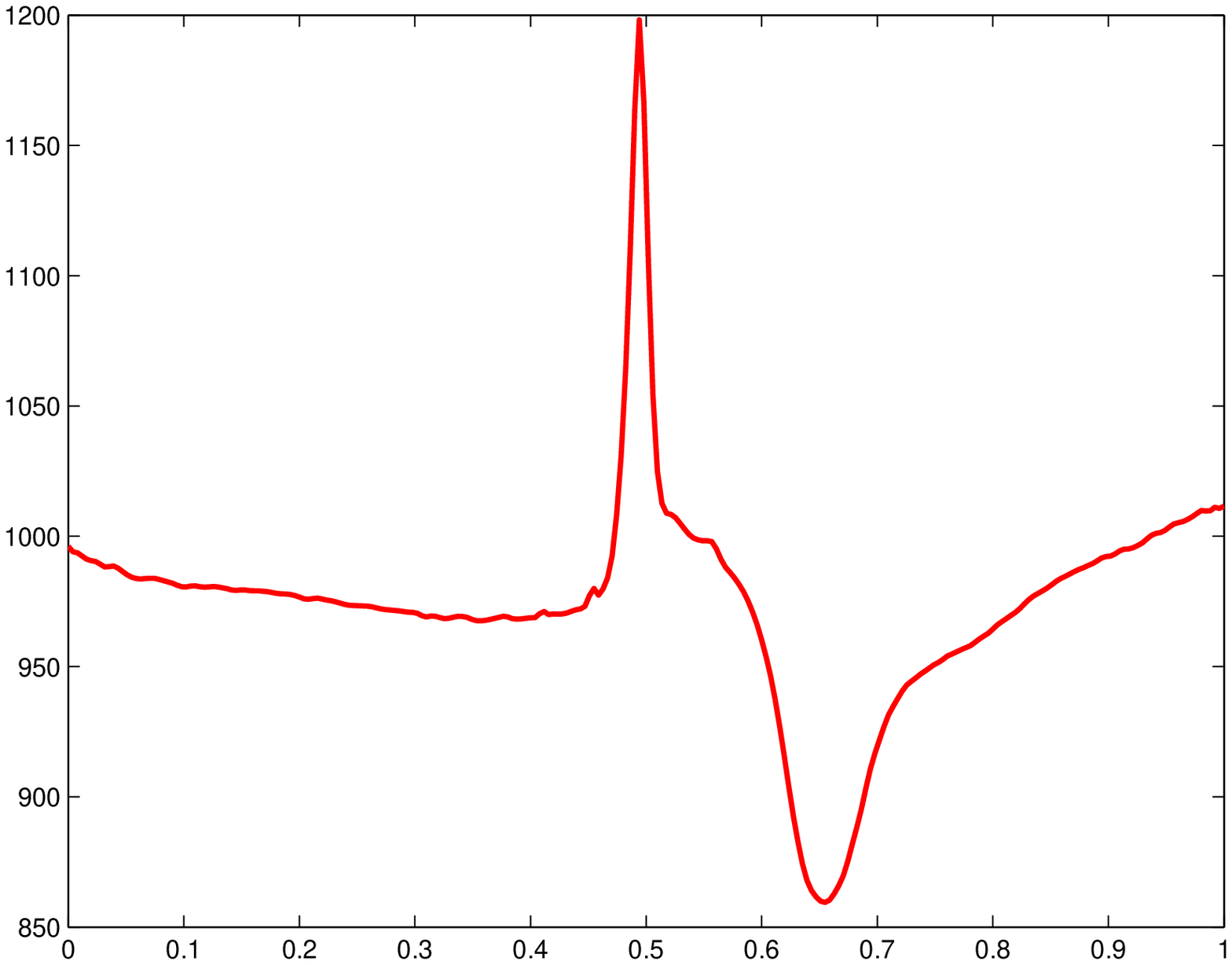} }
\hspace{0.2cm}
\subfigure[]{ \includegraphics[width=3.5cm]{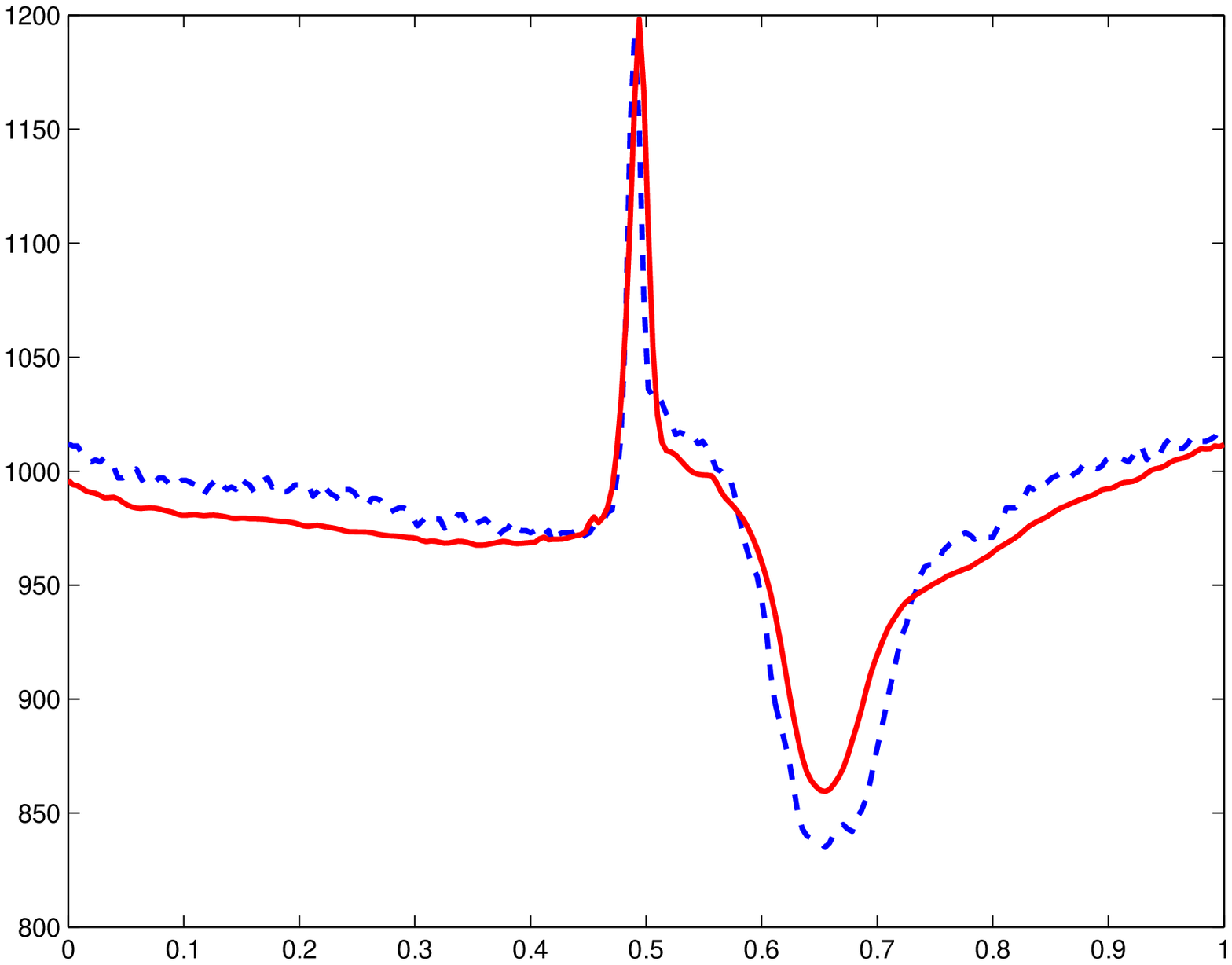} }
\subfigure[]{ \includegraphics[width=3.5cm]{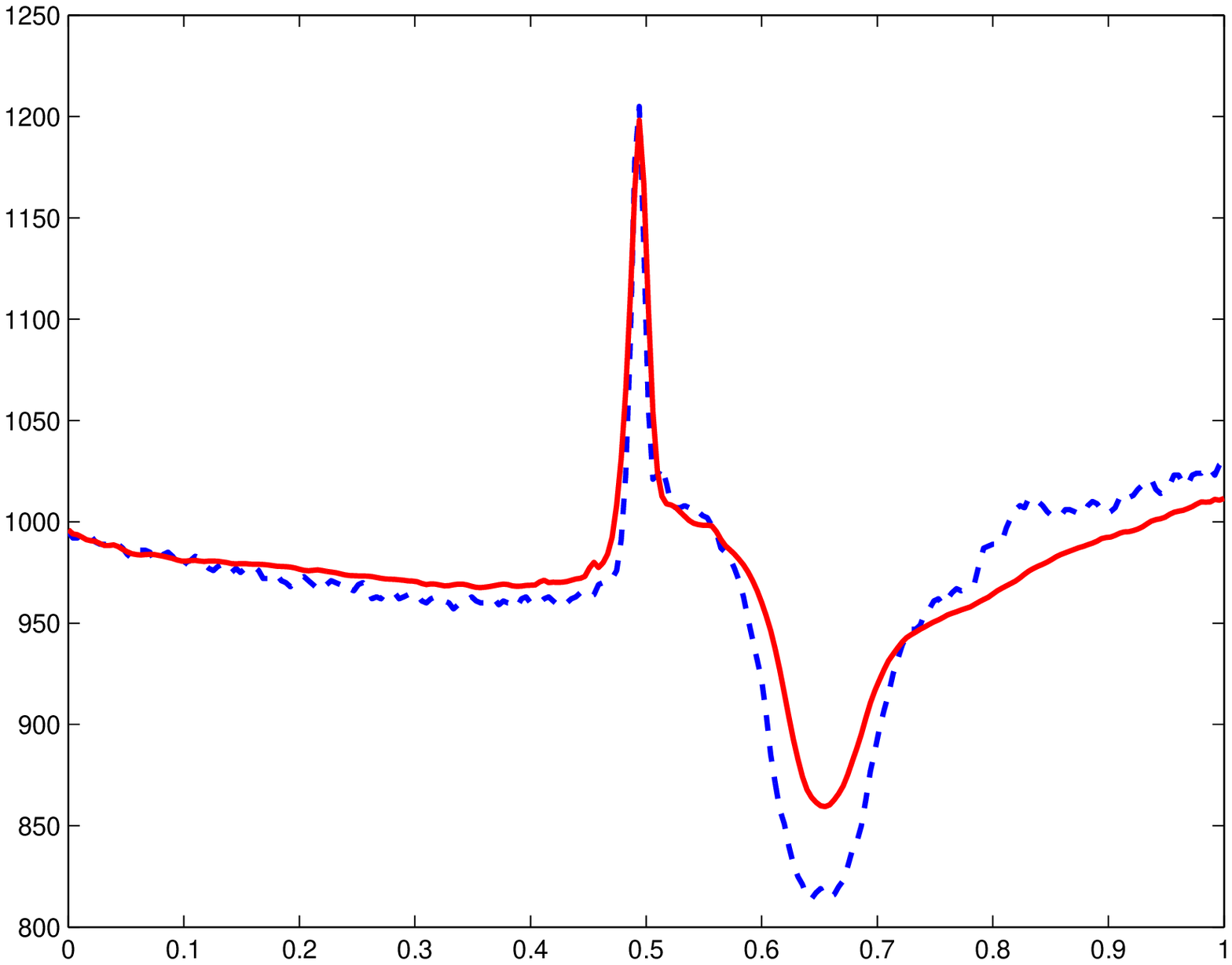} }
\subfigure[]{ \includegraphics[width=3.5cm]{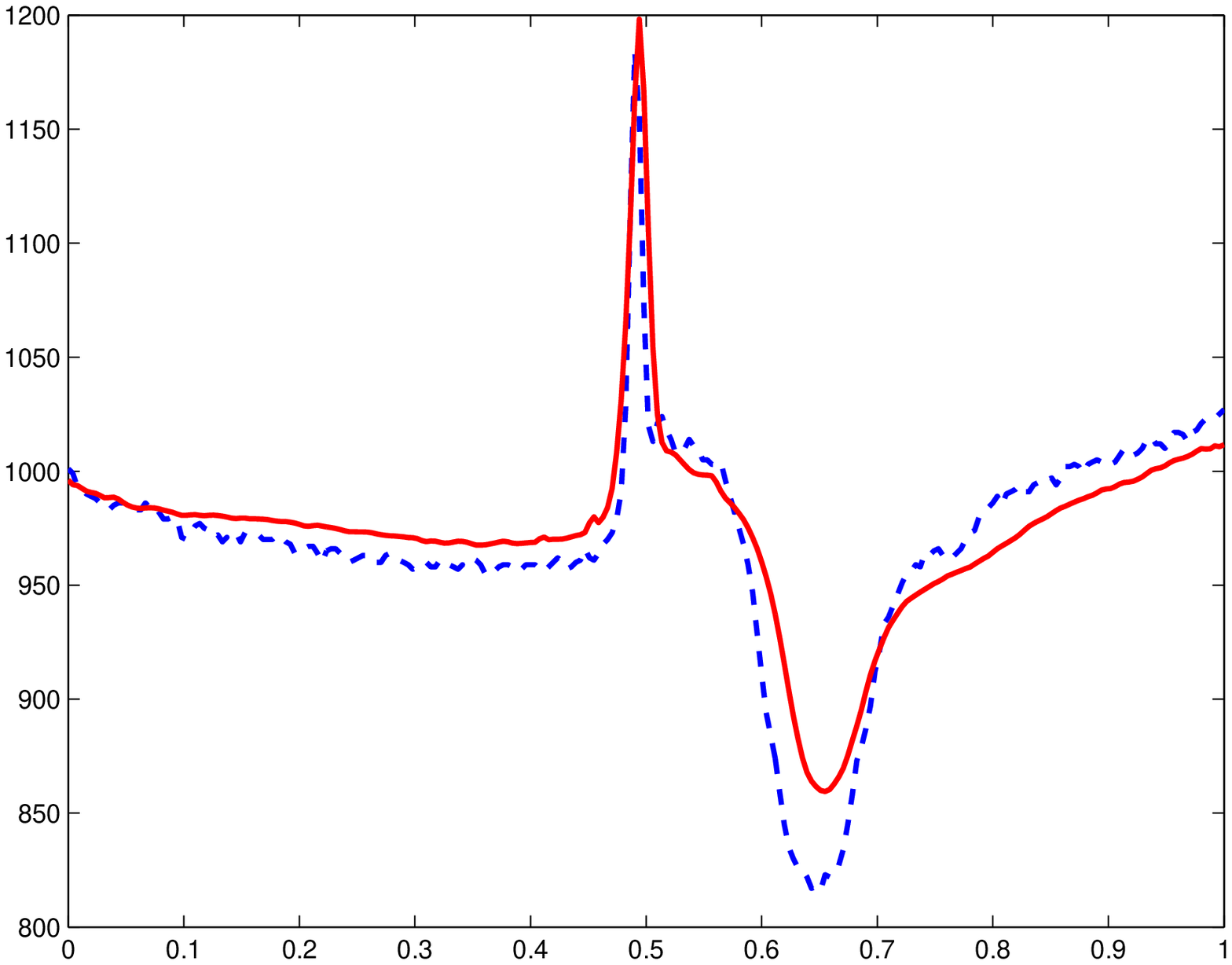} }

\hspace{4.1cm}
\subfigure[]{ \includegraphics[width=3.5cm]{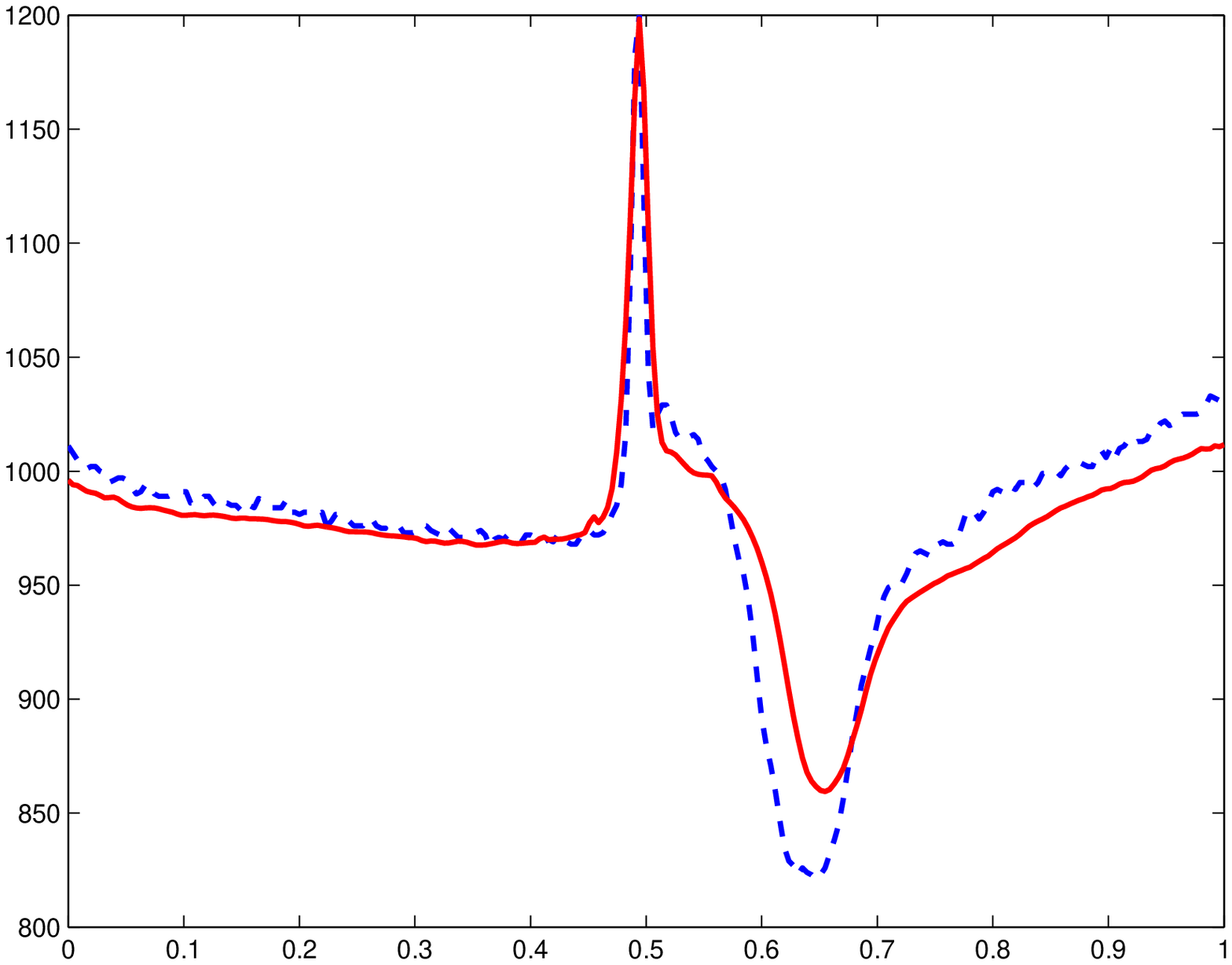} }
\subfigure[]{ \includegraphics[width=3.5cm]{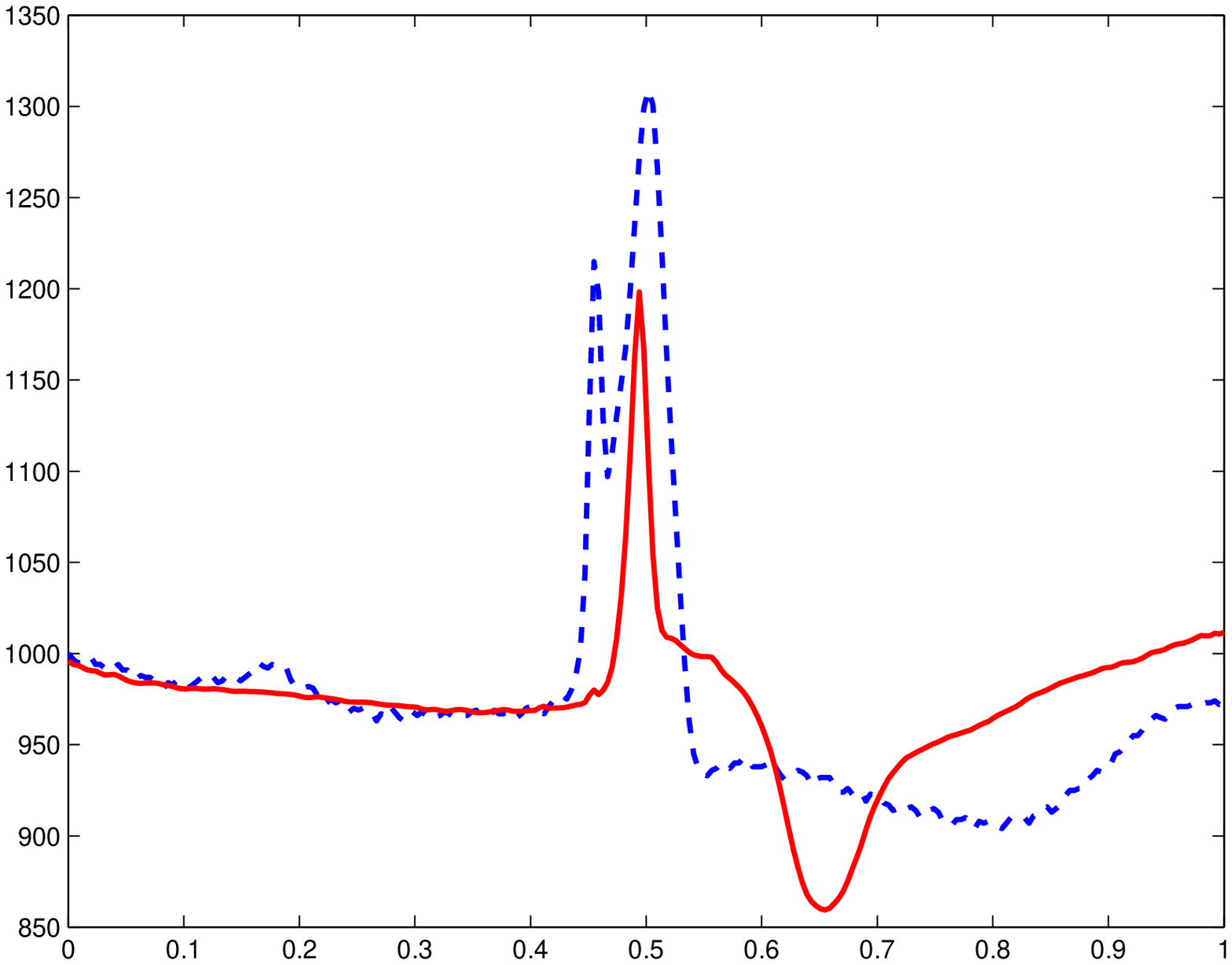} }
\subfigure[]{ \includegraphics[width=3.5cm]{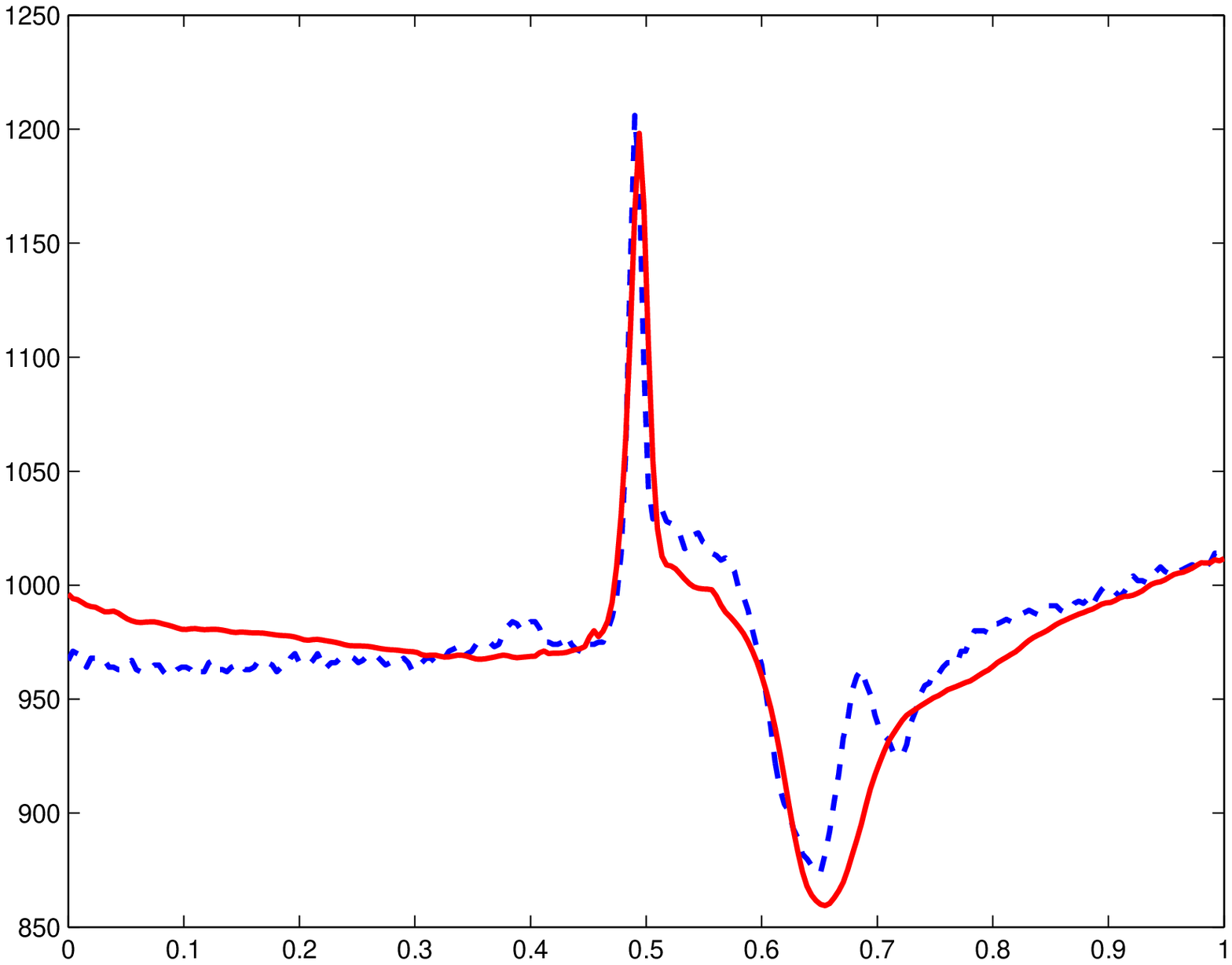} }

\caption{Case of cardiac arrhythmia: (a) Fr残het mean using non-rigid operators  operators of the $J=72$  signals after segmentation of the ECG record. (b)-(g) Superposition of six signals containing a single QRS complex (dashed curves) with the Fr残het mean using non-rigid operators (solid curve).  } \label{fig:Arrhythmia:diffeo}
\end{figure}

The computation of smoothed Fr残het means using  non-rigid operators in the normal and cardiac arrhythmia cases are displayed in Figure \ref{fig:Normal:diffeo} and Figure \ref{fig:Arrhythmia:diffeo}. In the normal case, the result is very similar to the one already obtained by the Fr残het mean using  translation operators, as shown by the comparison of Figure  \ref{fig:Normal:shift} and Figure \ref{fig:Normal:diffeo}. In the case of cardiac arrhythmia, the results displayed in Figure  \ref{fig:Arrhythmia:shift} and Figure \ref{fig:Arrhythmia:diffeo} clearly show the improvements obtained by using non-rigid deformation operators. The shape of the mean heart cycle in  Figure \ref{fig:Arrhythmia:diffeo}(a) is  a better estimation of the typical shape of the signals displayed in Figure \ref{fig:Arrhythmia:diffeo}(b)-(g). In particular, there is no low-pass filtering effect around the time point $t \approx 0.45$ in the shape of the Fr残het mean. 

\section{Discussion and conclusion}

We have presented a new algorithm for aligning heart beats extracted from an ECG record. Our approach is based on the notion of smoothed Fr残het means of curves using either translation or non-rigid deformation operators. In the case of a normal ECG obtained from a healthy subject, our approach yields a satisfactory mean heart cycle. Note that using Fr残het mean with translation operators is very similar  to the computation of a mean heart cycle  by aligning the data using cross-correlation which is the widely used approach in ECG data analysis  for signal averaging. When using non-rigid operators to align hearts beats having a high shape variability, with peaks showing an important variability in lag and duration from one pulse to another, our approach yields significant improvements over signal averaging by cross-correlation. The benefits of our procedure has been demonstrated for an ECG recording of a subject showing evidence of significant arrhythmia. We hope that the methods presented in this paper will stimulate further investigation into the development of better alignment procedures that take into account local shape variability in heart beats extracted from ECG records.

\bibliography{SignalAveragingECGData}
\bibliographystyle{alpha}

\end{document}